\documentclass[journal,12pt,onecolumn,draftclsnofootl, compsoc]{IEEEtran}
\usepackage{arxiv}
\usepackage[utf8]{inputenc} 
\usepackage[T1]{fontenc}    
\usepackage{hyperref}       
\usepackage{url}            
\usepackage{booktabs}       
\usepackage{amsfonts}       
\usepackage{nicefrac}       
\usepackage{microtype}      
\usepackage{cleveref}       
\usepackage{lipsum}         
\usepackage{graphicx}
\usepackage{acro}
\usepackage{graphicx}
\usepackage{caption}
\usepackage{etoolbox}
\usepackage{hyperref}
\usepackage{color}
\usepackage{wrapfig}
\usepackage{blindtext}
\usepackage{titlesec}
\usepackage[numbers]{natbib}
\usepackage{doi}

\makeatletter
\patchcmd{\theacronym}{\item[\@aclab]}{\item[\textbf{\@aclab}]\hfill}{}{}
\makeatother

\title{Everything You Wanted to Know About Consumer Light Management in Smart Energy}

\newif\ifuniqueAffiliation
\uniqueAffiliationtrue

\ifuniqueAffiliation 
\author
{
	\href{https://orcid.org/0000-0002-5653-3602}{\includegraphics[scale=0.06]{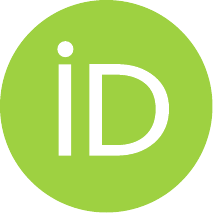}\hspace{1mm}Prajnyajit ~Mohanty} \\
	Dept. of Electronics and Communication  Engineering\\
	National Institute of Technology Rourkela\\
	India, 769008 \\
	\texttt{prajnyajitmohanty@gmail.com} 	
	\And
	\href{https://orcid.org/0000-0001-9805-25430}{\includegraphics[scale=0.06]{orcid.pdf}\hspace{1mm}Umesh C.~Pati} \\
	Dept. of Electronics and Communication  Engineering\\
	National Institute of Technology Rourkela \\
	India, 769008 \\
	\texttt{ucpati@nitrkl.ac.in} 	
	 \And
	\href{https://orcid.org/0000-0003-4917-7088}{\includegraphics[scale=0.06]{orcid.pdf}\hspace{1mm}Kamalakanta ~Mahapatra} \\
	Dept. of Electronics and Communication  Engineering\\
	National Institute of Technology Rourkela\\
	India, 769008 \\
	\texttt{kkm@nitrkl.ac.in} 
	\And
	\href{https://orcid.org/0000-0003-2959-6541}{\includegraphics[scale=0.06]{orcid.pdf}\hspace{1mm}Saraju P. ~Mohanty} \\
	Dept. of Computer Science and Engineering\\
	University of North Texas, USA\\
	\texttt{saraju.mohanty@unt.edu} 	 
	}

\fi

\renewcommand{\shorttitle} {\small Everything You Wanted to Know About Consumer Light Management in Smart Energy}

\hypersetup{
pdftitle={A template for the arxiv style},
pdfsubject={q-bio.NC, q-bio.QM},
pdfauthor={David S.~Hippocampus, Elias D.~Striatum},
pdfkeywords={First keyword, Second keyword, More},
}

\DeclareAcronym{ICT}{short = ICT, long= Information and Communication Technologies}
\DeclareAcronym{E-CPS}{short = E-CPS,long= Energy Cyber-Physical System}
\DeclareAcronym{IoT}{short = IoT, long= Internet of Things}
\DeclareAcronym{LED}{short = LED, long= Light Emitting Diode}
\DeclareAcronym{QoS}{short = QoS, long= Quality of Service}
\DeclareAcronym{AI}{short = AI, long= Artificial Intelligence}
\DeclareAcronym{ANN}{short = ANN, long= Artificial Neural Network}
\DeclareAcronym{CCT}{short = CCT, long= Correlated Color Temperature}
\DeclareAcronym{ABC}{short = ABC, long= Artificial Bee Colony}
\DeclareAcronym{UV}{short = UV, long= Ultraviolet}
\DeclareAcronym{Li-Fi}{short = Li-Fi, long=Light Fidelity}
\DeclareAcronym{LoRaWAN}{short = LoRaWAN, long= Long Range Wide Area Network}
\DeclareAcronym{DALI}{short = DALI, long= Digital Addressable Lighting Interface}
\DeclareAcronym{Wi-Fi}{short = Wi-Fi, long= Wireless Fidelity}
\DeclareAcronym{LoRa}{short = LoRa, long= Long Range}
\DeclareAcronym{GPRS}{short = GPRS, long= General Packet Rario Service}
\DeclareAcronym{WSN}{short = WSN, long= Wireless Sensor Network}
\DeclareAcronym{PWM}{short = PWM, long= Pulse Widht Modulation}
\DeclareAcronym{RF}{short = RF, long= Radio Frequency}
\DeclareAcronym{NFC}{short = NFC, long= Near Field Communication}
\DeclareAcronym{MPPT}{short = MPPT, long= Maximum Power Point Tracking}
\DeclareAcronym{PO}{short = PO, long= Perturb and Observe}
\DeclareAcronym{FOCV}{short = FOCV, long= Fractional Open-Circuit Voltage}
\DeclareAcronym{NN}{short = NN, long= Neural Network}
\DeclareAcronym{GA}{short = GA, long= Generic Algorithm}
\DeclareAcronym{PVD}{short = PVD, long= Power Voltage Detector}
\DeclareAcronym{SF}{short = SF, long= Spreading Factor}
\DeclareAcronym{RFID}{short = RFID, long= Radio Frequency Iddentification}
\DeclareAcronym{TEG}{short = TEG, long= Thermo Electric Generator}
\DeclareAcronym{KEH}{short = KEH, long= Kinetic Energy Harvester}
\DeclareAcronym{EER}{short = EER, long= Equal Error Rate}
\DeclareAcronym{WCET}{short = WCET, long= Worst Case Execution Time}
\DeclareAcronym{AET}{short = ART, long= Actual Execution Time}
\DeclareAcronym{CPU}{short = CPU, long= Central Processing Unit}
\DeclareAcronym{RL}{short = RL, long= Reinforcement Learning}
\DeclareAcronym{DAG}{short = DAG, long= Directed Acyclic Graph}
\DeclareAcronym{EDF}{short = EDF, long= Earliest Deadline First}
\DeclareAcronym{LSTM}{short = LSTM, long= Long Short Term Memory}
\DeclareAcronym{MAC}{short = MAC, long= Media Access Control}
\DeclareAcronym{EWMA}{short = EWMA, long= Exponentially Weighted Moving Average}
\DeclareAcronym{NARNET}{short = NARNET, long= Nonlinear Autoregressive Neural Network}
\DeclareAcronym{GPS}{short = GPS, long= Global Positioning System}
\DeclareAcronym{UN}{short = UN, long= United Nation}
\DeclareAcronym{SDG}{short = SDG, long= Sustainable Development Goal}
\DeclareAcronym{DVFS}{short = DVFS, long= Dynamic Voltage Frequency Scaling}
\DeclareAcronym{NB-IoT}{short = NB-Iot, long= Narrow Band Internet of Things}
\DeclareAcronym{BLE}{short = BLE, long= Bluetooth Low Energy}
\DeclareAcronym{DL}{short = DL, long= Deep Learning}
\DeclareAcronym{ML}{short = ML, long= Machine Learning}
\DeclareAcronym{CAGR}{short = CAGR, long= Compound Annual Growth Rate}

\begin{document}
\maketitle

\begin{abstract}
Consumer lighting plays a significant role in the development of smart cities and smart villages. With the advancement of (IoT) technology, smart lighting solutions have become more prevalent in residential areas as well. These solutions provide consumers with increased energy efficiency, added convenience, and improved security. On the other hand, the growing number of IoT devices has become a global concern due to the carbon footprint and carbon emissions associated with these devices. The overuse of batteries increases maintenance and cost to IoT devices and simultaneously possesses adverse environmental effects, ultimately exacerbating the pace of climate change. Therefore, in tandom with the principles of Industry 4.0, it has become crucial for manufacturing and research industries to prioritize sustainable measures adhering to smart energy as a prevention to the negative impacts. Consequently, it has undoubtedly garnered global interest from scientists, researchers, and industrialists to integrate state-of-the-art technologies in order to solve the current issues in consumer light management systems making it a complete sustainable, and smart solution for consumer lighting application. This manuscript provides a thorough investigation of various methods as well as techniques to design a state-of-the-art IoT-enabled consumer light management system. It critically reviews the existing works done in consumer light management systems, emphasizing the significant limitations and the need for sustainability. The top-down approach of developing sustainable computing frameworks for IoT-enabled consumer light management has been reviewed based on the multidisciplinary technologies involved and state-of-the-art works in the respective domains. Lastly, this article concludes by highlighting possible avenues for future research.
\end{abstract}



\section{Introduction}
 
Smart city and Smart energy are interrelated concepts designed to enhance the efficiency and sustainability of urban areas. The idea of smart city represents the extensive integration of \ac{ICT} in the daily lives of human beings in urban areas \cite{intro-2}. Implementing these technologies aims to elevate the living standards of residents by optimizing the effectiveness of services and infrastructure associated with cities. Smart city also refers to a city that integrates the physical infrastructure, information technology infrastructure, social infrastructure, and business infrastructure through \ac{IoT} to harness the collective intelligence of its community \cite{intro-1}. On the other hand, smart energy encompasses a much broader scope than conventional energy. It can be perceived as a model akin to the “Internet of Energy” that deals with smart power generation, smart energy management,  smart energy storage, and smart energy consumption. Energy refers to the characteristics of an object or system that determine its capacity to perform work. It exists in multiple forms, including potential energy, kinetic energy, chemical energy, and thermal energy. It is worth highlighting that the wide range of energy sources includes solar, fossil fuels, electricity, vibration, biomass etc. The significance of smart energy in smart cities is fundamentally rooted in the rapid growth of smart cities and the subsequent exponential increase in energy supply demand. Smart cities are reported to improve energy efficiency, minimize electronic waste, and decrease carbon emissions through the use of smart energy. Moreover, all forms of traditional energy, clean energy, green energy, sustainable energy, and renewable energy, integrated with \ac{ICT} technology, constitute smart energy. Consequently, the integration of smart energy concept in various physical processess forms \ac{E-CPS} in smart cities. \ac{E-CPS} employs advanced sensors, communication networks, and control systems to enhance energy efficiency and minimize environmental effects. These systems can adeptly manage energy resources, forecast demand trends, and modify energy output appropriately by integrating real-time data and \ac{AI} algorithms. \\

Consumer lighting constitutes a crucial application in smart city and smart home frameworks. On the other hand, it is considered as one of the most energy-consuming applications, which accounts for 30\% to 40\% of the total energy demand of smart cities \cite{intro-2}. The global smart consumer lighting market is valued at roughly 16.25 billion USD, and it is anticipated to expand at a \ac{CAGR} of 20\%, potentially reaching over USD 83.81 billion by 2032 \cite{motivation}. Undoubtedly, this rapid expansion is primarily driven by the rising need for energy efficient lighting solutions and the escalating trend towards smart homes and smart cities. Smart consumer light management systems serve an instrumental role in converting traditional consumer lighting systems to smart lighting systems. The integration of \ac{IoT}, smart sensors, and robust control strategies for energy saving makes it a widely adopted solution. The number of features in smart consumer lighting continues to grow with the advancement in \ac{ICT} technology. Subsequently, energy consumption also increases with the number of features, leading to most state-of-the-art smart consumer light management systems being energy hungry, demanding average power in terms of hundreds of milli-watts \cite{ref93}. Conversely, sustainability is an essential aspect that must be integrated on a larger scale to comply with Industry 5.0 \cite{ref94}. Thus, it demands a focus on optimizing power consumption in order to feature smart consumer light management systems with energy autonomy using harvesters with small form factor.

\begin{figure}[h!]
	\centering
	\includegraphics[scale=0.51]{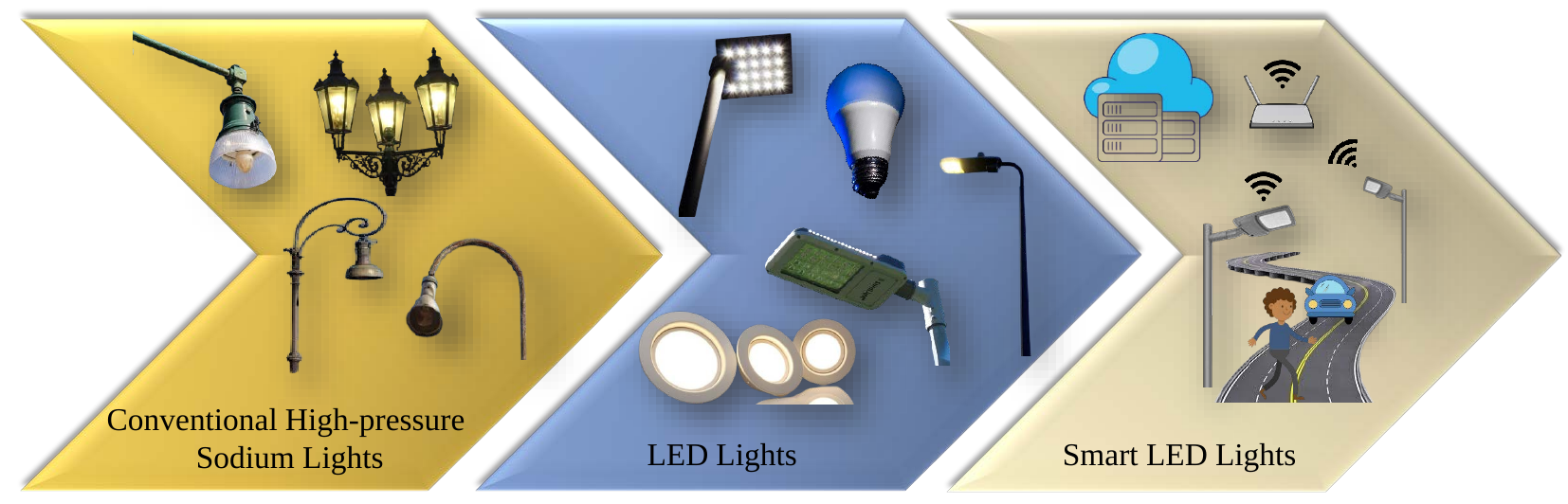}
	\caption{Evolution of Consumer Lighting Technology}
	\captionsetup{justification=centering}
	\label{ch01:cl-gen}
\end{figure}

Consumer light management system helps in reducing energy consumption and increases total lighting efficiency via smart sensors and advanced data analytics techniques. It can automatically adjust lighting configurations according to various factors such as occupancy, ambient light, and weather conditions, aiding cities in minimizing their carbon emission and reducing energy expenses. It augments the safety and security of residents by ensuring adequately illuminated streets, public areas, and indoor environments. Moreover, the application has extensively improved both energy efficiency and quality of life in smart cities. The evolution of consumer lighting application over the past decades has been presented in Fig. \ref{ch01:cl-gen}. \\

The remainder of this survey is structured into nine sections. Section \ref{contri} discusses the related prior work and contributions made in this manuscript. Section \ref{char-adv} elucidates smart consumer light management, its significance, and its attributes. Section \ref{sust-cl} presents the need for sustainability and various elements of smart consumer light. Section \ref{pm-cl} illustrates power management strategies in consumer lighting management systems. Section \ref{eh} delineates energy harvesting, power conditioning, and storage techniques. Section \ref{sust-comp} presents the sustainable computing architectures for smart consumer light management systems. Task scheduling and energy optimization techniques of these systems have been detailed in Section \ref{task-sc}. The open research directions have been highlighted in Section \ref{or}. The manuscript has been concluded in Section \ref{concl}. An appendix containing a list of acronyms used is provided at the conclusion of the manuscript. 

\section{Contributions of the Current Paper}
The contribution made in the manuscript can be understood by reviewing the relevant literature and potential limitations existing in related works.  
\label{contri}
\subsection{Related Prior Work}
Numerous studies delve into different facets of smart lighting management strategies for consumer applications. An in-depth analysis has been conducted on the significance of street lighting in smart cities, and the overall architectural design of smart street lights is thoroughly examined \cite{ref118}. In addition, a comprehensive investigation is conducted on different control strategies for street lights \cite{ref114}. This survey examines the essential hardware components for designing smart street lights, including various lamps and sensors in state-of-the-art smart street lighting systems. The control algorithms that are discussed in the paper focus on advanced techniques, including video processing, \ac{AI}, fuzzy logic, and web-based techniques. A thorough review of these algorithms is provided. Further, an extensive investigation of the framework used in the design of an \ac{IoT}-enabled smart public lighting system has been conducted \cite{ref116}. The technology involved in each layer of the framework has been discussed. Furthermore, a detailed comparison study of existing \ac{IoT}-enabled street lighting systems is presented \cite{ref119}. The comparison is based on three types of street lights such as centralized, decentralized and hybrid. Subsequently, a broad overview of smart street lights and its implementation explicitly in industrial environment along with consumer applications such as smart office lighting, smart street lighting, smart home lighting have been explored in details \cite{ref117}. Additionally, the implementation and impacts of street lights on order picking in warehouses have been explored. Further, the role of smart street lights in smart cities has been explained through a communication review, and security aspects are presented in addition to various control strategies \cite{ref115}. Subsequently, the required drivers, protocols, technologies, communication networks, and applications for implementing smart \ac{LED} lighting systems in smart buildings have been reviewed \cite{ref121}. In addition, the control methodologies for operating \ac{LED} lighting in buildings have been systematically overviewed. Subsequently, the design methodologies for highway lighting have been explored, in which the advantages, disadvantages, practical research challenges involved in the design process have been highlighted \cite{ref122}. Further, state-of-the-art smart lighting platforms designed for indoor uses in residential and office environments are comprehensively reviewed \cite{ref120}. It examines various aspects, including autonomous control algorithms, connectivity, applications, and the associated benefits and barriers. Table \ref{survey-comparision} summarises the comparison of the proposed survey with previously published survey articles.
\begin{table}[!h]
	\caption{Comparision of Previous Survey Articles on Smart Consumer Lighting Management Systems}
	\resizebox{0.99\textwidth}{!}{
			\begin{tabular}{l|l|l|l|l|l}
					\hline
					\textbf{} & \textbf{IoT} & \textbf{Energy Saving} & \textbf{Power} &  & \textbf{Task}  \\
					\textbf{Research Works} & \textbf{Framework} & \textbf{Control Strategies} &  \textbf{Management} & \textbf{Sustainability} & \textbf{Scheduling} \\
					\hline
					Chew et al., 2017 \cite{ref120}	& \checkmark \checkmark & \checkmark \checkmark & $\times$ & $\times$ & $\times$ \\
					\hline
					Mukta et al., 2020 \cite{ref122} 	& \checkmark \checkmark & \checkmark & \checkmark & $\times$ & $\times$ \\	
					\hline
					Chinchero et al., 2020 \cite{ref121} & \checkmark \checkmark & \checkmark & $\times$ & $\times$ & $\times$ \\
					\hline
					Mahoor et al., 2020 \cite{ref115}	& \checkmark \checkmark & \checkmark \checkmark & $\times$ & $\times$ & $\times$ \\
					\hline
					Füchtenhans et al., 2021 \cite{ref117}		& \checkmark  & \checkmark & $\times$ & $\times$ & $\times$ \\
					\hline
					Omar et al., 2022 \cite{ref119}	& \checkmark \checkmark & $\times$ & $\times$ & $\times$ & $\times$ \\
					\hline
					Manyake et al., 2022 \cite{ref116} & \checkmark \checkmark & \checkmark & $\times$ & $\times$ & $\times$ \\
					\hline
					Agramelal et al., 2023 \cite{ref114} & $\times$  & \checkmark \checkmark & $\times$ & $\times$ & $\times$ \\
					\hline
					Khemakhem et al., 2024 \cite{ref118}	& \checkmark \checkmark & $\times$ & $\times$ & $\times$ & $\times$ \\
					\hline
					This Paper	& \checkmark  & \checkmark & \checkmark \checkmark & \checkmark \checkmark & \checkmark \checkmark \\
					\hline
				\end{tabular}	
		}
	\label{survey-comparision}
	`\checkmark \checkmark' represents indepth analysis, `\checkmark' refers general analysis and `$\times$' indicates no analysis. 
\end{table}

The limitations and unaddressed issues in previous published survey articles are outlined below.
	\begin{itemize}
			\item In accordance with Industry 4.0, energy-autonomous and battery-less devices are prioritized over traditional \ac{IoT} devices. This shift is driven by primary concerns surrounding power consumption, carbon footprints, and environmental impacts associated with \ac{IoT} devices. Most of the state-of-the-art smart consumer light management systems lack to address the issue of sustainability. The strategies for achieving sustainability in consumer light management systems have not been discussed in any previously published survey articles.
			\item The management system is the \ac{IoT} device that features the light with \ac{IoT} compatibility. Power consumption is a vital performance metrics in case of any \ac{IoT} device. The methodologies and techniques to optimize power consumption in management systems have not been discussed in previously published related review articles.
		\end{itemize}
\subsection{Contributions and Novelty of Current Paper}
The contributions of the manuscript have been outlined below.
\begin{itemize}
	\item Most survey articles typically focus on reviewing the \ac{IoT} framework of smart light management systems, which comprises communication protocols, sensor technologies, cloud integration, and various control strategies adopted in it. In contrast, the proposed survey briefly overviews its \ac{IoT} framework, particularly in the context of its application in indoor as well as outdoor environments. It emphasizes on detail reviewing of the sustainable aspect of smart consumer light management systems.
	\item Unlike previous surveys, the manuscript highlights the importance as well as the necessity of sustainability in consumer light management systems. The current issues in the existing power management of smart consumer light management systems have been critically investigated and potential solutions are proposed.
	\item The primary distinction of the manuscript lies in an extensive review of the multidisciplinary technologies involved in the design of smart consumer light management framework using concepts of smart energy such as \ac{IoT}, energy harvesting, power management, sustainable computation techniques, power consumption optimization, and task scheduling. To the best of authors' knowledge, this is the first survey article that reviews the entire framework involved in the design process of sustainable smart consumer light management system.
	\item The manuscript encompasses the comprehensive review of energy harvesting enabled \ac{IoT} devices developed for a wide range of applications. Further, it presents five advanced state-of-the-art sustainable computing architectures with their technical advantages and limitations that can be adopted for designing smart consumer light management systems with energy-autonomous capability.
	\item Majority of the state-of-the-art sustainable \ac{IoT} devices suffer through the potential threat of power failure during unavailability of energy. In this survey, the aforementioned issue has been addressed; subsequently, the importance of optimization in power consumption in consumer light management systems without affecting the real-time operation and \ac{QoS} has been highlighted. Thus, the potential optimization techniques and task scheduling methodologies are reviewed.
\end{itemize}

\section{Smart Consumer Light Management}
\label{char-adv}
\begin{figure}[!h]
	\centering
	\includegraphics[scale=0.38]{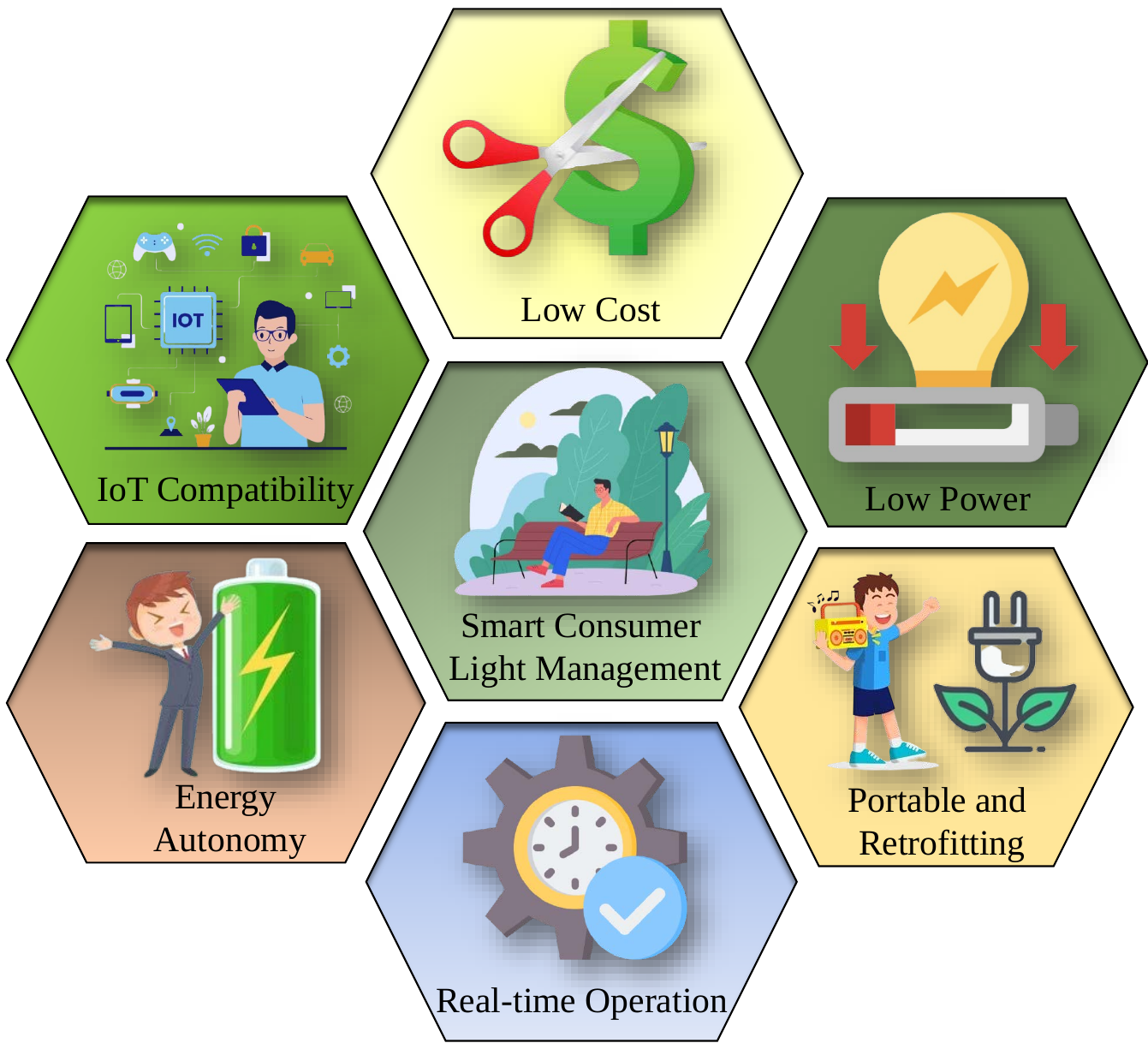}
	\caption{Characteristics of Smart Consumer Light Management System}
	\label{ch01:characteristics}
\end{figure}

\begin{figure}[h!]
	\centering
	\includegraphics[scale=0.22]{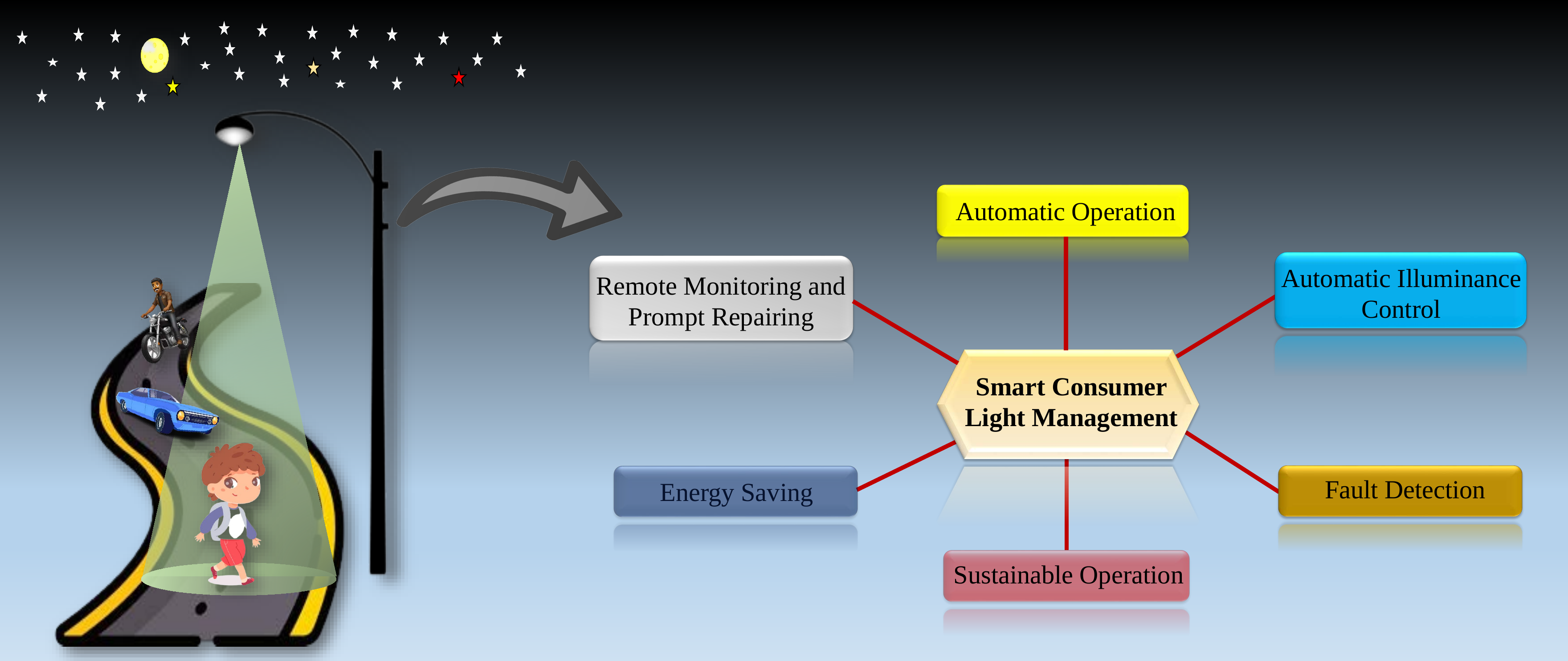}
	\caption{Advantage of Smart Consumer Light Management over Conventional Light Management}
	\label{ch01:Advantages}
\end{figure}

Smart consumer light management is often referred as consumer lighting \ac{E-CPS}. It represents an advanced integration of smart lighting with cyber-physical technologies, enabling consumers to effectively control, monitor, and optimize the energy consumption associated with lighting applications. It integrates sensors, controllers, and advanced control strategies with the physical lighting infrastructure to provide energy savings, convenience, and customizable lighting environments. The advancements in this technology have been remarkable over the past decade. \ac{LED} technology has completely transformed the lighting industry, providing long-lasting lighting options. In addition, the involvement of \ac{IoT} in consumer lighting has brought a significant improvement in the technology that has changed the user experience, making it an energy-efficient and sustainable solution. The use of smart consumer lighting systems enables users to conveniently monitor and control their lights from a distance using smartphone applications or voice commands. A smart consumer light management system possesses essential characteristics, as shown in Fig. \ref{ch01:characteristics}. 
These characteristics make them efficiently operate the street light. These characteristics are interdependent and combinedly offer `smart' tag to consumer light management systems. \ac{IoT} compatibility is the most significant characteristic that makes it smart in the true sense. It involves the integration of intelligent sensors, actuators, and communication modules with \ac{LED} lighting to facilitate various functions, including the automatic control of illuminance and remote monitoring. The automatic control feature regulates the illuminance of \ac{LED} based on several factors, including movement around the light, ambient light level, and weather conditions. Conversely, remote monitoring enables easy supervision of the operational status of the lights for the administrators and responds promptly if any faults are detected. The firmware must ensure that the management system exectute its tasks adhering to firm deadlines in order to maintain real-time functionality. Further, the device must be energy-autonomous in order to ensure ultra low maintenance, reduced carbon footprints, and sustainable operation. Low power consumption is an essential design metric of any \ac{IoT} device, and it plays an instrumental role specifically in the case of energy-autonomous devices. Low power consumption signifies the high lifespan, low form factor, and low manufacturing cost of the device. The device should be developed with the objective of integrating it into the current \ac{LED} lighting infrastructure. Therefore, low-form factor is an essential feature that the device must have, as it affects its portability. The advantages of smart consumer light management over traditional systems have been illustrated in Fig. \ref{ch01:Advantages}. 

\subsection{Architecture of Smart Consumer Light Management Systems}
\begin{figure}[!h]
	\centering
	\includegraphics[scale=0.51]{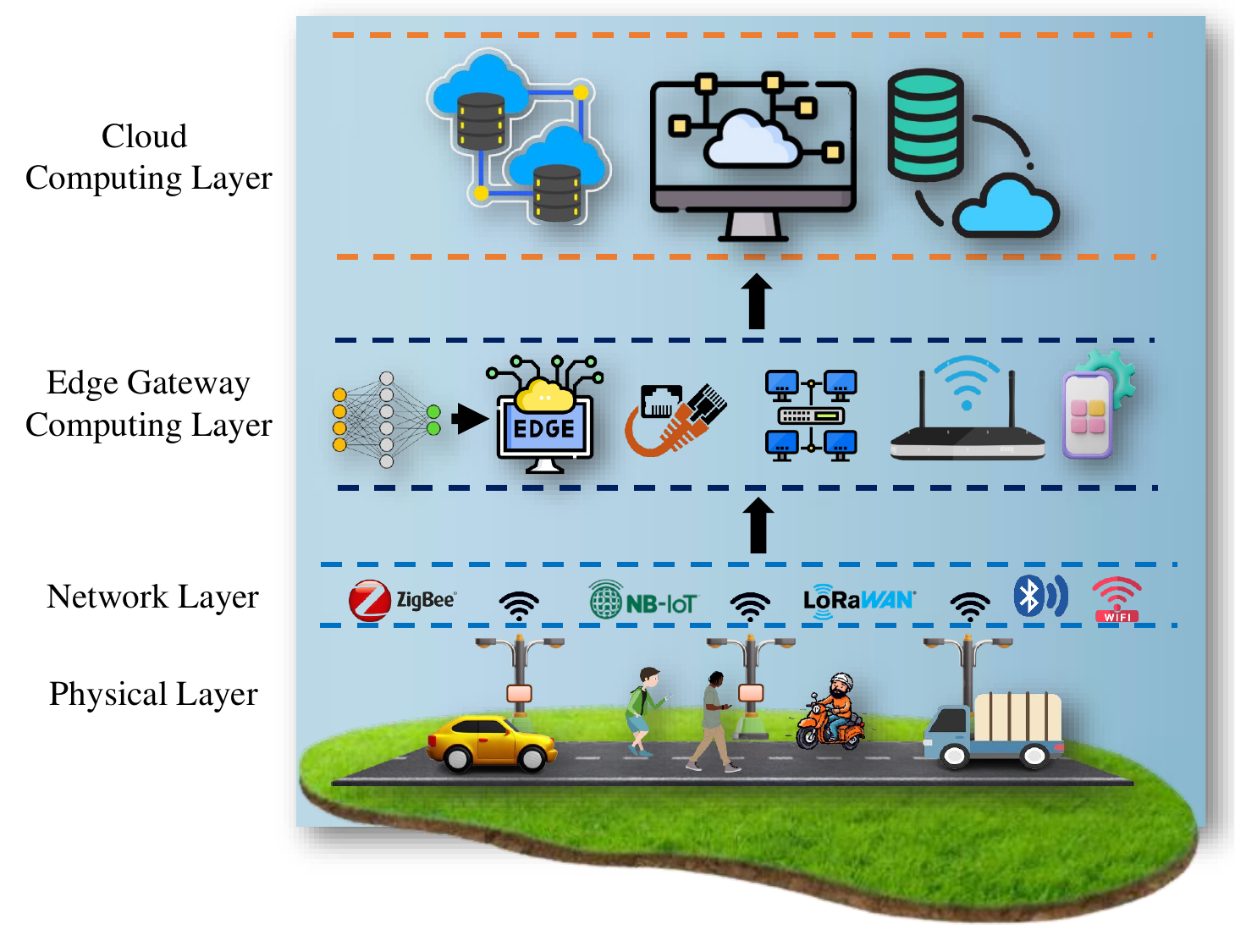}
	\caption{Architecture of Smart Consumer Light Management System}
	\label{ch01:architecture}
\end{figure}

The architecture of smart consumer light management system has been shown in Fig. \ref{ch01:architecture}. The architecture is comprised of four layers such as physical layer, network layer, edge-gateway computation layer, and cloud computation layer. The functions of each layer have been outlined below.
\begin{itemize}
	\item \textit{\textbf{Physical Layer}}: It comprises physical elements such as sensors, along with conditioning circuits, actuators, microcontrollers or microprocessors, and communication units. It additionally encompasses the firmware necessary for the operation of the end device. In this layer, sensors are used to measure various physical parameters, including occupancy, light intensity, temperature, humidity, and pressure. In this layer, low-level computations like illuminance control and daylight detection are carried out using microcontroller or microprocessor. The control actions, including automatic turn on or turn off and adjustment of light intensity, are taken using actuator. The measured data is communicated through the communication module integrated with the physical device.
	
	\item \textit{\textbf{Network Layer}}: This layer functions as a connectivity layer and typically includes a \ac{WSN}. It collects data from various physical layers and transmits it to the higher layers for processing and analysis. It serves as a conduit between the physical layer and subsequent layers in the architecture. It plays a crucial role in managing the communication between the physical device and the central processing unit, facilitating real-time monitoring and control of the system.
	
	\item \textit{\textbf{Edge-Gateway Computing Layer}}:  This layer serves a dual purpose, acting both as a gateway and an edge computing device. As an edge computing device, it processes data closer to the source, reducing latency and improving overall network efficiency. This layer offers various significant features such as fault detection and illuminance control through scheduling techniques to street light management systems through edge computing. It connects physical devices with the cloud server as a gateway, allowing for data transfer and communication. This dual role renders this layer crucial for the smooth functioning of an interconnected network. 
	
	\item \textbf{\textit{Cloud Computing Layer}}: This layer represents the highest tier in the bottom-up architecture of smart consumer light management system. This layer often exists in data centers and is accessible globally via the internet. It is responsible for processing and analyzing the large amounts of data acquired from the sensors in the physical layers. It can make real-time adjustments to lighting systems based on energy consumption goals. Additionally, this layer allows for remote monitoring and control of the entire lighting system, providing a seamless user experience for smart consumer light management.
\end{itemize}
\subsection{Application of Smart Consumer Light Management Systems}
\begin{figure}[!h]
	\centering
	\includegraphics[width=\textwidth]{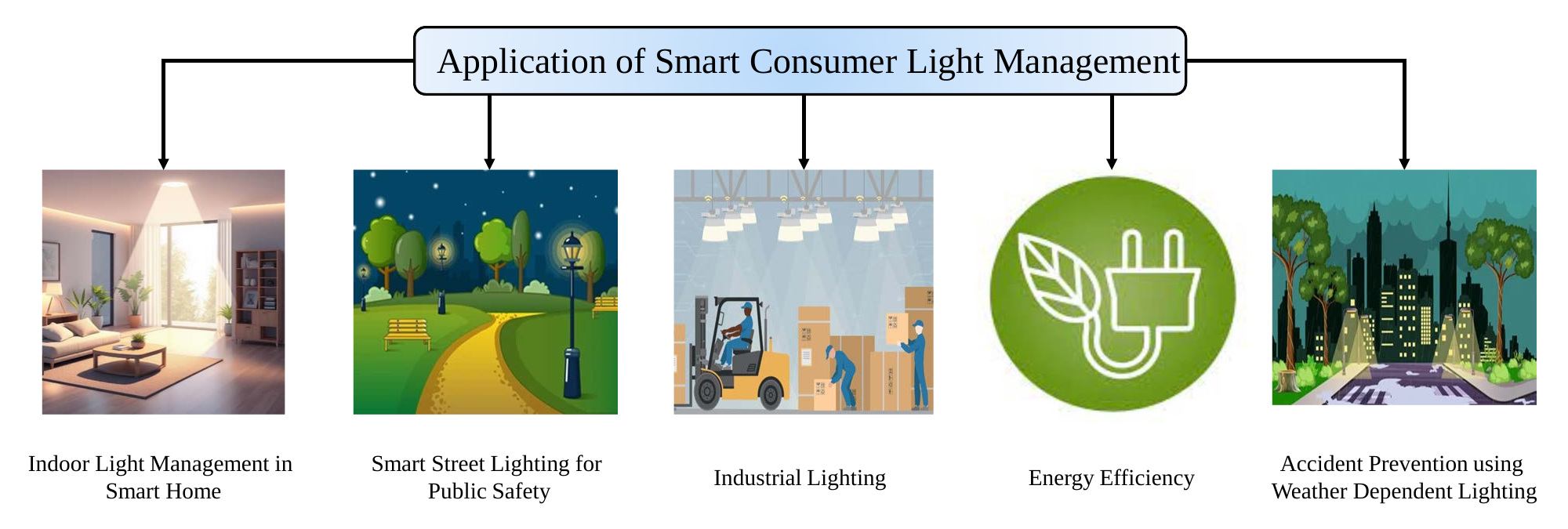}
	\caption{Applications of Smart Consumer Light Management System }
	\label{ch01:application}
\end{figure}

The aforementioned architecture has often been deployed in outdoor environments and indoor environments. The applications of smart consumer light management have been illustrated in Fig. \ref{ch01:application}. In outdoor environments, it is often incorporated in street lights, while in indoor environments, it is implemented for indoor light management, particularly in commercial buildings like offices, shopping malls, and museums. 
The challenges and solutions are different in both of these environments. A smart and efficient \ac{LED} lighting system has been proposed using \ac{LED} technology, sensor application and smart control strategy \cite{ref1}. A probabilistic approach has been incorporated to analyze traffic patterns over different time intervals in order to control the illuminance of the light. Smart street light management systems have been reported to significantly impact accident prevention   \cite{ref2}.  Studies in psychology indicate that variations in color temperature can have a profound impact on the human circadian rhythm.  Therefore, the implementation of illumination control based on \ac{CCT} has been incorporated, enhancing visibility for drivers in areas with low light. Weather conditions, such as heavy rain and fog, can significantly reduce visibility on the road \cite{ref3}. An \ac{IoT}-based framework has been proposed to address the visibility challenges caused by harsh weather conditions. It can control the illuminance of the \ac{LED} light detecting rain in order to  solve the visibility issue during harsh weather. Further, an energy-saving scheme has been proposed to save electricity consumption of the street light \cite{ref4}. In this work, the illuminance of the \ac{LED} light has been controlled based on ambient light conditions and the speed of the vehicles. The \ac{LED} light is configured to illuminate with high intensity in case of vehicles moving at a faster velocity. Subsequently, it switches the \ac{LED} to low-intensity mode. The proposed scheme has been claimed to be efficient than the conventional ambient light intensity based illuminance control of \ac{LED} lights. This scheme has been reported to reduce energy by 40\%. Subsequently, \ac{IoT}-enabled system has been developed to control the illuminance of the street lights in accordance with two parameters, i.e. occupancy and light intensity \cite{ref5}. \ac{IoT} connectivity provides an advantage to the admin in remote monitoring, controlling, and faulty detection in street lights. \\

Further, similar framework has been developed that can control and monitor the operation of street light energy efficiently \cite{ref6}. The street light operating firmware has been developed using \ac{ABC} optimization algorithm. It has been implemented and reported to save 12,615.635 kWh of energy during one year of experimentation. Furthermore, a novel street lighting scheme has been proposed that adjusts the brightness of streetlights based on traffic conditions \cite{ref7}. It utilizes the signatures of received signal strengths to extract traffic parameters, subsequently used to determine the appropriate lighting level. This scheme effectively controls the illuminance streetlight by adapting to the behaviors of vehicles and pedestrians on the roads. The proposed system has been observed to enhance the performance of existing street light management systems by demonstrating 95\% accuracy in detecting road users. On the other hand, it requires 10.5\% of the electrical energy compared to other existing methods. In \cite{ref8}, a novel architecture of smart \ac{IoT} platform that makes use of street lights as computing nodes and incorporates a prediction model for workload management in smart campus environments. \ac{ANN} learning algorithms have been used to analyze the network and determine the resource requirements of each network node. These algorithms enable the nodes to function as a unified network resource allocation service. In addition, it also works as a micro weather station that measures temperature, wind speed, \ac{UV} index, etc. It also helps in intrusion detection. Subsequently, a weather adaptive control mechanism for street light management has been proposed \cite{ref9}. It controls the illuminance and \ac{CCT} of the light based on temperature, humidity and movement of the objects around the road. Furthermore, \ac{AI} technique has been implemented with street light to detect various object classes such as pedestrian, bicycle, motorbike, and vehicle \cite{ref15}. The illuminance of the street light is controlled in accordance with these object classes.\\

Another critical aspect of \ac{IoT} enabled consumer light management sytsem is communication protocol. Several advanced communication protocols such as \ac{NB-IoT}, 5G, \ac{Li-Fi}, \ac{LoRaWAN}, and \ac{BLE} have been used in state-of-the-art \ac{IoT} devices. Efforts have been made to improve communication in street light systems by implementing \ac{NB-IoT} technology to enhance security and communication range at low-power requirements \cite{ref10}. Wireless communication based on standard IEEE 802.15.4 has been incorporated into public lighting management systems \cite{ref11}. This work implements \ac{DALI} protocol, which uses bidirectional communication to enable remote monitoring and controlling. The system is reportedly manufactured at low-cost and can be seamlessly integrated with high-pressure sodium lamps to \ac{LED} lamps. Subsequently, ZigBee protocol is considered as one of the most popular, low-power, low-cost communication protocol that has been used in street light operating systems for short-range communication \cite{ref12} -- \cite{ref13}. Furthermore, efforts have been made to address the short communication range problem in street light control systems by integrating ZigBee with \ac{GPRS} technology \cite{ref14}. In order to ensure higher communication range, \ac{Wi-Fi} has been used in designing intelligent automatic street light management system \cite{ref16}. On the other hand, \ac{LoRa} communication has been integrated into \ac{IoT}-assisted fog and edge-based smart lamp posts \cite{ref17}. The proposed system also has the ability to monitor environmental factors like temperature, humidity, and CO levels aside from its lighting capabilities. In addition, the device utilizes \ac{Wi-Fi} to transfer the collected data to a cloud server. \\

Lighting management in residential is also considered as a critical research application as buildings in the USA consume approximately 40\% of total energy, among which lighting for residential buildings accounts for approximately 10\%, while commercial buildings exhibit a higher proportion of 20\%. In \cite{ref18}, a light management system has been proposed considering the spatial characteristics and patterns of occupant behavior to control the lighting parameters effectively. Further, a novel approach has been proposed to manage the color of a multi-channel \ac{LED} lighting system in a smart home context, implementing camera of smartphone \cite{ref19}. The algorithm aims to enhance the output spectrum of the luminaires, resulting in light that can be adjusted for \ac{CCT}, accurate color, and high color rendering index. Subsequently, a control scheme with intelligent power gateway has been introduced for \ac{LED} based light management in office \cite{ref20}. The system has been designed with adaptive middleware, which can be modified according to the requirements of the users. The results show a significant potential for power savings, with a maximum reduction of approximately 58\%. Furthermore, efforts have been made to improve the energy efficiency of \ac{LED} light while prioritizing user satisfaction \cite{ref21}. The system enables regulating the intensity of the light according to user satisfaction. It has been reported to reduce the power consumption by 21.9\%. Subsequently, an IEEE 802.15.4-based \ac{WSN} has been designed to integrate the \ac{DALI} protocol for lighting automation in buildings \cite{ref22}. The system is reported to be cost-effective and fully centralized. An energy-saving, easy-to-install, wireless, low-cost \ac{IoT}-based device has been designed for office light management \cite{ref23}. It can be easily manufactured with a 10\% cost of installing a smart system. Additionally, it provides retrofitting option to consumers. The proposed system has been tested with a commercial 15 W T8 \ac{LED} tube to save energy maximum up to 28.13\%. Further, attempts have been made to design cost-effective indoor lighting management system which is tested in an office \cite{ref24}. The system controls the intensity of the light depending on the motion and user satisfaction. It has been reported to reduce the energy consumption of the light in the range of 55\% to 69\%.

\subsection{Control Strategies for Smart Consumer Light Management Systems}
Consumer light management system helps in conserving energy by controlling the illuminance according to number of factors. Various illuminance control strategies have been adopted in state-of-the-art consumer lighting applications, which significantly minimize energy consumption associated with the light. A comparison of prior works on the basis of energy saving as the impact of several control strategies is presented in Table \ref{cons-energysaving}.
\begin{table}[!h]
	\centering
	\caption{Related Prior Work on Various Control Strategies and Energy Saving}
	\resizebox{0.99\textwidth}{!}{
		\begin{tabular}{l|l|l|l|l}
			\hline
			\textbf{} &  \textbf{Sensing} & \textbf{Illuminance}  & \textbf{Energy} & \textbf{System } \\	
			\textbf{Research Works} &  \textbf{Parameters} & \textbf{Control Strategy} & \textbf{Saving} & \textbf{Implementation} \\	
			\hline
			&   Occupancy and &  &     & Installed in \\
			Byun et al., 2013 	\cite{ref21} &  Light intensity & Step dimming &  21.9\% &  office\\
			\hline
			&   Occupancy and &  Step dimming   &     & Installed in \\
			Tan et al., 2013 \cite{ref148}  &  Light intensity & and Scheduling &  44\% &  office\\
			\hline
			&   Occupancy and  & &  &    Installed in\\
			Nagy et al., 2015 \cite{ref145}  & Light intensity & Two steps &  37.9\% -- 73.2\%  & office \\
			\hline
			&   Occupancy and  & &  &    Installed in \\
			Higuera et al., 2015 \cite{ref146} & Light intensity & Step dimming &  13.4\% -- 43\%  &  office\\
			\hline 
			& Occupancy and & Zoning and & &  \\
			Lau et al., 2015 \cite{ref138} & Light intensity & Step dimming & upto 37\% &  Simulation \\
			\hline
			&   Occupancy and  & &     & Installed in \\
			Chew et al., 2016 \cite{ref147}  &  Light Intensity & Step dimming &  55\% -- 62\%  &  Classroom\\
			\hline
			& & Step dimming and &  & Installed with  \\
			Shahzad et al., 2016 \cite{ref141} & Light intensity & Group control & 68\% -- 80\% & Street light  \\
			\hline
			& Occupancy and & Group control and & & Installed with \\
			Atis et al., 2016 \cite{ref142} & Light intensity & Step dimming & 33\% & Street light  \\
			\hline
			& & Zoning and & & Installed with \\
			Juntunen et al., 2018 \cite{ref140} & Occupancy & Two step & 60\% -- 77\% & Street light  \\
			\hline
			& Vision, Occupancy & & & Installed with  \\
			Petritoli et al., 2019 \cite{ref144} & and Inductive loops & Step dimming & 59\% & Street light \\
			\hline
			Bonomolo et al., 2020 \cite{ref143} & Light intensity & Step dimming & Upto 70\% & Simulation \\
			\hline
			& Occupancy and & Step dimming and & & Installed with \\
			Sifakis et al., 2021 \cite{ref139} & Light intensity & Group control & 56\% & Street light  \\
			\hline
	\end{tabular}}
	\label{cons-energysaving}
\end{table}
There are five types of illuminance control strategies implemented in state-of-the-art smart consumer lighting application as outlined below.
\begin{itemize}
		\item \ac{CCT} Method: \ac{CCT}-based illuminance control denotes a system that modulates lighting according to correlated color temperature and illuminance levels. This method is often used in intelligent lighting systems to improve comfort, energy efficiency, and the visual appeal of environments. This method is reported to be specifically useful in case of accident prevention during low visibility.
	
		\item Two steps Method: This method is typically implemented to automatically turn on and turn off features in light. It switches the illuminance of the light between 0 and 100\%. In this method, the illuminance is regulated according to either daylight or motion. This method is specifically useful in turning off the street lights during day hours, eliminating the need for manual operation and consequently conserving energy. 
	
		\item Step dimming Method: This is a simple and economical illuminance control method. It utilizes \ac{PWM} technique to regulate the illuminance between 0 to 100\%. Step dimming allows for easy adjustment of light levels without the need for complicated controls or systems. It is a reliable option for spaces that require regular changes in lighting intensity.
	
		\item Zoning Method: According to this method, an area gets partitioned into several zones, each possessing specific lighting needs and controls. This method allows more effective illumination regulation according to the distinct requirements of various zones, improving both functionality and energy efficiency. This method is particularly advantageous in large areas such as offices, conference rooms, or retail establishments, where different lighting needs may exist within the same area. 
	
		\item Scheduling Method: This method ensures that the lighting in a space is adjusted based on the time of day, helping to conserve energy and create a comfortable environment for occupants. It is particularly useful in indoor applications such as offices, schools, museums, etc.
\end{itemize}
 
\section{Sustainable \ac{IoT} Framework in Smart Consumer Light Management}
\label{sust-cl}
\begin{figure}[!h]
	\centering
	\includegraphics[scale=0.52]{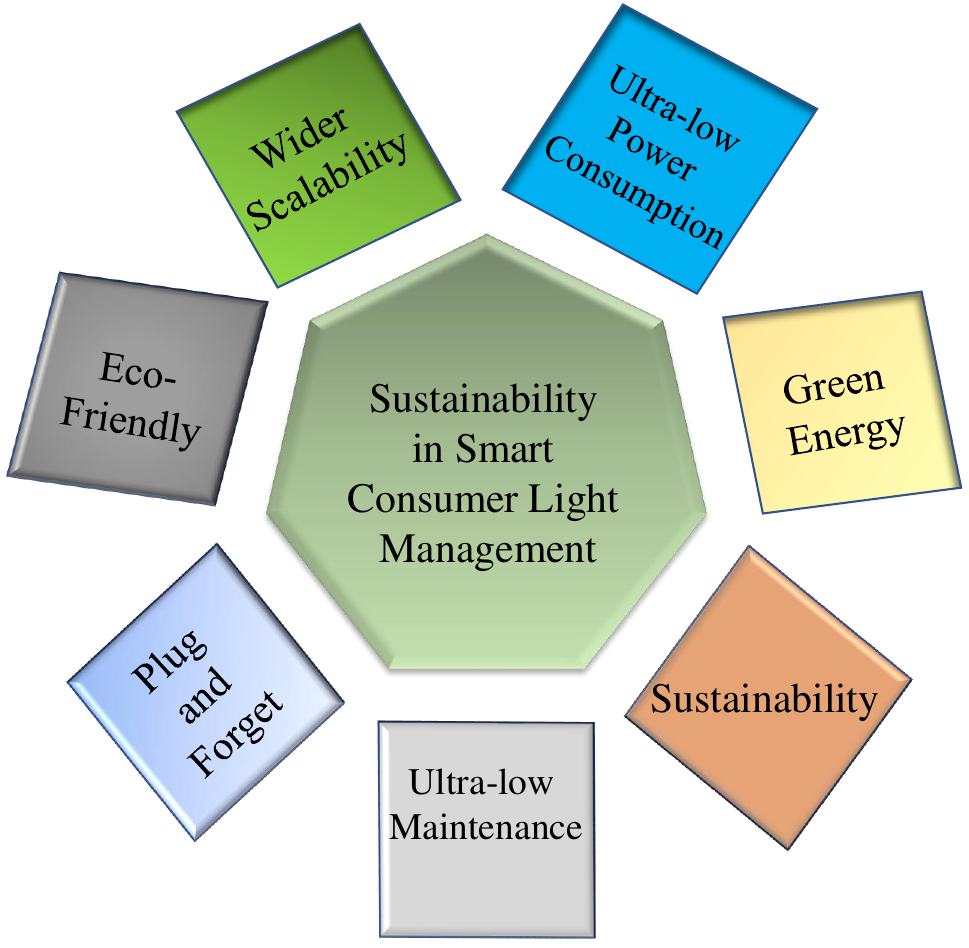}
	\caption{Advantages of Integrating Sustainability into Consumer Light Management System}
	\label{ch01:advantage-sustainability}
\end{figure}

The recent developments in smart energy and consumer light technology have not only enhanced the efficiency and convenience of lighting solutions but also contributed to a greener environment by reducing energy consumption and waste. Nevertheless, the increase in the use of \ac{IoT} devices also raises concerns about powering such a massive number of devices, explicitly considering the wide use of batteries. The global number of active \ac{IoT} enabled devices deployed on field is estimated as 16.7 billion by 2023 \cite{stat-3}. The number is predicted to reach 29 billion by 2027, according to the same report, due to the extensive use of \ac{IoT} technology. Batteries have a relatively short lifespan, which can result in the need for frequent replacements and higher expenses, implying more maintenance efforts. In addition, improper disposal of used batteries can have detrimental effects on the environment. It has been reported that the increasing reliance on batteries without considering battery life is anticipated to result in a staggering 78 million dead batteries every day by 2025  \cite{eu-res}. Therefore, adhering to sustainability in \ac{IoT} technology in tandem with \ac{SDG} 7 of \ac{UN} has become an impetus. The implementation of \ac{E-CPS} in consumer lighting is fundamentally supported by sustainable \ac{IoT} technology. It possesses the advantage of being easily installed on any existing \ac{LED} based light, allowing for more efficient energy management. The energy harvesting capability allows it to harness energy from various renewable sources, including artificial light, solar power, thermal energy, and wind power. It eliminates the need for relying solely on the power supply of the connected light source. Furthermore, this characteristic of sustainable \ac{IoT}-enabled consumer light management system allows it for seamless integration with any \ac{LED} light without the need for extra circuitry or any impact on the power source of existing lights. Thus, it can be immensely beneficial in converting the conventional lighting infrastructure not only in the case of street lighting but also in offices, smart homes, and commercial buildings. Designing a sustainable alternative for smart consumer light management offers various benefits, as shown in Fig. \ref{ch01:advantage-sustainability}. 

\begin{figure}[h!]
	\centering
	\includegraphics[scale=0.69]{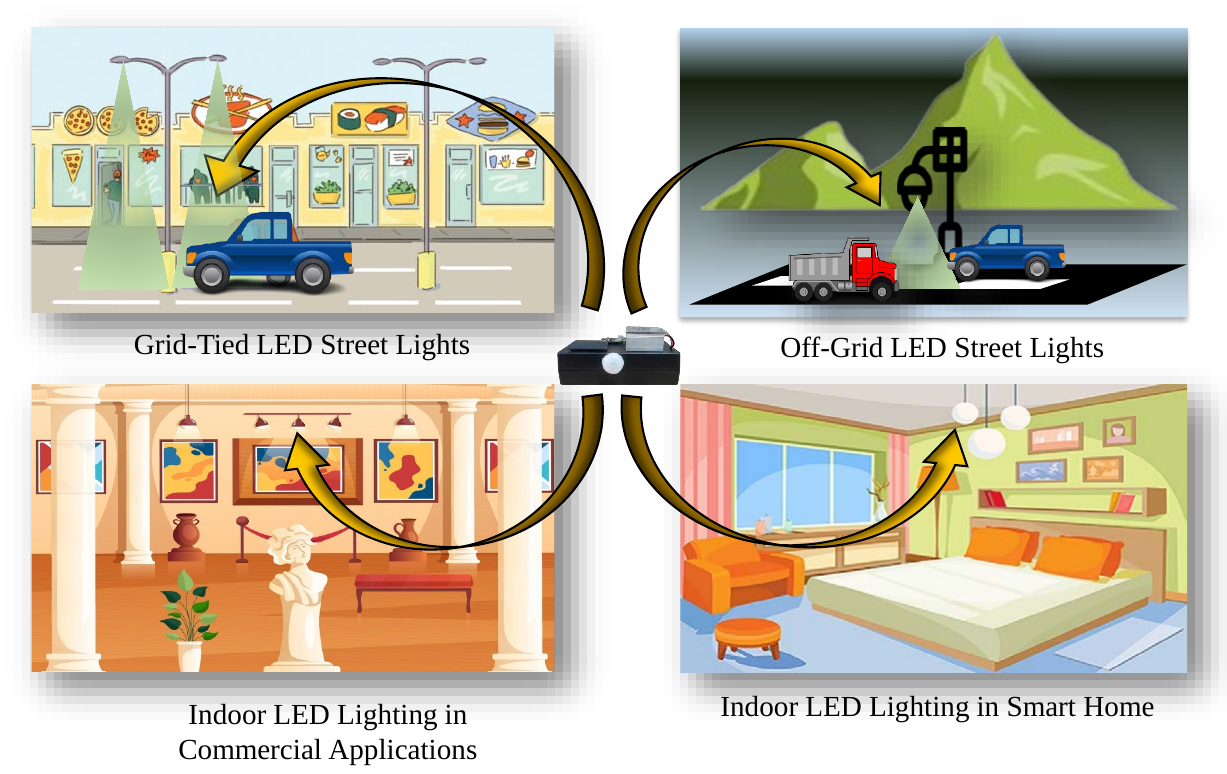}
	\caption{Vision of Sustainable Consumer Light Management System}
	\label{ch01:motivation}
\end{figure}

It primarily overcomes the scalability issues of \ac{IoT}-enabled light management systems, making it a plug-and-forget device applicable to any type of \ac{LED}-based light, such as standalone, on-grid, and hybrid. The inclusion of smart energy technology, specifically the energy harvesting feature, power consumption optimization techniques, and advanced energy storage elements such as supercapacitors and ultra-capacitors, makes it an eco-friendly solution, significantly reducing the maintenance requirement and carbon footprint of the device. According to recent studies, the global count of street lamps is projected to reach 352 million by 2025 \cite{stat-1}. By the end of 2022, the global number of smart street lights has been estimated as 23 million. This number is expected to increase to 63.8 million by 2027, accounting for less than 19\% of the total number of street light lamps \cite{stat-2}. Thus, there is a substantial demand for a solution that transforms traditional lights into intelligent lights. Smart consumer light management system is instrumental in achieving the aforementioned goal. The majority of the works conducted on consumer light management systems are in two directions, such as developing control strategies to reduce energy consumption and adding features to \ac{LED} lights to expand functionality. However, limited research has been conducted on promoting sustainability in \ac{IoT}-enabled consumer light management systems. Fig. \ref{ch01:motivation} portrays the objective and importance of a consumer light management system enabled by sustainable \ac{IoT} technology.

\begin{figure}[h!]
	\centering
	\includegraphics[scale=0.4]{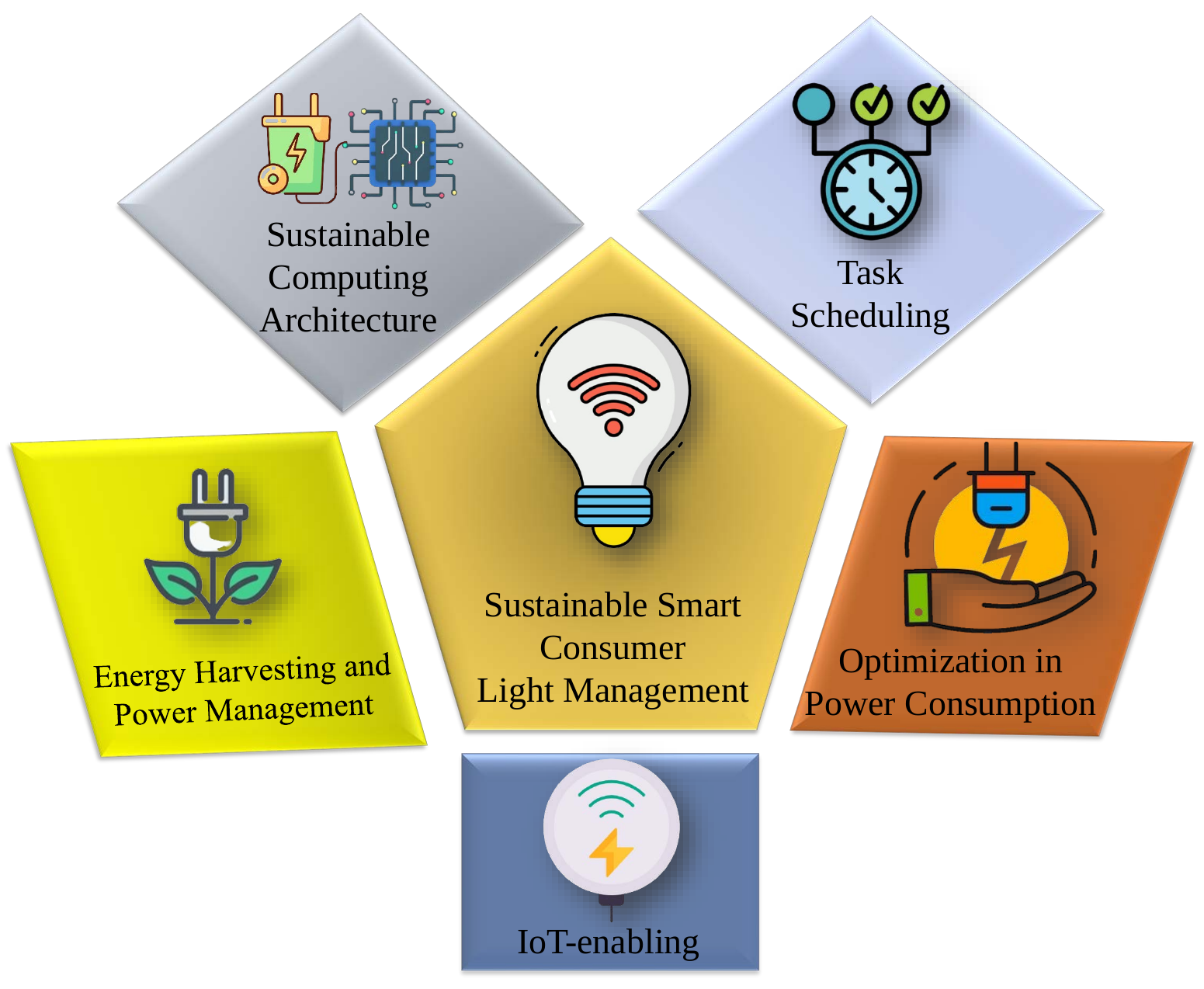}
	\caption{Elements of Sustainable Smart Consumer Light Management System}
	\label{ch01:features-sustainability}
\end{figure}

Sustainability in consumer light management systems involves several advanced interdisciplinary technologies, as shown in Fig. \ref{ch01:features-sustainability}. 
The design process of state-of-the-art sustainable \ac{IoT} framework necessitates in-depth knowledge and skills that align with the specific requirements of the application. The investigation of the energy source should be conducted within the framework of consumer lighting management. The primary performance parameters that must be considered throughout the design process of \ac{IoT} devices are size, cost, and power consumption. An ideal \ac{IoT} device should possess the qualities of being cost-effective, energy-efficient, and small. Therefore, in order to design \ac{IoT} devices that fulfill these requirements, it is necessary to possess a thorough understanding of power management and optimization. Furthermore, the design of such devices requires a thorough understanding of communication protocols. Designers can design a sustainable \ac{IoT} framework by involving all these techniques that not only fulfill the requirements of the application but also reduce adverse environmental effects, promoting the use of green and sustainable energy.

\section{Power Management in Consumer Light Management System}
\label{pm-cl}
\begin{figure}[h!]
	\centering
	\includegraphics[scale=0.5]{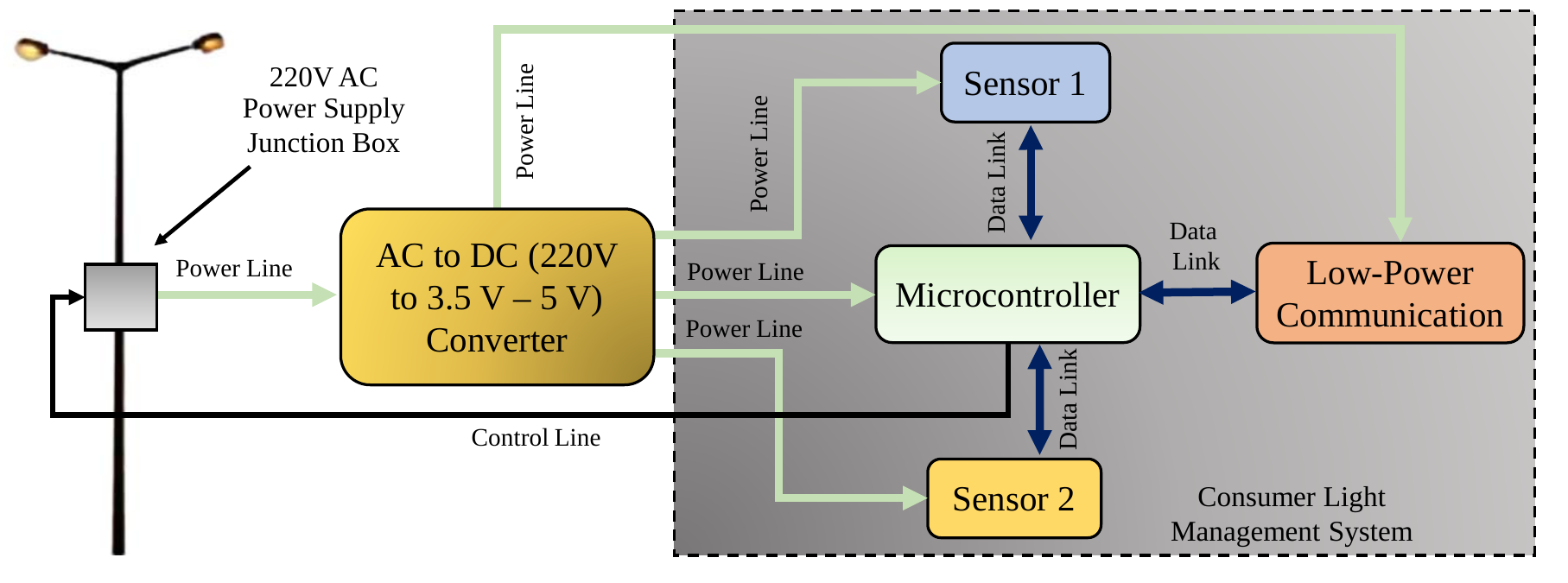}
	\caption{Power Management Scheme for Grid-Tied or Hybrid Street Light Management System}
	\label{PM1-CLMM}
\end{figure}

Effective power management is a crucial consideration when incorporating smart energy into \ac{IoT}-enabled devices. Optimal power management is essential for ensuring that the devices can function for long periods without frequent maintenance. Power management in consumer lighting applications involves more than just supplying power to the management system. It also includes protecting it from fluctuating power supply, preserving its lifetime through over-voltage and under-voltage protection, providing backup energy, and optimizing the power supply. There are three power management techniques implemented in state-of-the-art consumer light management systems. Fig. \ref{PM1-CLMM} shows the power management scheme implemented in most of the state-of-the-art related works. \\

This technique is simple yet widely applicable. The management system relies on the power supply of the \ac{LED} for energy. The power supply of \ac{LED} is mostly 220V AC, which gets converted to 3.5V to 5V DC through a converter mechanism that is compatible with \ac{IoT} components. The advantage of this technique is that it doesn't include battery in the management. In \cite{ref2}, the aforementioned technique has been incorporated into the street light management system. The power supply circuit is specifically designed to convert 220 V AC into three separate DC supplies: 12V, 5V, and 3V. This conversion is achieved by combining a converter, switching regulator, and linear regulator. A similar power management scheme has been implemented in \cite{ref6}, where the supply voltage is 230V AC. Subsequently, the management system of the street light is powered from the power supply line using a switched mode power supply \cite{ref11}. The device is designed to convert an input voltage of 230V AC to an output voltage of 3.3V DC. It achieves an accuracy of $\pm$2\% and operates with an efficiency of 65\%. Further, a similar conversion mechanism has been used in \cite{ref111}, which is used to convert the single-phase 220V AC to 5V DC to power the sensor node. However, two major limitations have been observed in this power management technique, as outlined below.
\begin{itemize}
	\item There is a significant amount of power leakage during the process of conversion from AC to DC. It leads to increased energy wastage as well as power consumption, ultimately resulting in a decrease in the overall efficiency of the light management system. Additonally, this scheme is less preferred in standalone or off-grid lights due to its higher power consumption. 
	\item This technique is relatively obsolete as it does not eliminate the dependability on traditional grid power promoting sustainability in \ac{IoT}. 
	\item Power consumption has not been considered for optimization as the availability of power is not a constraint here. Nevertheless, it holds significant value as a design metric within the realm of \ac{IoT}. 
\end{itemize}

\begin{figure}[h!]
	\centering
	\includegraphics[scale=0.43]{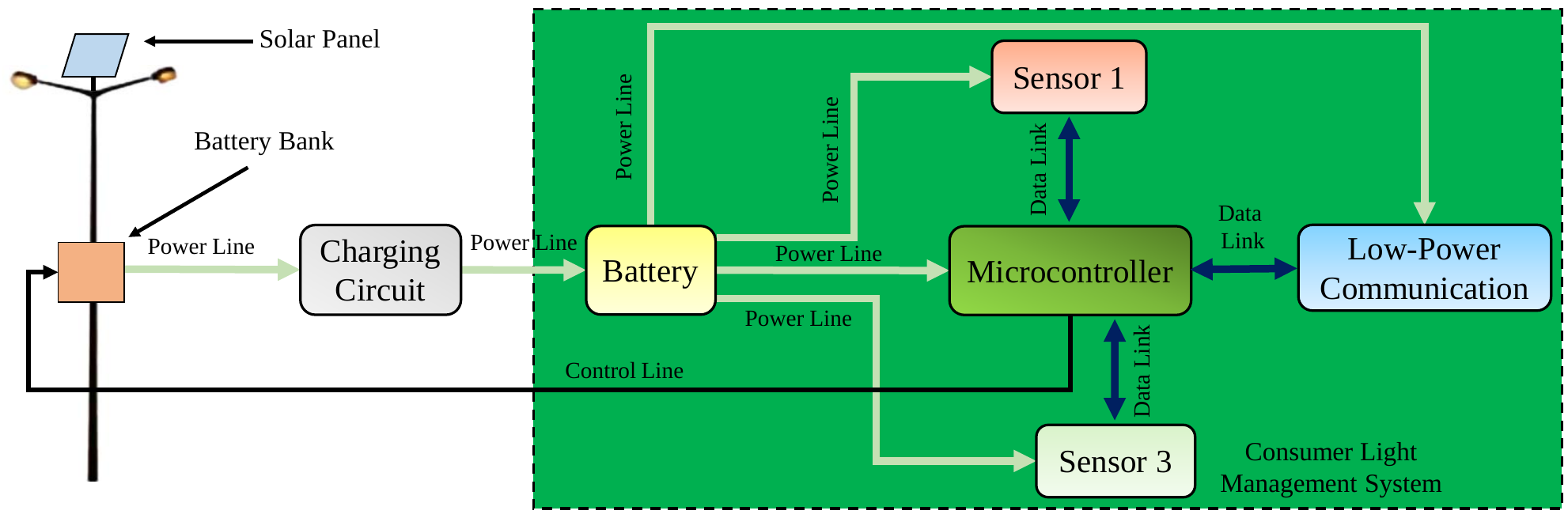}
	\caption{Power Management Scheme in Standalone for Off-grid Street Light Management System}
	\label{PM2-CLM}
\end{figure}

Fig. \ref{PM2-CLM} depicts the block diagram of another power management scheme used in state-of-the-art consumer light management systems. This scheme has also been used in standalone solar street lights where power is constrained. It includes a rechargeable battery, mostly lithium-ion type. The charging circuit drags power from the battery of the solar light to charge the mini battery of the management system. The charging circuit typically includes a DC-DC converter.
However, the management system depends on the energy storage unit of the light for power, which does not make it a sustainable device. The prototypes of \ac{LED} light management system developed in \cite{ref112} and \cite{ref113} have been powered by the solar panel of the light following the scheme depicted above. The major drawback of this power management scheme is the inclusion of batteries. Additionally, the issue of high power consumption has not been addressed in the case of consumer light management systems that follow this power management scheme. \\

\begin{figure}[h!]
	\centering
	\includegraphics[scale=0.42]{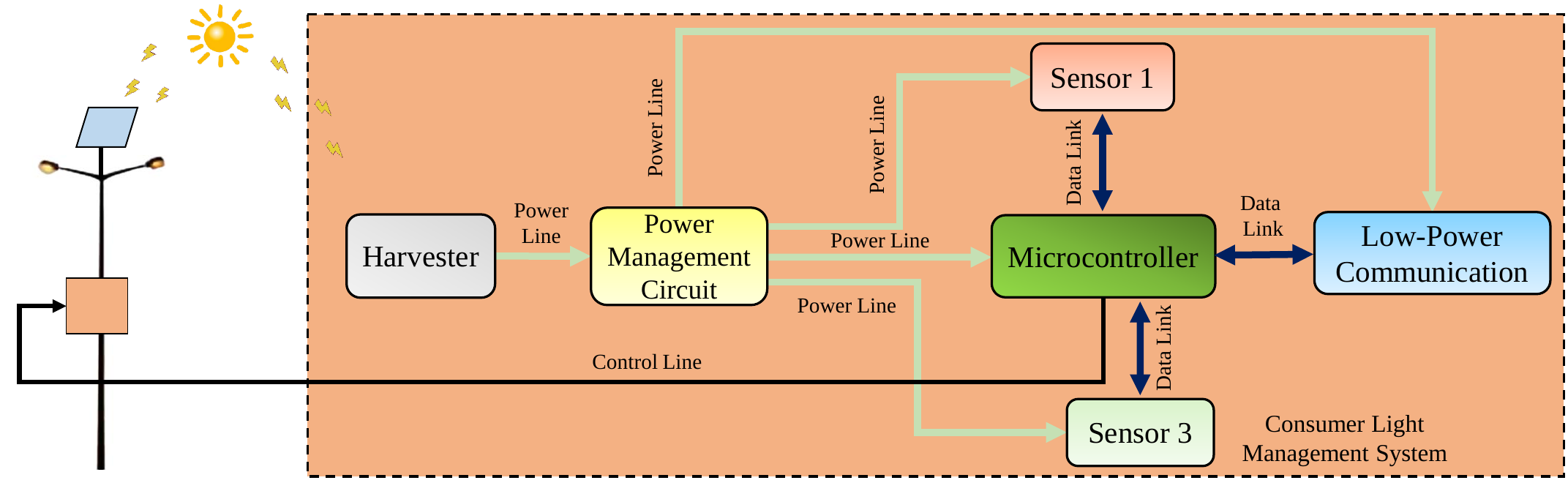}
	\caption{Energy-Autonomous Power Management System for Consumer Light Management}
	\label{PM3-CLM}
\end{figure}

Fig. \ref{PM3-CLM} illustrates the block diagram of a sustainable consumer light management system. 
The drawback cited in the aforementioned power management schemes has been solved using this technique. It can be implemented in any \ac{LED} light. It harvests energy from various renewable sources such as solar, thermal, artificial light, etc., using its energy harvester to power the sensor node, achieving sustainability. This feature makes it an ideal solution to retrofit it to existing \ac{LED} lighting systems without disrupting the entire infrastructure. It enhances energy efficiency and enables \ac{IoT} compatibility in consumer light. 

Efforts have been made to develop self-powered street light management system that utilizes a smaller solar panel to collect solar energy \cite{mohanty2}, \cite{ref75}. It is energy-autonomous and operates the \ac{LED} light to bring energy efficiency. Further, attempts are made to reduce the size of the solar cell by optimizing the power consumption of the management system to achieve the same goal \cite{ref93}. In this work, the device harvests energy from artificial light sources in order to perform its tasks autonomously. Duty cycle optimization technique has been employed to minimize power consumption. It incorporates battery-less framework. Subsequently, hybrid power management system for harvesting energy from solar and solar thermal energy has been proposed \cite{ref94}. It can power the management system in order to operate the street light energy efficiently.

\section{Energy Harvesting and Power Management Technology in Smart Energy}
\label{eh}
Energy harvesting and power management techniques are the core of smart energy. Designing an efficient energy harvesting subsystem to power the \ac{IoT} devices through ambient energy sources is the primary design step towards achieving sustainability. Subsequently, implementing optimized power management techniques that effectively regulate and distribute energy across a smart energy network possesses equal importance. Thus, the combination of energy harvesting and power management technologies is crucial for developing a reliable and effective smart energy system capable of fulfilling the needs of sustainable \ac{IoT}. It deals with harnessing energy from natural sources, conditioning it according to the requirement of the application, and storing it for powering the electronics components of the device. It helps a device to sustain its operation uninterruptedly without the requirement of maintenance. 

\subsection{Energy Harvesting Technology}
There are two important terminologies associated with energy harvesting technology that are important to be understood, i.e., energy scavanging and energy harvesting. Energy scavenging includes the search for reusable sources. It is obtained from natural sources and the untapped portion of the discarded energy \cite{scavange}, \cite{scavange-2}. It helps alleviate a certain amount of strain on the primary system in terms of activity and cost, serving a specific purpose to promote sustainability. Energy scavenging is a perfect solution for sensors located near machinery, like industrial equipment, household appliances, and vehicles. Table \ref{energy-source} presents the energy sources available for powering \ac{IoT} devices.
\begin{table}[h!]
	\centering
	\label{energysource}
	\caption{Sources of Energy used to Power \ac{IoT} devices}
	\label{energy-source}
		\begin{tabular}{l|l}
			\hline
			\textbf{Type of Energy}	& \textbf{Source of Energy} \\
			\hline
			Mechanical Energy & Vibration or Deformations \\
			\hline
			Solar Energy or Light Energy & Sun or Artificial light \\
			\hline
			Thermal Energy & Temperature difference \\
			\hline
			Acoustic Energy & Vibration due to sound wave\\
			\hline
			Radio Frequency & Electromagnetic radiation \\
			\hline			
		\end{tabular}	
	\end{table}
	
Energy harvesting is a comprehensive process that involves a range of activities aimed at utilizing and storing energy scavenged from renewable energy sources. The size of the global energy harvesting system market has been assessed at 0.6 billion USD in 2023, with an estimated growth to 0.9 billion USD by 2028 \cite{stat-4}. It indicates a steady increase at a compound annual growth rate of 10\% throughout the forecast period. As technology advances, integrating renewable energy sources into \ac{IoT} devices will become even more widespread, further promoting a greener and more sustainable future. Table \ref{energy-source2} summarises energy sources with harvesting methods used in state-of-the-art energy-autonomous \ac{IoT} devices. Light, vibration, heat, \ac{RF}, and air/water flow are the most common energy sources used for harvesting energy for \ac{IoT} devices \cite{ref25}. \\
	\begin{table}[h!]
		\centering
		\caption{Characteristics of Ambient Energy Sources used to Power \ac{IoT} Devices}
		\resizebox{0.99\textwidth}{!}{
			\begin{tabular}{l|l|l|l|l|l}
				\hline
				\textbf{}	& \textbf{Energy} & \textbf{Harvesting} &  &  &   \\
				\textbf{Research Works}	& \textbf{Source} &  \textbf{Method} & \textbf{Power Density} & \textbf{Merit} & \textbf{Demerit}  \\
				\hline
				Orrego et al., 2017 \cite{ref32} &  &   & &   &    \\
				Wang et al., 2017 \cite{ref33}  &  & Piezo  & & Predictable and even low-  & Not Steadily   \\
				Sardini et al., 2011 \cite{ref34} & Wind & Turbine & 4--50 $\mu$W/$cm^2$ \cite{ref32} & speed is sufficient to power  &  Available   \\
			
				\hline
			
				&  &   &  &   & Not steadily    \\
				Rossi et al., 2015 \cite{ref31} &  &  & 0.5–1000 $\mu$W/$cm^2$ (Indoor) \cite{ref31} &    & Available and    \\ 
				Russo et al., 2017 \cite{ref30} & Light & Photovoltaic  & 5–100 mW/$cm^2$ (Solar) \cite{ref30} &  Predictable and Mature  & Expensive   \\
				
				\hline		
				
				Song et al., 2016 \cite{ref26} &  & &  &  &   \\				
				Monti et al., 2014 \cite{ref27} &  & Electromagentic,  &  & Highly efficient   & Risk of Material    \\
				Goudar et al., 2014 \cite{ref28} & Mechanical & Piezoelectric  & 0.819 $\mu$W/$cm^2$ \cite{ref26} &   and Controllable  & Breakage   \\	
				
				\hline
				
				Visser et al., 2008 \cite{ref35} &  &   &  & Continuous Available   &     \\
				Shah et al., 2016 \cite{ref36}, &  &   &  &  and Process Information  & Short Operating   \\	
				Xiao et al., 2016 \cite{ref37} & RF & Rectenna  & 0.01–0.3 $\mu$W/$cm^2$ \cite{ref35} &  and Simultaneously  & Range   \\	
				
				\hline	
			
				Yuan et al., 2014	\cite{ref38} &  &  &  &  &   \\	
				Kumar et al., 2012	\cite{ref39}, &  & Coherence  &  & Harvestable at  & Not Steadily   \\	
				Yuan et al., 2017 \cite{ref40} & Sound & Resonance  & 6.02 $\mu$W/$cm^2$ \cite{ref38} & Low Sound Level  & Available   \\
				\hline			
			\end{tabular}	
		}
		\label{energy-source2}
	\end{table}
	
		Energy source for powering electronics devices typically depends on the application. Solar, wind, and thermal are commonly chosen sources for energy harvesting enabled devices used in outdoor environments. Harvesting energy from mechanical vibration, such as wind and human movement, has been reported to feature energy autonomy to \ac{IoT} devices deployed in various crucial applications \cite{wind-selfpowered}, \cite{vibration-selfpowered}. Solar energy has been recognized as a proven energy source for powering energy-autonomous \ac{IoT} devices deployed in a wide range of applications such as health care \cite{wu-etal}, environmental monitoring \cite{ref96}, agriculture \cite{ref78}, solid waste management \cite{ref76}, air pollution monitoring \cite{ref79}, and underground water monitoring \cite{ref77}. It is considered as the most convenient renewable energy source in street light management systems as solar energy is a mature technology with high power density. A self-powered street light management system has been designed that harvests energy from sunlight \cite{ref75}. Furthermore, the utilization of thermal energy generated by direct sunlight has been harnessed as an energy source for powering a smart consumer light management system \cite{ref94}. On the otherhand, solar energy is not applicable in case of indoor applications. \ac{RF} and artificial light are widely utilized to power \ac{IoT} devices in indoor environments. Energy harvesting that implements \ac{NFC} has been used in various energy-autonomous devices deployed in indoor environments such as livestock farming \cite{NFC-livestock}, pH measurement of drinking water \cite{NFC-livestock2}, soil health monitoring in indoor plants \cite{NFC-livestock3}, freshness monitoring in fishes and packed food \cite{NFC-livestock4}, \cite{NFC-livestock5}. Subsequently, there is potential for energy harvesting from artificial light using highly efficient thin-film solar cells. It has been particularly used for batteryless devices, which require energy uninterruptedly. The ability to harness energy from sunlight and artificial light using a single, specially designed harvester is a significant advantage of light as an energy source. Therefore, several energy-autonomous devices that depend on solar energy are also capable of harnessing energy from artificial light sources \cite{wu-etal}, \cite{ref96}. It has also been used for sustainable consumer light management systems \cite{ref93}, \cite{ref94}. \\
	
\begin{figure}[h!]
		\centering
		\includegraphics[scale=0.33]{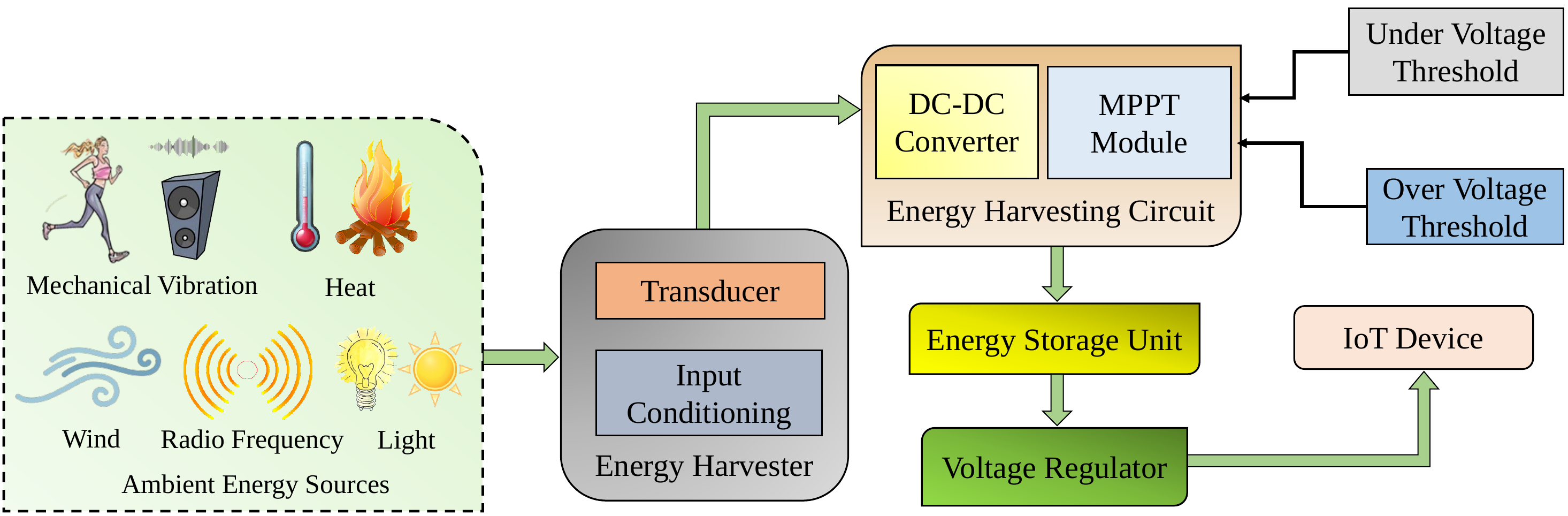}
		\caption{Comprehensive Block Diagram of Energy Harvesting and Power Management Technology}
		\label{EH-sources}
	\end{figure}

		The energy acquired then needs to be converted into electrical energy to be compatible with the intended use. The conversion process requires power conditioning circuits. In certain instances, the conversion process is straightforward, while for other resources, it necessitates complex circuitry and machinery to become functional. Fig. \ref{EH-sources} illustrates the comprehensive block diagram of energy harvesting and power management technology used to power \ac{IoT} devices.\\
	
	\subsection{Power Management and Energy Storage Technology}
	DC-DC conversion mechanism is used to convert the harvested energy through energy harvested into electrical energy. Specifically, it involves taking the variable voltage and current generated by the energy harvester and converting it into a stable and regulated voltage suitable for powering \ac{IoT} devices. The DC-DC conversion mechanism is essential for maximizing the efficiency of the energy harvesting system and ensuring a consistent power supply for the connected devices \cite{DC-DC}. There are mainly three types of converter mechanisms used in state-of-the-art energy-autonomous \ac{IoT} devices, which are boost converter, buck converter, and buck-boost converter.

	\begin{itemize}
	\item Boost Converter: It is used to step up the voltage obtained from the harvester. The increase in voltage is crucial to guarantee the efficient storage or utilization of the generated energy for devices that demand a higher voltage. It optimizes the efficiency of the energy harvesting process, enabling the efficient use of the harvested energy by increasing the voltage. It is highly adaptable and easily handles input voltages in a range for which it can generate output voltage compatible with \ac{IoT} devices, i.e., 0V to 5V, making it an essential component in energy harvesting systems.
	
	\item Buck Converter: It is used to lower the output voltage of an energy harvester. This type of converter is known for its high efficiency and is frequently used in energy harvesting systems to control voltage levels to power low-power devices. By reducing the voltage, the buck converter guarantees that the energy obtained from various sources like solar panels or vibration sensors is converted and stored at the most suitable voltage for the specific application. Furthermore, buck converter can prevent battery overcharging and protects sophisticated electronics from being exposed to excessive voltage levels.
	
	\item Buck-Boost Converter: It is used in a wide range of applications where there is a substantial variation in the input voltage, such as in the case of solar energy harvester. It converts efficiently and regulates the output voltage, even when the input voltage fluctuates, ensuring the smooth operation of the overall system. In addition, it has the ability to adjust voltage levels for various components in a circuit, making them a crucial and adaptable part of today's electronics.
	\end{itemize}

	In addition to DC-DC converter, \ac{MPPT} controller serves vital role in managing the process of enery harvesting in power management unit. \ac{MPPT} is used with variable power sources to maximize energy extraction as weather conditions vary. It is specifically useful in cases of devices that rely on solar energy. MPPT technology can increase energy output to a great extent compared to traditional fixed voltage systems by continuously adjusting the electrical load to match the optimal operating point of the power source. This not only maximizes the energy production of renewable sources but also improves the overall performance and longevity of the power system. \ac{MPPT} controller can be designed using several techniques such as hill-climbing, \ac{PO}, incremental conductance, \ac{FOCV}, fuzzy logic control, constant voltage, \ac{NN}, and \ac{GA}. Subsequently, the harvested energy gets stored in energy storage unit after being conditioned. There are several energy storage elements, such as fuel cells, non-rechargeable batteries, solid-state batteries, rechargeable batteries, capacitors, supercapacitors, and lithium-ion capacitors, that are used depending on the specifications of applications. A comparison between energy storage units is presented in Table \ref{storage-comp}.  
	\begin{table}[h!]
		\centering
		\caption{Comparision of Energy Storage Units used in Energy-Autonomous \ac{IoT} Devices}
		\begin{tabular}{l|l|l}
			\hline
			\textbf{Energy Storage Element}	& \textbf{Advantages}  & \textbf{Limitations}   \\
			\hline
			Fuel Cell & Reliable and Portable & Expensive  \\
			\hline	
			&  & Limited charge cycle, \\	
			& Rechargeable			& Less power density, \\
			Lithium-ion Battery &  and Cost effective & Harmful to environment \\	
			\hline
			Solid State Battery	& Thin and Flexible & Less power density \\
			\hline
			& Environmental friendly, &   \\
			& Huge charge-discharge cycle, & Affected by self-discharge, \\
			Supercapacitor & High power density &  Low energy density\\
			\hline
			& Low cost, Compact size,   &     \\
			& Long life span,    &     \\
			Capacitor & High power density   &   Low energy density  \\
			\hline
			& High power density, & Costly,   \\
			& Better energy density & Limited energy density \\
			Lithium-ion Capacitor & compared to Supercapacitor & compared to Battery \\		
			\hline
		\end{tabular}	
		\label{storage-comp}
	\end{table}

\section{Sustainable Computing Architectures for Consumer Light Energy Cyber Physical Systems}
\label{sust-comp}
\begin{figure}[h!]
	\centering
	\includegraphics[scale=0.45]{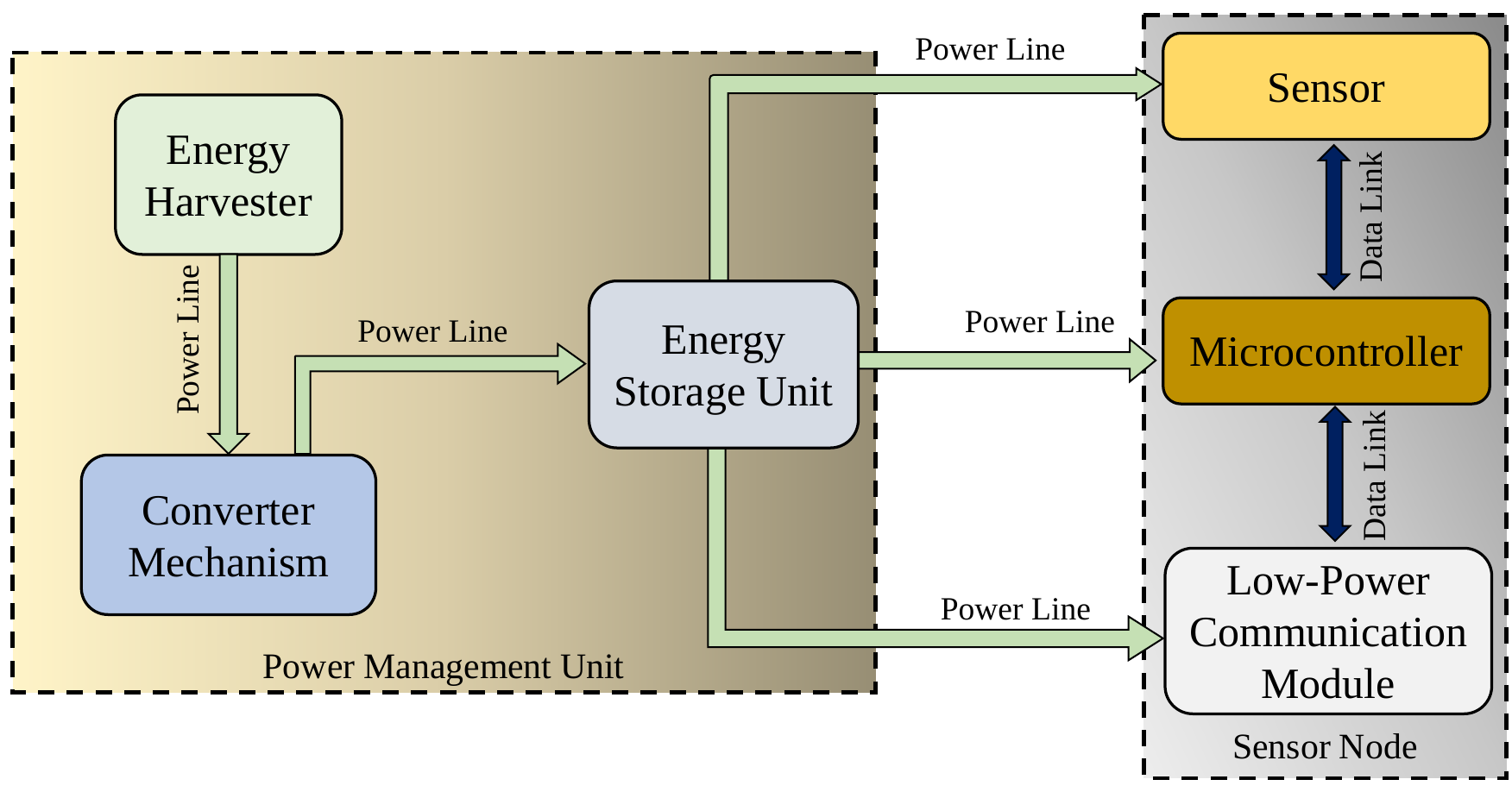}
	\caption{Sustainable Computing Architecture for Battery Driven Energy-Autonomous \ac{IoT} Devices Deployed in Outdoor Applications}
	\label{sust-framework1}
\end{figure}

Sustainable computing architecture is the basic building block of smart consumer light management system. It is responsible to minimize the energy consumption of the consumer light management systems thereby contributing to reduction in the size of energy harvesters, storage elements, device footprint, and overall manufacturing costs. Sustainable computing architectures reduce and optimize the energy consumption of the device by means of efficient use of hardware and software combinations. Consumer light management systems have been reported to attain enhanced energy efficiency with reduced ecological footprint through the implementation of various sustainable computing architectures. Recent development, testing, and validation of many sustainable \ac{IoT} devices have exhibited their ability to provide satisfactory functionality when used in real-world scenarios. These devices have been proven reliable, effective, and environment friendly, making them highly suitable for various applications. Fig. \ref{sust-framework1} shows the block diagram of a sustainable computing architecture, which is simple yet mostly opted in various devices.
In this architecture, energy can be harnessed using a suitable harvester based on application and then conditioned with a power management unit. The harvested energy is typically stored in an energy storage element, such as battery or supercapacitor bank. The sensor node draws power from the energy storage unit. Therefore, the storage unit effectively stores sufficient energy to help the sensor node to sustain its operation for longer duration. This architecture is commonly utilized for energy-autonomous \ac{IoT} devices that are deployed in outdoor applications and incorporate batteries as energy storage. The higher energy density of batteries compared to other energy storage elements enables the storage of harvested energy for extended durations, allowing the sensor node to operate uninterruptedly when the energy source is unavailable. The architecture has been used in a self-powered \ac{IoT}-enabled street light management system with energy-autonomous capability has been developed \cite{ref75}. A single PV harvester has been used to harvest energy, and a lithium-ion battery of 1200 mAh has been incorporated as an energy storage unit. The system manages the operation of \ac{LED} based street light. In \cite{ref78}, an \ac{IoT} system has been designed to monitor soil conditions continuously. The proposed system includes end nodes that are powered from solar energy. These units can be easily deployed on the field for long period. The system measures temperature, moisture, electrical conductivity, and carbon dioxide of the soil and transmits the information along with the geolocation of the end node using \ac{LoRaWAN}. The manufacturing cost of an end node is claimed to be 179.34 USD. Similarly, in \cite{ref76}, a solar-powered energy-autonomous device has been proposed incorporating the aforementioned architecture for trashbin monitoring. \ac{LoRaWAN} has been used for low-power communication. Power consumption has been optimized using duty cycle technique. The proposed system operates in real-time to assist the authority in effectively managing solid waste. Further, an \ac{IoT}-compatible low-cost sensor system has been designed to monitor air pollution \cite{ref79}. The device relies on solar energy with a rechargeable battery for power, and it is designed to withstand various weather conditions. It measures carbon monoxide, nitrogen dioxide, PM10, temperature and humidity. \ac{ANN} has been implemented to calibrate the sensor.  \\
\begin{table}[h!]
	\centering
	\caption{State-of-the-art Energy-Autonomous \ac{IoT} Devices incorporating Sustainable Computing Architecture}
	\resizebox{0.9\textwidth}{!}{
		\begin{tabular}{l|l|l|l|l}
			\hline
			\textbf{} & \textbf{Energy} & \textbf{Energy} & \textbf{Target} & \textbf{Average}     \\
			\textbf{Reference Works}	& \textbf{Source} &  \textbf{Storage} & \textbf{Application} & \textbf{Consumption}    \\
			\hline
			&  & Rechargeable  & Soil Health &  \\
			Zhang et al., 2019 \cite{ref81} & Sunlight & Battery, 10000 mAh & Monitoring & N/A  \\
			\hline
			&  & Rechargeable  & Soil Health &  \\
			Sadowski et al., 2020 \cite{ref82} & Sunlight & Battery, 6600 mAh & Monitoring & 29.33 mW  \\
			\hline
			&  & Rechargeable  & Air Quality &  \\
			Bhusal et al., 2020 \cite{ref80} & Sunlight & Battery, 2000 mAh & Monitoring & 7 mA  \\
			\hline
			&  & Rechargeable  & Street Light &  \\
			Mohanty et al., 2021 \cite{ref75} & Sunlight & Battery, 1200 mAh & Management & N/A  \\
			\hline
			&  & Rechargeable  & Soil Health &  \\
			Kombo et al., 2021 \cite{ref77} & Sunlight & Battery, 5000 mAh & Monitoring & 104.081 mW  \\
			\hline
			&  & Rechargeable  & Soil Health &  \\
			Ramson et al., 2021 \cite{ref78} & Sunlight & Battery, 2500 mAh & Monitoring & 13 mA  \\
			\hline
			&  & Rechargeable  & Air Pollution & $\approx$ 14.306 mW  -- \\
			Ali et al., 2021 \cite{ref79} & Sunlight & Battery, 4000 mAh & Monitoring & 14.97 mW  \\
			\hline	 
			&  & Rechargeable  & Soil Health &  \\
			Houssaini et al., 2020 \cite{ref86} & Sunlight & Battery, 6600 mAh & Monitoring & 29.33 mW  \\
			\hline		
			&  & Rechargeable  & Solid Waste &  \\
			Ramson et al., 2022 \cite{ref76} & Sunlight & Battery, 2500 mAh & Management & 1.5 mA  \\
			\hline
		\end{tabular}	
	}
	\label{sust-arch-comp1}
\end{table}

A cost-effective and efficient wireless sensor network has been designed to monitor groundwater levels in Zanzibar, Tanzania \cite{ref77}. It can harvest energy from sunlight and reserves the energy using lithium-ion battery. This network provides timely data to aid in making decisions regarding groundwater resource management. The manufacturing cost of the system ranges from 350 USD to 400 USD. Duty cycle technique has been implemented to optimize power consumption, keeping it below 1\%. In contrast, similar system has been designed for the same application, which is capable of harvesting energy from artificial light in addition to sunlight \cite{ref80}. One of the limitations of state-of-the-art energy-autonomous systems is limited communication range. Efforts have been made to enhance the communication range by incorporating \ac{LoRa} and \ac{NB-IoT} technology into a soil health monitoring system \cite{ref81}. The system harvests energy using a solar panel and relies on recharagble battery as energy storage unit. It has been reported that the communication range has been increased up to 1.6 KM in complex environments. The system requires an average of 2 mA to achieve this range with a packet loss rate of approximately 3\%. Subsequently, further attempts have been made to enhance communication in solar energy powered \ac{IoT} devices using three major \ac{IoT} communication protocols such as  Zigbee, \ac{LoRaWAN}, and \ac{Wi-Fi} \cite{ref82}. The proposed device in this work has undergone experimental testing in an application related to monitoring soil health. A state-of-the-art comparison among the devices designed with the aforementioned architecture has been provided in Table \ref{sust-arch-comp1}. \\

\begin{figure}[h!]
	\centering
	\includegraphics[scale=0.49]{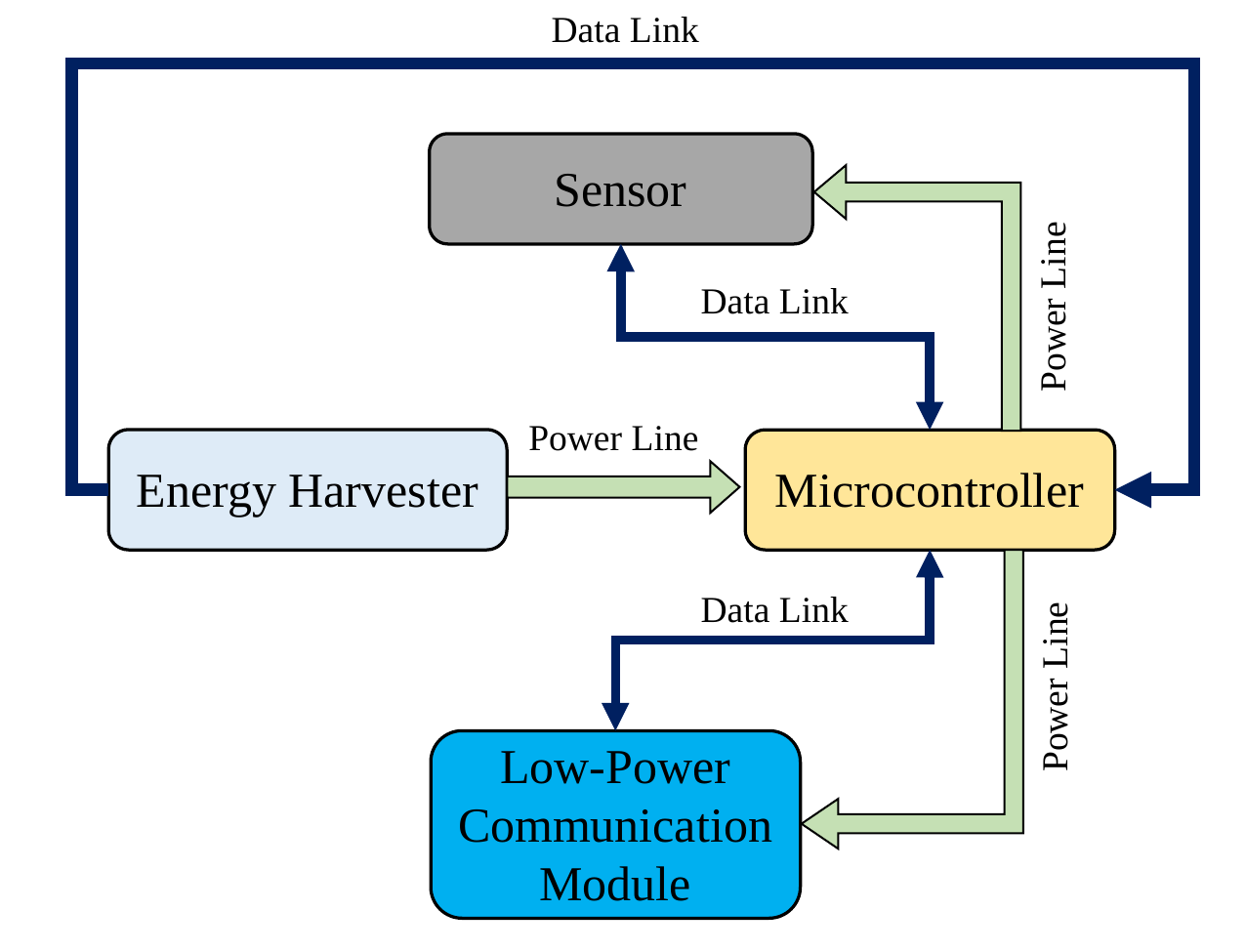}
	\caption{Sustainable Computing Architecture Utilizing Harvester as Transducer}
	\label{sust-framework2}
\end{figure}

There is a growing focus on reducing reliance on battery technology due to its negative environmental impact. It has become a significant area of research. Therefore, supercapacitors, capacitors, and lithium-ion capacitors have emerged as the leading alternatives to rechargeable batteries. Nevertheless, the energy density of these elements is significantly lower than that of batteries, making it extremely difficult to utilize them in \ac{IoT} devices. In order to achieve sustainability in a battery-less system, it is crucial to ensure continuous energy harvesting. Fig. \ref{sust-framework2} illustrates the advanced architecture of battery-less sustainable \ac{IoT} devices. 
In this architecture, the harvester serves a dual purpose by functioning as a transducer within the system while also harvesting energy. Furthermore, there is no use of an external complicated power management unit. Instead, it utilizes the internal power management of the microcontroller called \ac{PVD} to accurately determine the output voltage level of the harvester. This architecture utilizes an in-built low-power timer along with \ac{PVD} for measurement operations. This architecture provides the benefit of reducing the number of sensors in a system and eliminating the need for additional power management, resulting in a simpler design. An ambient sunlight-powered wireless sensor platform using a single photovoltaic transducer has been designed to measure the same ambient light \cite{ref83}, \cite{ref84}. This wireless system offers a convenient and efficient solution. Additionally, it can measure temperature and humidity. The device is \ac{BLE}-enabled and has been implemented in measuring indoor light intensity up to 200 lux. Furthermore, the system has undergone thorough testing to demonstrate its compatibility with smart agriculture applications \cite{ref85}. Furthermore, a system has been developed based on the aforementioned architecture to measure electrical impedance for evaluating the water stress level in plants \cite{ref88}. Experimental tests demonstrated a reading error rate of less than 15\% when working with impedance modules up to 180 k$\Omega$. The major limitation in these systems is the communication range due to the use of \ac{BLE} as communication module. Therefore, \ac{LoRaWAN} has been implemented as a communication unit to address the mentioned problem \cite{ref87}. The proposed system can communicate a maximum of 30 bytes subject to the \ac{SF}, and the bandwidth of LoRa is set at 7 and 125 KHz, respectively. The implementation of such architecture has also been achieved in the healthcare system, specifically in monitoring human metabolism through glucose monitoring \cite{ref89}. \\ 

\begin{figure}[h!]
	\centering
	\includegraphics[scale=0.47]{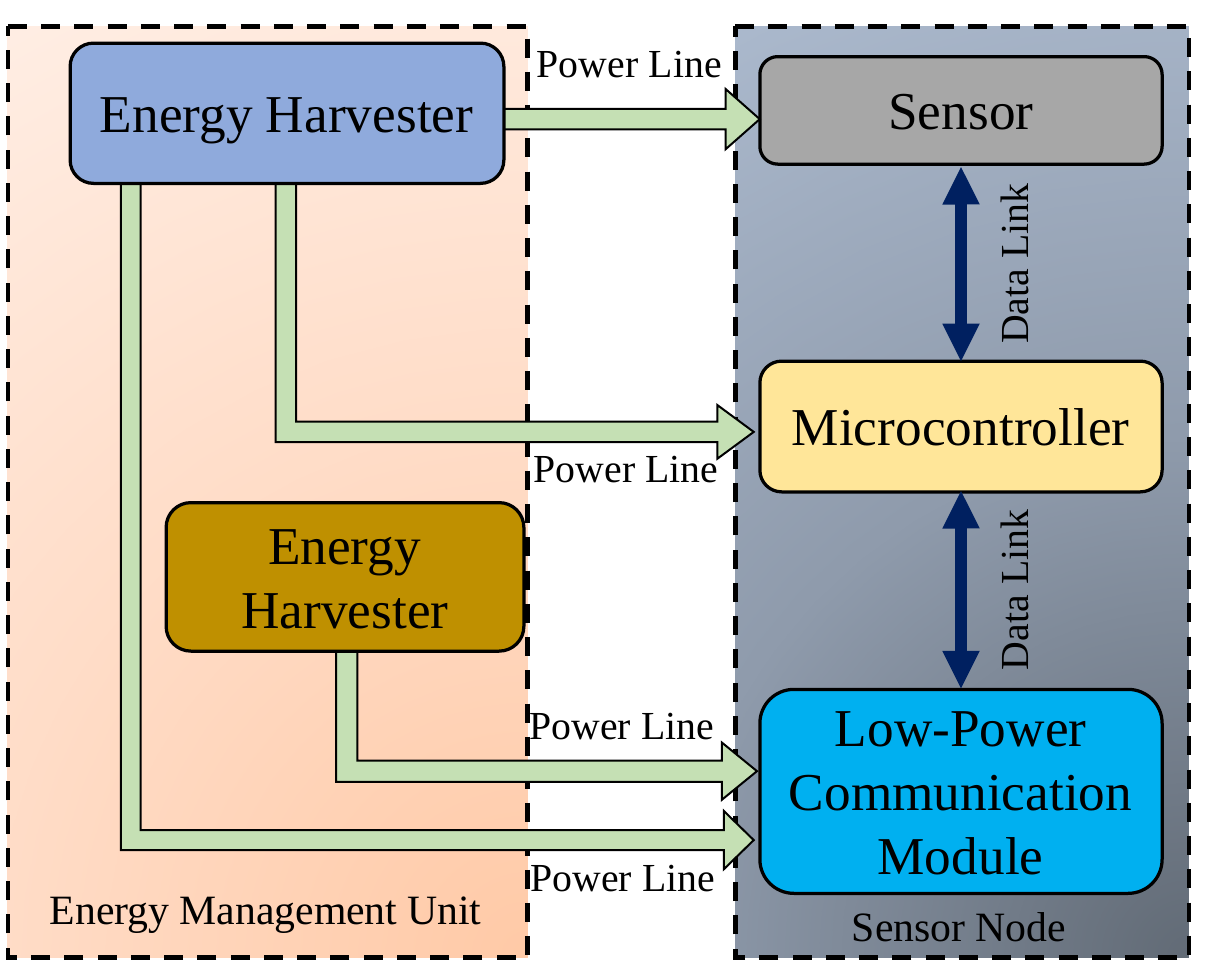}
	\caption{Sustainable Computing Architecture Leveraging Multiple Harvesters to Overcome Communication Range Issue in Energy-Autonomous IoT Devices}
	\label{sust-framework-3}
\end{figure}

Communication unit in an \ac{IoT} device accounts for over 50\% of the energy consumption of the device. It is a significant issue, especially in \ac{IoT} devices that rely on energy harvesting. The availability of energy is crucial in these devices. Therefore, a range of low-power communication protocols, including \ac{LoRaWAN}, \ac{BLE}, SigFox, and \ac{NB-IoT}, alongside suitable power optimization techniques, are incorporated into these devices to minimize overall power consumption. It can be achieved with systems capable of harnessing energy from renewable sources with a high power density, such as sunlight and wind. Conversely, when addressing energy sources characterized by low power density such as \ac{RF}, acoustic, and vibration, there can be drawbacks. These drawbacks may result in decreased communication range and lower \ac{QoS}. Fig. \ref{sust-framework-3} shows a state-of-the-art architecture used in energy harvesting enabled \ac{IoT} devices to overcome the aforementioned issue.
In \cite{ref90}, a battery-less \ac{RFID} wireless sensor has been designed that is tested with environment temperature monitoring and golf ball tracker application. The device primarily harvests energy from \ac{RF} source in backscattering principle, and additionally, thin-film solar cell has been incorporated to harvest energy from artificial light to power the device to enhance communication range. The reported range has been improved to 15 meters to 20 meters with a variation of +/-2 meters, achieved through a frequency sweep from 800 to 1000 MHz. The performance is significantly higher compared to passive \ac{RFID}, with a boost of 6-10 times. Subsequently, a perovskite solar cell has been fabricated of 10.1\% efficiency to additionally power the \ac{RFID} sensor \cite{ref91}. \\

\begin{figure}[h!]
	\centering
	\includegraphics[scale=0.38]{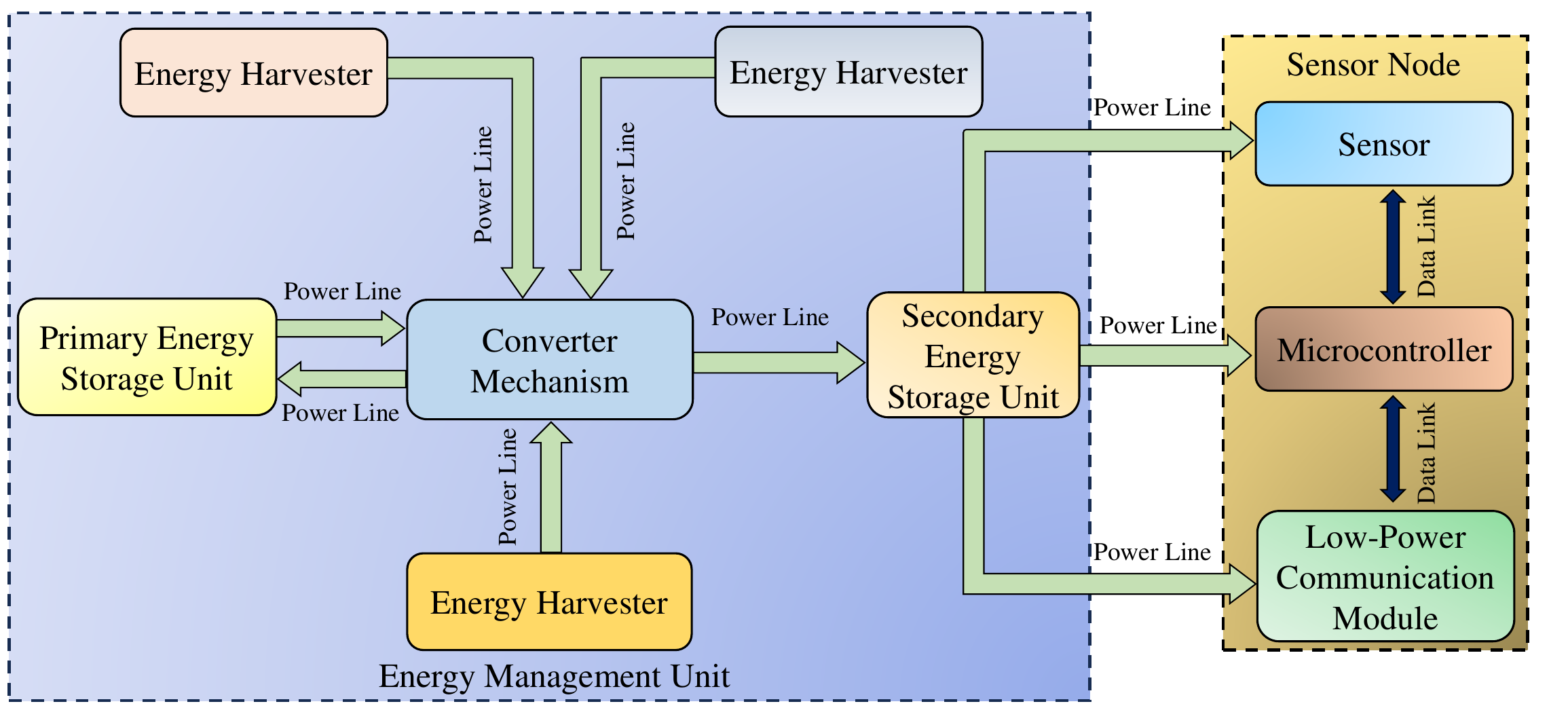}
	\caption{Sustainable Computing Architecture Based on Hybrid Power Management in Energy-Autonomous IoT Devices}
	\label{sust-framework4}
\end{figure}

\begin{figure}[h!]
	\centering
	\includegraphics[scale=0.58]{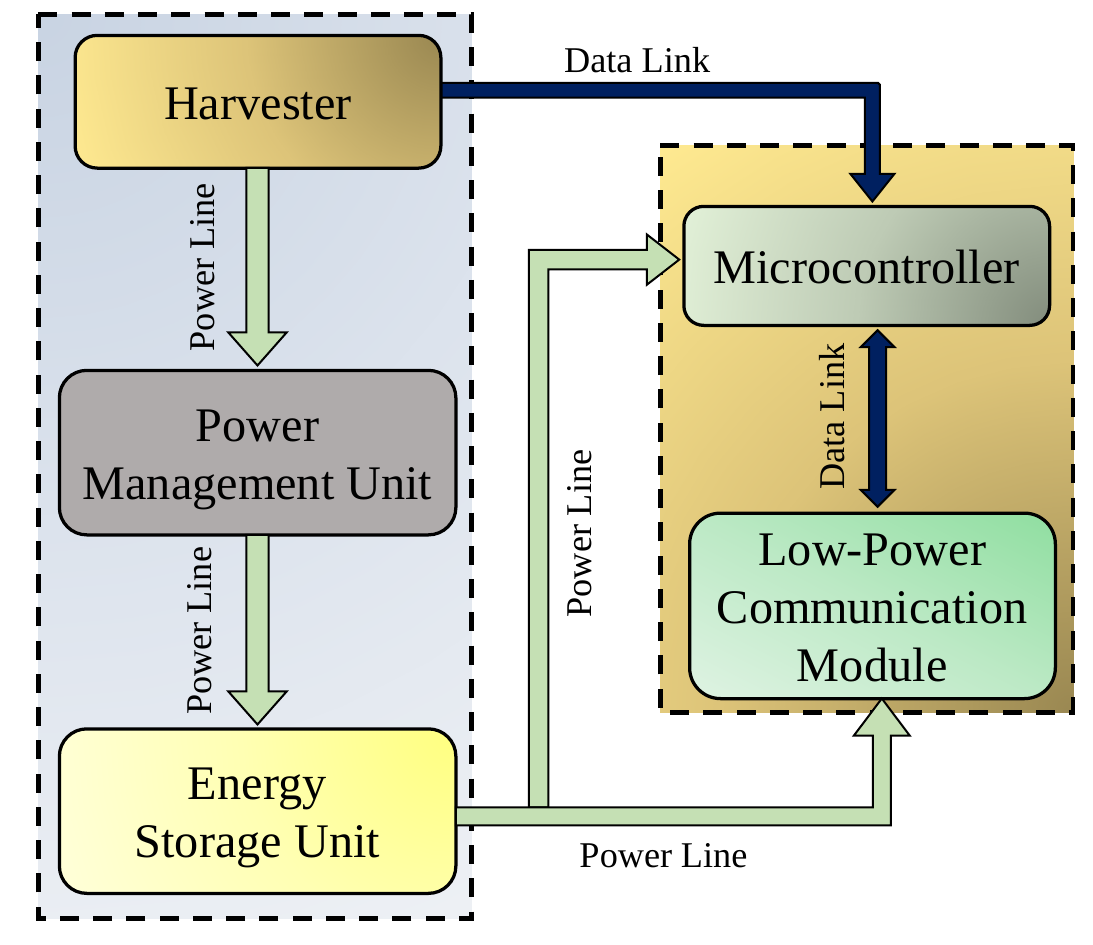}
	\caption{Sustainable Computing Architecture Employing Harvester as Transducer in Addition to Powering the Device}
	\label{sust-framework5}
\end{figure}

One of the main challenges faced by battery-less \ac{IoT} devices is their extremely limited energy storage capacity. Supercapacitors and capacitors, unlike batteries, have lower energy storage density. Therefore, if the energy source experiences fluctuations or interruptions, the performance of the device is often affected due to insufficient energy. This issue has been solved in two approaches as outlined below in state-of-the-art energy-autonomous \ac{IoT} devices. 
\begin{itemize}
	\item Primary energy storage elements, along with secondary storage elements, can be utilized to ensure a continuous power supply in situations where energy availability from renewable sources is limited. Consider, for instance, the crucial importance of energy availability during the rainy season for a solar energy powered \ac{IoT} device. Therefore, the primary energy storage component guarantees an uninterrupted power supply to the sensor node, resulting in unaffected performance of the device. 
	\item Integration of multiple harvesters allows for continuous energy harvesting from the same or various energy sources, regardless of the time. Thus, the device can ensure that the sensor node gets power uninterruptedly, irrespective of time.
\end{itemize}
Fig. \ref{sust-framework4} illustrates the architecture of a state-of-the-art energy-autonomous \ac{IoT} device that uses a hybrid storage element to overcome the issues reported in battery-less devices.
A dual energy harvesting enabled battery-less street light management system has been proposed to operate street lights energy efficiently \cite{ref93}. It incorporates a single thin-film solar cell that can harvest energy under sunlight as well as artificial light, i.e., from the street light. It incorporates a supercapacitor as an energy storage unit and utilizes \ac{LoRaWAN} for long-range communication. Subsequently, a similar system has been proposed that can harvest energy from solar and solar thermal energy \cite{ref94}. This work addresses the issues encountered in \cite{ref93}. Thus, dual-energy storage elements, including a capacitor and lithium-ion capacitor, have been incorporated into the proposed hybrid energy management scheme. The system employs the capacitor to provide power to the sensor node during day and relies on the lithium-ion capacitor for power during night. Furthermore, a battery-less \ac{LoRaWAN} enabled \ac{IoT} device has been developed and evaluated for monitoring environmental parameters \cite{ref96}. It can harvest energy from solar and artificial light. This device implements a supercapacitor as its energy storage unit, and the size of the supercapacitor has been optimized to meet the specific requirements of the application. A hybrid energy storage and management solution has been proposed for a \ac{LoRaWAN} enabled sensor \cite{ref92}. It harvests energy from sunlight using a 330 mW PV panel. Supercapacitor has been used along with lithium-ion batteries as storage units. Battery has been used as a secondary energy storage element for power backup. The device has undergone thorough testing and validation for a specific application related to monitoring environmental parameters like temperature, humidity, and pressure on \ac{LoRaWAN}. Furthermore, a dual thermal energy harvesting scheme has been proposed for water flow rate measurement device \cite{ref95}. It employs a specially designed energy management unit to manage energy consumption effectively. It incorporates ultrasonic sensors, a task-based computing scheme, and a \ac{LoRa} module to sense and report the flow rate independently. In \cite{ref97}, a \ac{LoRaWAN} enabled environmental data and water quality monitoring device has been proposed. In this work, energy has been harvested combinedly from sunlight and thermal differences created between the water surface and materials exposed to sunlight through \ac{TEG} for powering \ac{LoRa} enabled device. There has been an ongoing trend toward utilizing the output signals from energy harvesters to extract contextual information instead of using specific sensors. The harvester provides valuable contextual information about the deployment environment due to the diverse nature of the harvested energy. Fig. \ref{sust-framework5} shows the typical sustainable architecture energy-autonomous \ac{IoT} devices in which the harvester has been used as a transducer in addition to powering the device.
The architecture has the potential to provide a cost-effective design, minimal use of electronics, simple design, and a smaller form factor. In this direction, \ac{TEG} has been used to harvest energy from heat to power the device and additionally provide temperature information to the same \cite{ref100}, \cite{ref101}, \cite{ref98}. It has proven to be valuable in a range of important applications, including detecting human activity, monitoring chemical reactions, and measuring water flow. On the other hand, in addition to measuring light intensity through PV harvester \cite{ref83}, \cite{ref84}, monitoring or detecting human activity and gesture recognition has been accomplished in PV harvester integrated devices designed using the aforementioned sustainable architecture \cite{ref104}, \cite{ref103}, \cite{ref102}. \\

Subsequently, a system has been proposed utilizing \ac{RF} harvesting to monitor hand hygiene in humans \cite{ref105}. It shows results in detecting hand gestures, with an error rate of less than 8\%. The device has the potential to be highly beneficial in healthcare applications. Furthermore, efforts have been made to recognize hand gestures in humans using \ac{RF} harvester through the backscattering principle \cite{ref106}. A battery-less \ac{RF} device has been designed in \cite{ref107}, which employs an \ac{RF} harvester to detect touch and gesture applicable to various real-world scenarios. Further, \ac{KEH} has been used for transportation mode detection in addition to power the device. It has the ability to detect and analyze the vibrations that passengers experience while using various modes of transportation. Further, \ac{KEH} has been used for transportation mode detection in addition to powering the device \cite{ref108}. It has the ability to detect and analyze the vibrations that passengers experience while using various modes of transportation. The device demonstrates an impressive overall accuracy of over 92\% in classifying five modes while significantly reducing system power consumption by more than 34\% compared to traditional accelerometer-based approaches. Battery-less energy-autonomous wearable device incorporating kinetic energy harvester has been designed to detect human activity with 95\% accuracy \cite{ref109}. It has been reported that the device consumes 57\% less power than the conventional motion sensor-based approaches. Further efforts have been made toward gait detection using \ac{KEH} enabled energy-autonomous device \cite{ref110}. This technique of gait detection has been proven more energy efficient, specifically achieving 82.5\% less energy consumption compared to the accelerometer-based approach. The device detects gait against spoofing attackers, achieving minimum \ac{EER} of 11.2\%. A summary of the latest advancements in the field has been compiled in Table \ref{eh-comp}.  
\begin{table}[h!]
	\centering
	\caption{Energy-Autonomous IoT Devices Designed using Harvester as Sensor or Transducer}
	\begin{tabular}{l|l|l}
		\hline
		\textbf{Research Works} & \textbf{Energy Source} & \textbf{Target Application} \\
		\hline
		Campbell et al., 2014 \cite{ref100} & Thermal & Water and applience metering \\
		\hline		
		Zarepour et al., 2017 \cite{ref101} & Thermal & Detection of chemical reaction \\
		\hline	
		Proto et al., 2018 \cite{ref98}  & Thermal & Activity detecion of human \\
		\hline
		Varshney et al., 2017 \cite{ref104} & Sunlight & Hand gesture recognition of human \\
		\hline
		Ma et al., 2019 \cite{ref103} & Sunlight & Gesture recognition of human \\		
		\hline	
		Sandhu et al., 2021 \cite{ref102} & Sunlight & Activity detection of human \\
		\hline
		Khamis et al., 2020 \cite{ref105} & RF & Hand hygiene monitoring in human\\			
		\hline	
		Wang et al., 2018 \cite{ref106} & RF & Hand gesture recognition of human \\			
		\hline		
		Pradhan et al., 2017 \cite{ref107} & RF & Touch detection of human \\			
		\hline
		Lan et al., 2020 \cite{ref108} & Kinetic & Transport mode detection \\			
		\hline
		Lan et al., 2020 \cite{ref109} & Kinetic & Activity detection of human \\			
		\hline
		Xu et al., 2019 \cite{ref110} & Kinetic & Activity detection of human \\			
		\hline
	\end{tabular}	
	\label{eh-comp}
\end{table}

\section{Task Scheduling and Energy Optimization Techniques for Consumer Light Energy Physical Systems}
\label{task-sc}
A significant limitation in state-of-the-art \ac{IoT} enabled consumer light management systems is the lack of efforts provided to power consumption. Power consumption is currently a major focus of research for designers and scientists due to its impact on the overall performance and lifespan of \ac{IoT} devices. Addressing power consumption issues in \ac{IoT} devices demands a focus on developing energy-efficient algorithms and optimizing hardware design. Devices can operate for longer periods of time without needing to be recharged or have their batteries replaced by reducing power consumption. As a result, it lessens the environmental impact of discarded batteries and enhances user experience. In addition, it offers designers a competitive advantage by allowing them to seamlessly integrate energy harvesting technology with \ac{IoT}. Further, power consumption in \ac{IoT} devices is influenced by their task scheduling to a great extent. Optimizing task scheduling can decrease power consumption by ensuring that devices are only active when needed. \ac{IoT} devices can enhance their operational efficiency and prolong battery lifespan through strategic management of task completion. In addition, implementing a well-organized task scheduler can effectively prioritize essential tasks and optimize resource allocation, resulting in significant long-term power savings. There are three most standard energy optimization and task scheduling strategies implemented in state-of-the-art \ac{IoT} devices as outlined below.
\begin{itemize}
	\item \ac{DVFS}
	\item Decomposing and Combining of Tasks 
	\item Duty cycle
\end{itemize}

\begin{figure}[h!]
	\centering
	\includegraphics[scale=0.4]{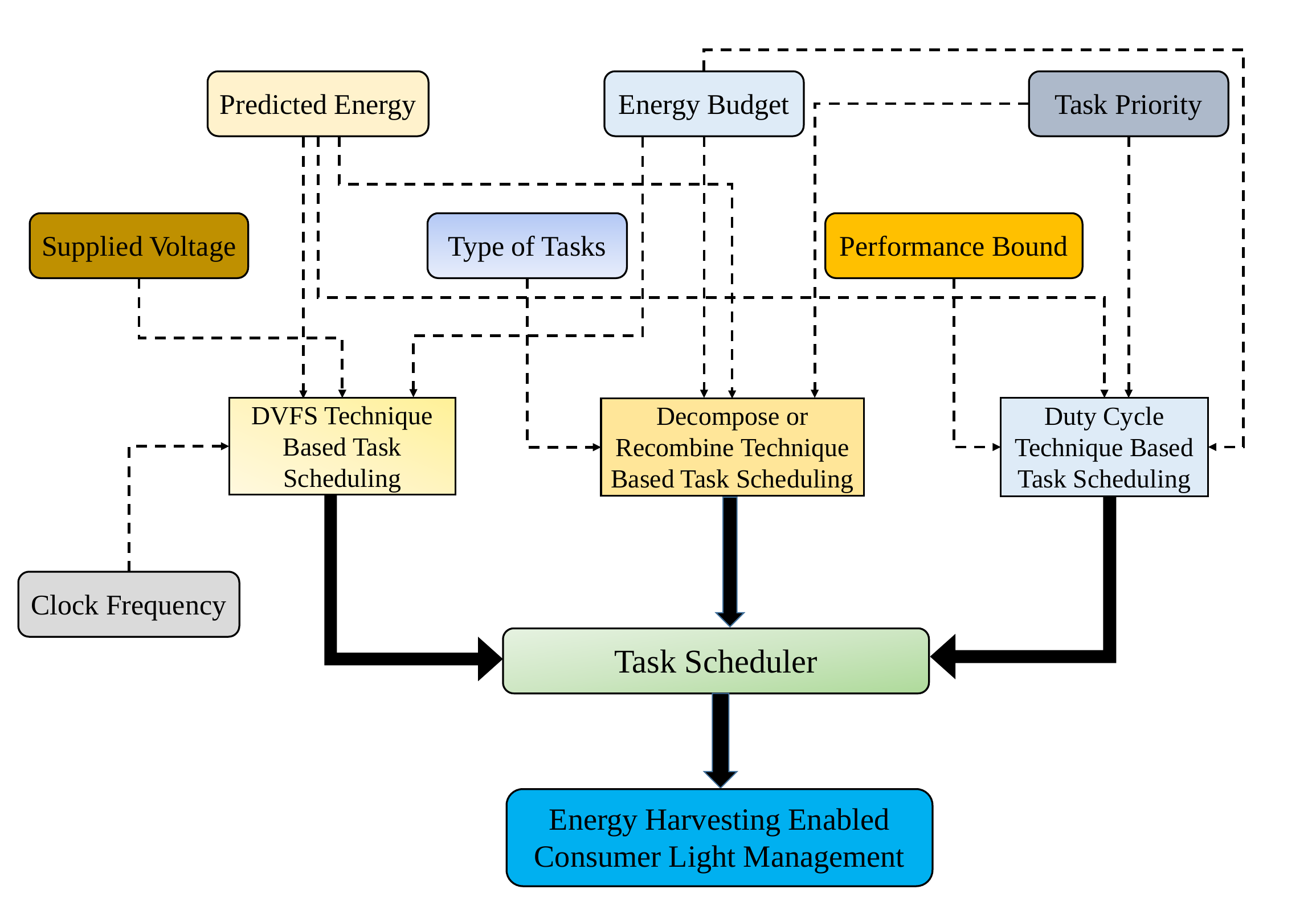}
	\caption{Strategies for Power Consumption Optimization and Task Scheduling in \ac{IoT}-enabled Consumer Light Management System}
	\label{task-scheduling-schemes}
\end{figure}

Fig. \ref{task-scheduling-schemes} illustrates strategies for optimizing energy consumption and scheduling in consumer light management systems. Implementing these techniques is not only restricted to help optimize energy consumption but also enables the designer to maintain high \ac{QoS} in the device by executing tasks as per priority level. 

\subsection{Dynamic Voltage and Frequency Scaling (DVFS)}
The power consumption of the sensor nodes is impacted by factors such as the supplied voltage, current flow, and the operating frequency, specifically the clock frequency. Thus, it can be optimized by adjusting both of these parameters in real time. The objective \ac{DVFS} technique is to either optimize performance within the constraints of a finite energy budget or to reduce energy consumption within a performance bound. This technique offers a dynamic approach to lower power consumption of the microprocessor or microcontroller in the sensor node by adjusting the supply voltage and frequency according to the specific performance needs of the application. There are certain applications in which the system achieves high performance for a very short period. Thus, it presents an opportunity to save a significant amount of energy \cite{ref41}. Real-time applications may exhibit variations in \ac{AET}, which often finish earlier than the estimated \ac{WCET}. \ac{DVFS} technique takes advantage of these variations in the workload to adaptively modify the supply voltage and frequency of the system \cite{ref42}. The relation between dynamic power consumption ($P_{dyn}$) of the sensor node with clock frequency and supplied voltage can be expressed as:
\begin{equation}
	{{P}_{dyn}}\alpha {{f}_{clock}}{({V_{supply}})^{2}}
	\label{dvfs-eqtn};
\end{equation}

$P_{dyn}$ decreases quadratically upon supply voltage ($V_{supply}$) gets reduced linearly. On the other hand, a decrease in $V_{supply}$ also leads to a decrease in the switching speed of the CMOS transistors. As a result, the maximum clock frequency of the microprocessor or microcontroller eventually gets reduced. The clock frequency below which the microprocessor or microcontroller is not capable of executing tasks is called critical frequency. Thus, it automatically adjusts its frequency to a higher level if the requested frequency is unavailable. It is crucial for the microprocessor or microcontroller to be aware of the duration of the idle period in order to select the most effective low-power state. \ac{DVFS} has been gradually incorporated into most smart handheld devices, and it has been widely adopted by laptop computers, server processors, and mobile devices to conserve energy. \\

In \cite{ref43}, a method has been proposed combining \ac{DVFS} and workload scheduling. The method has been implemented in a heterogeneous multicore \ac{WSN} device. The workload of tasks has been adjusted with the power mode of the hardware in order to satisfy the varied real-time power budget. \ac{ML} based \ac{DVFS} framework has been proposed in \cite{ref44}. The framework decides the voltage-frequency scaling suitable to the characteristics of the task and the configuration of the processor. It strikes out the requirement of explicit modeling of processor configuration and is capable of suiting itself to various environments without the support of any external supervisor due to the learning ability. Further, \ac{DVFS} has been implemented on ARM-based off-the-shelf microcontroller \cite{ref45}. It has been reported that the aforementioned approach is better in terms of energy saving than \ac{DVFS} being applied only for \ac{CPU}. Attempts have been made to implement \ac{DVFS} technique for ultra-low power embedded devices \cite{ref46}. The proposed method ensures that the system can seamlessly adjust energy levels based on load requirements in real time. It enables the system to regulate its operating frequency based on computational demand, operating at lower levels when demand is low and scaling up for maximum efficiency when more power is required. The overall energy consumption of the system has been reduced from 24.74\% to 47.74\%. \\

Efforts have been made to minimize the power consumption of wireless sensor nodes by meticulously selecting low-power microcontrollers and implementing \ac{DVFS} technique in the power management scheme \cite{ref47}. The implementation of the \ac{DVFS} technique is not only restricted to \ac{WSN} or \ac{IoT} devices but is also used in cloud computing. In \cite{ref48}, supply voltage and frequency for servers in cloud computing is tunned incorporating \ac{DVFS} technique. As a result, the energy consumption of the server has been reported to be reduced significantly, maximum up to 25\% during the period when the server handles light workloads. Further, \ac{DVFS} has been used for scheduling tasks in various \ac{IoT} devices. In \cite{ref49}, \ac{DVFS} has been explored for task scheduling in a power management framework. The proposed algorithm efficiently schedules tasks based on forecasted energy, energy budget, and task deadlines. A task scheduling algorithm has been proposed, which has the capability to decide whether the tasks should be offloaded to a cloud server or executed locally \cite{ref50}. This algorithm utilizes \ac{DVFS} to adjust the frequency of the cores in a multicore heterogeneous processing architecture of mobile devices. \ac{RL} based \ac{DVFS} method has been proposed in \cite{ref51} for task scheduling. This approach has been found to reduce power consumption in multicore embedded systems while maintaining reliability for executing sporadic tasks and meeting deadlines. \\

\ac{DVFS} algorithm is not suitable for scheduling tasks on energy harvesting-based battery-less sensors because of the limitations of resource-constrained hardware. The energy harvesting circuits are intentionally designed to be simple in order to minimize energy losses. However, this simplicity may limit the availability of different voltage levels for performing tasks on the sensor node. As a result, alternative task scheduling algorithms can be used to seamlessly minimize energy consumption in resource-constrained sensor nodes without any additional overhead in terms of energy and resources.

\subsection{Decomposing and Combining of Tasks}
\begin{figure}[!h]
	\includegraphics[scale=0.6]{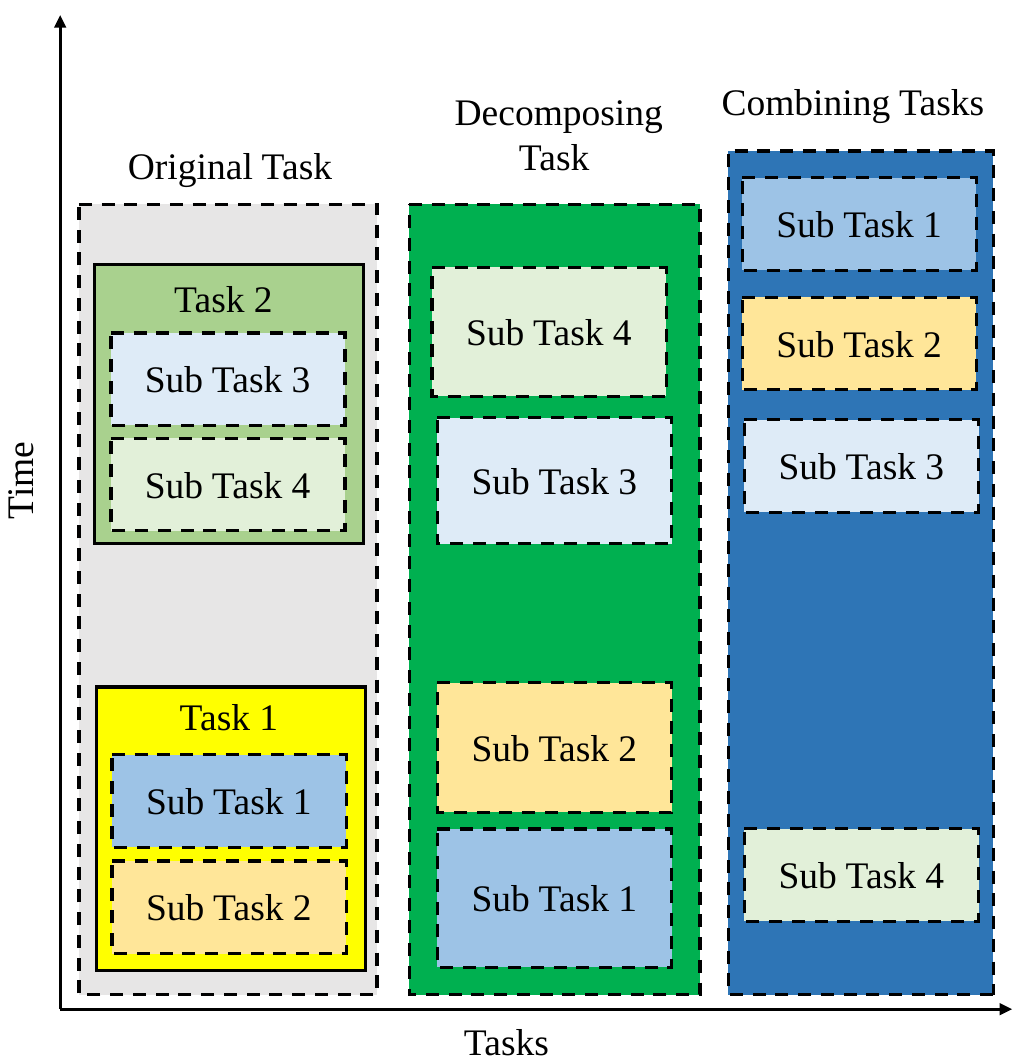}
	\centering
	\caption{Decomposing and Recombining Tasks}
	\label{Task-scheduling}
\end{figure}

This algorithm breaks down the energy-intensive tasks into smaller subtasks, resulting in reduced energy consumption during their execution. Typically, this technique involves four phases that involve decomposing and combining, such as decomposition, combining, admission control, and optimization. The functioning of this technique has been graphically presented in Fig. \ref{Task-scheduling}.
\begin{itemize}
	\item Decomposition: Energy-intensive tasks are broken down into various subtasks based on their potential to be combined with other subtasks to save energy. If the energy harvested is insufficient to sustain continuous high-powered tasks, the subtasks can still be carried out using the limited energy available. 
	\item Combining: This phase involves the integration of various subtasks that can be performed on a single processor in order to reduce energy usage. Furthermore, there are certain tasks that can be simultaneously executed and reduce the idle time of the processor. Simultaneous execution offers the benefit of decreased delay and lower latency in task execution. However, it also necessitates more energy, which is only available intermittently in energy harvesting enabled \ac{IoT} sensor nodes.
	\item Admission Control: During this phase, the tasks are sorted based on their priority and the energy consumed when executed. While efforts are made to optimize the energy consumption of the system; still, harvested energy is very often insufficient to ensure that all the ready-to-go tasks are executable. Thus, the admission controller takes into account the priority of tasks, the amount of harvested energy available, and the energy consumption of tasks to filter the tasks. It considers two major aspects of tasks in order to arrange the tasks, i.e., task priority and task deadline. Task priority has significant importance specifically in time-critical real-time applications. Task deadline is categorized into two types such as hard deadlines and soft deadlines. Hard deadlines are very critical, and missing hard deadlines can cause major losses to the system and often can be fatal. On the other hand, soft deadlines are critical for the application; however, unlike hard deadlines, violation of soft deadlines does not create severe consequences.
	\item Optimization: In this phase, the execution of tasks is optimized by considering the available energy, the number of task executions, and the energy consumption of each task. This phase further filters the tasks to ensure the efficient utilization of the harvested energy. The availability of additional energy is crucial in determining the implementation of time-sensitive tasks in the future. The system needs to ensure sufficient energy is stored in the storage unit in order to ensure the execution of future tasks with hard deadlines.
\end{itemize}

Multicore processors are widely used in various computing systems, including general-purpose computers and real-time embedded systems. In \cite{ref52}, a decomposition-based scheduling algorithm has been proposed for real-time tasks. The proposed technique showcases a solution to address the issue of dividing subtasks into multiple parts, resulting in a reduction of the overall density of the \ac{DAG} tasks following decomposition. In \cite{ref53}, a scheduling technique for parallel real-time tasks has been proposed. The decomposition algorithm takes the structural characteristics of each task into account with the objective of enhancing schedulability. Schedulability tests have been done for the global \ac{EDF} scheduling algorithm using decomposition techniques. A combination of Global \ac{EDF} with restricted migration and density separation has been reported to perform better in terms of schedulability. Further, a method based on granularity is proposed for task decomposition analysis for manufacturing systems \cite{ref54}. In this method, small subtasks are recombined and organized based on the desired level of detail, resulting in the decomposition of the manufacturing tasks. This work introduces a mathematical model and an algorithm to solve the problem of subtask scheduling efficiently. Subsequently, multi-granularity task decomposition and hierarchical task scheduling have been investigated within a collaborative computing network spanning the cloud, edge, and end devices \cite{ref55}. \ac{LSTM} based resource prediction scheme has been used for task decomposition in this work. Nevertheless, in the case of energy harvesting-enabled and batteryless \ac{IoT} devices, it is challenging for the device to execute multiple tasks simultaneously rather than a specific portion of a task at a time due to the availability of limited energy \cite{ref56}. Thus, this technique is not typically considered to be incorporated for task scheduling in state-of-the-art energy-autonomous \ac{IoT} devices.

\subsection{Duty Cycle}
\begin{figure}[h!]
	\includegraphics[scale=0.4]{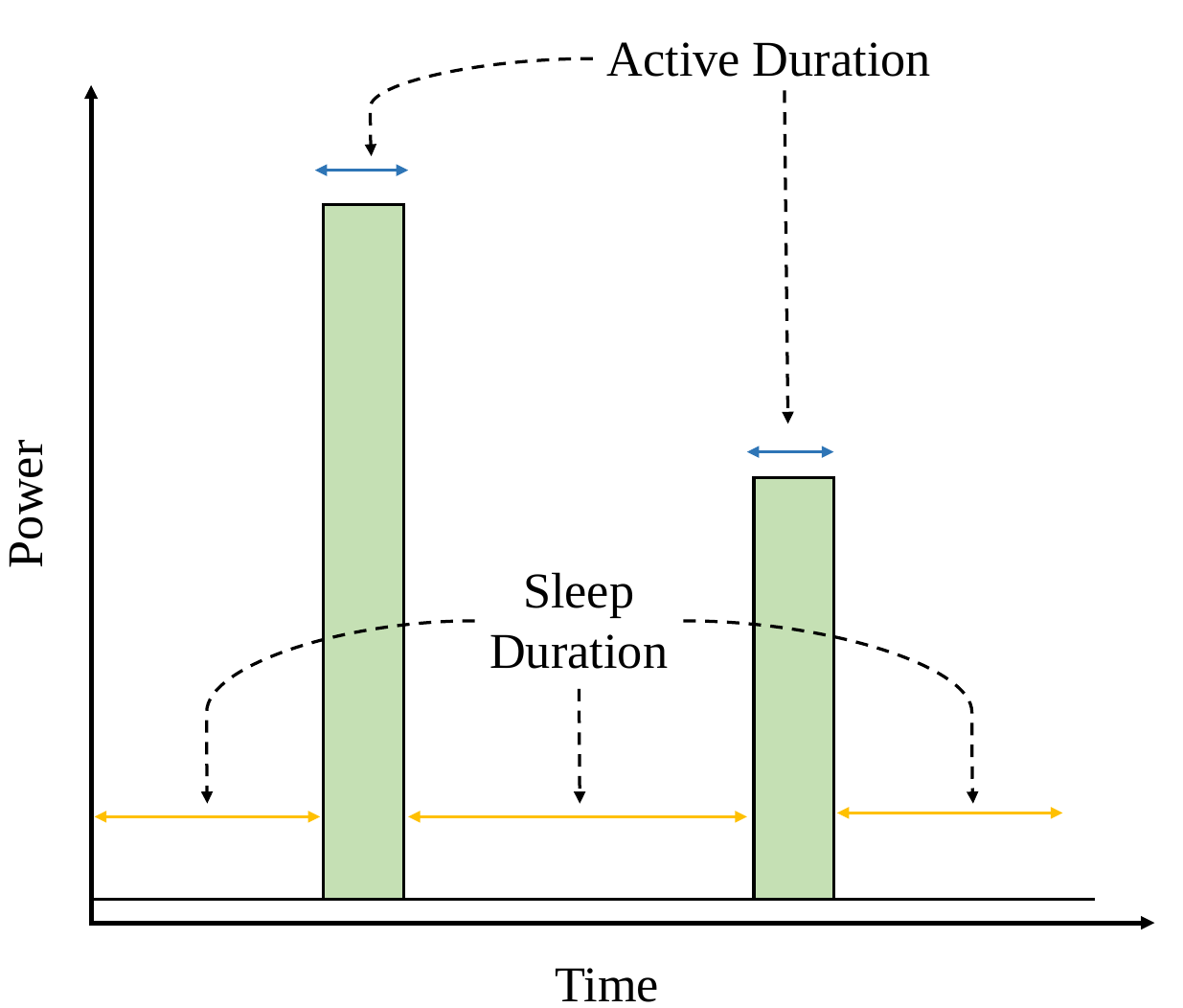}
	\centering
	\caption{Duty Cycle Based Power Consumption Optimization}
	\label{duty-cycle}
\end{figure}

One of the most effective approaches to reduce energy consumption and task scheduling in state-of-the-art energy-autonomous devices is implementing the duty cycling technique, which is widely recognized and employed. It involves periodically turning off the device to conserve energy when it is not actively used. The functioning of this technique has been graphically presented in Fig. \ref{duty-cycle}.
This technique enables the device to switch between different energy-saving modes. Primarily, it keeps the device in active mode for a short period, during which it executes energy-consuming tasks such as wireless communication, sensing, and control actions. Subsequently, the device goes into low-power modes for the rest of the time in an operation cycle. The device usually consumes power in the milli range when it is active and in the micro range when it is in low-power or sleep mode. Solar energy powered battery-driven \ac{LoRaWAN} based \ac{IoT} devices proposed in \cite{ref57} -- \cite{ref60}. implements the duty cycle technique for power consumption. On the other hand, battery-less \ac{IoT} devices proposed in \cite{ref61}, \cite{ref62} have incorporated the aforementioned technique for energy optimization and scheduling. \\

However, the duty cycle in state-of-the-art energy-autonomous \ac{IoT} devices is configured considering various factors such as number of tasks, energy consumption of tasks, and the remaining energy of the node. The common major drawback of these works is the variability of harvested energy caused by changes in environmental conditions, affecting ambient energy availability. Thus, the duty cycle of a node should be adjusted based on the amount of energy it receives to effectively utilize the energy for future tasks. The sleep period of a node can be regulated based on the amount of energy that will be harvested in the future in order to strategically manage the energy consumption for task execution. \\

\noindent Thus, duty cycle has been devided into two categories as outlined below.
\begin{itemize}
	\item Energy Budget-Based Duty Cycling
	\item Predicted Harvested Energy-Based Duty Cycling
\end{itemize}

\subsubsection{Energy Budget-Based Duty Cycling} The duty cycling mechanism in a sensor network relies on factors such as the amount of harvested energy, energy consumed, and the distance between nodes and the data receiver. In a typical \ac{IoT} system, nodes close to the central point often experience faster resource depletion due to the additional workload of relaying data from distant nodes through multiple hops. However, the nodes with energy harvesting capability can sustain and operate continuously. The duty cycle of a sensor node is determined by the average amount of energy harvested and energy consumed in both high power mode or active mode and low power mode or sleep mode. The duty cycle should be tuned to ensure the total energy consumption does not surpass the harvested energy. Correspondingly, harvesting more energy elevates the duty cycle to enhance performance while staying within the allocated energy budget. \\

In \cite{ref63}, a model for determining the appropriate size of an energy storage unit for sustainable embedded systems has been proposed. An optimization scheme based on the duty cycle technique has been implemented on an embedded device to validate its performance experimentally. Furthermore, a mathematical model is introduced for optimizing the power consumption and task scheduling in sensor nodes based on the available energy \cite{ref64}. An emperical model has been introduced for maximizing performance improvement and task scheduling in sensor nodes, considering the available energy. The model optimizes system performance by adapting to the dynamics of the renewable energy source, enabling energy-neutral operation. In this work, the harvestable energy in the future has been predicted using the \ac{EWMA} technique. In \cite{ref65}, a duty cycle-based \ac{MAC} protocol has been proposed to optimize energy usage in cooperative \ac{WSN}s. The proposed algorithm can schedule the active period and sleep period of the sensor node considering the remaining energy of the node, and data needs to determine when they should be active or in sleep mode. \\

Subsequently, duty cycle technique has been used in determining the optimal sensing scheduling policy for a sensing system with energy harvesting capability equipped with limited size battery \cite{ref66}. The proposed work dynamically chooses the next sensing epoch based on the battery level at the current sensing epoch. Furthermore, energy allocation schemes for energy harvesting-enabled sensing devices have been proposed to maximize the utilization of periodically harvested solar energy and minimize fluctuations in energy allocation \cite{ref68}. On the other hand, a novel online sensing scheduling policy has been proposed for an energy harvesting-enabled sensing system considering both finite and infinite battery cases \cite{ref67}. The policy selects the subsequent sensing epoch based on the battery level during the current sensing epoch. A duty cycling technique-based event-driven strategy is proposed for efficient power management in a device deployed on the roadside \cite{ref69}. The system harvests energy from sunlight and sends data packets using \ac{EDF} algorithm, depending on the traffic flow. The proposed approach results in a significant reduction in power consumption, implying maximizing the lifetime of the device.

\subsubsection{Predicted Harvested Energy-Based Duty Cycling}
\begin{figure}[h!]
	\includegraphics[scale=0.5]{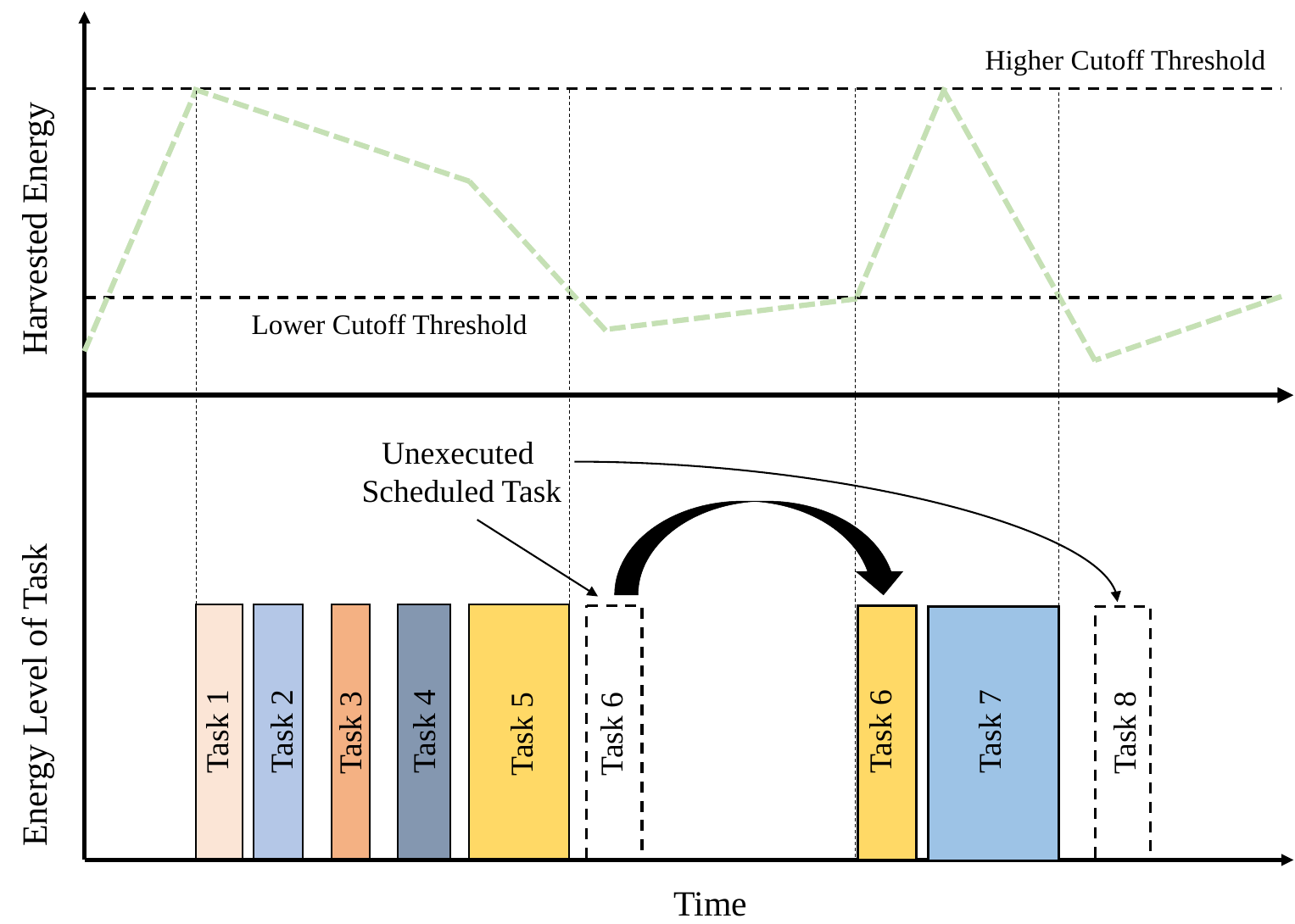}
	\centering
	\caption{Predicted Energy Based Duty Cycling for Power Consumption Optimization}
	\label{predicted-duty-cycle}
\end{figure}

The amount of harvested energy has often been reported as insufficient to enable the energy harvesting devices to execute high energy-consuming tasks. Thus, estimating the amount of harvestable energy in the future is crucial for effectively managing tasks during periods of energy scarcity. This allows the devices to schedule the tasks considering forecasted energy, ensuring deadlines are met without interruptions. Fig. \ref{predicted-duty-cycle} illustrates the working principle of the technique. The first five tasks are carried out when there is sufficient energy available. However, even though task 6 is scheduled, the level of harvested energy falls below the lower cutoff threshold. Therefore, it has been delayed until the energy level from harvesting is adequate. Since the scheme can predict future energy availability, task 6 is scheduled based on the predicted energy. Implementation of this approach results in more efficient energy utilization and a reduction in the number of missed task deadlines. \\

A fuzzy logic-based adaptive duty cycle-based task scheduling strategy has been developed \cite{ref70}. The proposed approach considers three input parameters such as current residual energy, predicted harvesting energy and predicted residual energy parameters for estimating the duty cycle of a sensor node. A prediction model has been used to forecast future available energy. The remaining energy has been estimated for future time slots depending on forecasted energy and energy consumption of the tasks. The amount of harvestable energy can be predicted using various techniques, including \ac{ML}, \ac{DL}, and statistical techniques \cite{syam}, \cite{mohantyifip}, \cite{mohanty1}. Further, a scheduling framework has been proposed that considers real-world dynamics to optimize long-term tracking performance, considering both energy and mobility factors \cite{ref71}. \\

The framework utilizes \ac{EWMA} filter to calculate a virtual energy budget for the remaining forecast period, enabling accurate energy predictions. Further, the proposed information-based \ac{GPS} sampling approach utilizes this virtual energy budget to estimate the current tracking error through dead-reckoning. It then schedules a new \ac{GPS} sample when the error surpasses a threshold. In \cite{ref72}, attempts have been made to efficiently schedule non-uniform samples, both in terms of time and space. It is essential to consider the estimated energy that will be harvested when developing a sampling method. It is particularly crucial when the estimated energy is low. This algorithm also considers the information from neighboring sources when determining the duty cycle. It enhances the duty cycle of nodes in the area surrounding a location where an event has taken place. The other nodes operate at their usual duty cycle to save power. A prediction-based adaptive duty cycle \ac{MAC} protocol has been proposed for energy harvesting enabled \ac{WSN} devices \cite{ref73}. \ac{NARNET} has been used to forecast future harvestable energy based on historical solar irradiance. The proposed approach efficiently manages the available energy through a duty cycle adjustment scheme. Further, a \ac{MAC} protocol based on a prediction-based adaptive duty cycle mechanism has been proposed \cite{ref74}. The adaptive duty cycle is determined by the predicted amount of energy harvested and the geographical distance to the cluster head or sink.

\section{Open Research Direction in Smart Consumer Lighting}
\label{or}
This section discusses the open research direction in smart consumer light management systems. These directions are outlined below.
\begin{itemize}
	\item AI techniques for forecasting traffic patterns and alleviating the load on lighting systems are expected to be a significant future trend. Various advanced \ac{AI} techniques, including \ac{DL} and \ac{ML}, can be implemented on cloud-based, cloud-edge-based, or cloud-fog-edge-based frameworks to evaluate real-time data from sensors and cameras to forecast traffic patterns and modify lighting parameters accordingly \cite{FS-3}.
	\item The increasing deployment of smart technologies emphasizes the critical significance of cybersecurity in smart lighting systems. Thus, integrating cybersecurity into the design of future smart lighting systems is a prominent trend. Securing smart lighting systems against malicious cyber attacks such as denial of service or unauthorized control is essential for safeguarding critical infrastructure, averting outages, and reducing the risks of criminal activities in a smart city environment \cite{FS-2}. Stringent encryption methods and comprehensive authentication mechanisms can be used to protect consumer light management systems against cyber-attacks \cite{FS-1}.
	\item Exploring energy harvesting sources such as vibration and wind for powering management systems integrated with street lighting presents a promising avenue for research. It has the potential to yield more sustainable and economically viable solutions for consumer light management systems.
	\item The battery serves as the primary energy storage component in off-grid street lights; hence, developing a battery health monitoring system for these lights presents a compelling research opportunity \cite{FS-5}. This monitoring system may facilitate the detection of possible battery concerns, such as degeneration or malfunction, prior to a total failure of the street light. Continuous monitoring of battery health enables proactive repair scheduling, minimizing downtime and assuring the operational status of street lights. Implementing a battery health monitoring system in off-grid street lighting might significantly enhance reliability and efficiency.
\end{itemize}

\section{Conclusion}
\label{concl}
Smart consumer lighting management serves a significant part in the development of smart cities. Over the last decade, it has been shown as a solution that enhances safety, security, and energy efficiency for consumers in public places and indoor environments. Despite ongoing research and development aimed at enhancing various facets of this application, more than 80\% of conventional public lights are yet to be replaced with smart solutions. In recent times, vigorous efforts have been undertaken to incorporate \ac{IoT} compatibility into consumer lighting. The pursuit of sustainability in \ac{IoT} devices has garnered worldwide interest since most \ac{IoT} devices rely on batteries for power. These batteries have been identified as the primary source of carbon emissions from \ac{IoT} nodes, resulting in several detrimental environmental impacts. This manuscript presents a comprehensive review of consumer lighting technology and emphasizes the importance of sustainability in it. The primary distinction of this manuscript lies in reviewing the essential technologies involved in featuring sustainability in consumer light management systems. It provides a comprehensive resource for researchers and industry experts working on consumer light management systems, addressing notable gaps in the current literature and providing a significant reference for the academic community. \\

The advancement in \ac{ICT} and Industry 5.0 is expected to introduce significant evolution in smart consumer light management, making it more adaptive, user-friendly, and sustainable, with an emphasis on human-centric technology. It focuses on customization, enabling lighting systems to adjust to individual preferences, needs, and even moods. It may signify lighting that modulates lighting parameters according to user behavior and preferences, improving well-being and productivity. In addition, the integration of \ac{AI} in smart consumer light management can ensure light management according to individual preferences for visual comfort, minimizing eye fatigue by implementing task-based illumination control for particular tasks such as reading, working, or relaxing according to the time of day or specific circumstances. The emphasis on sustainability in Industry 5.0 is going to shape the choice of materials for lighting products, leading manufacturers to adopt recyclable and eco-friendly materials, along with manufacturing processes designed to reduce environmental impact.

\bibliography{SampleReferences}

\begin{thebibliography}{100}

\bibitem{intro-2}
Saraju~P. Mohanty, Uma Choppali, and Elias Kougianos.
\newblock Everything you wanted to know about smart cities: The internet of
  things is the backbone.
\newblock {\em IEEE Consumer Electronics Magazine}, 5(3):60--70, 2016.

\bibitem{intro-1}
C.~Harrison, B.~Eckman, R.~Hamilton, P.~Hartswick, J.~Kalagnanam,
  J.~Paraszczak, and P.~Williams.
\newblock Foundations for smarter cities.
\newblock {\em IBM Journal of Research and Development}, 54(4):1--16, 2010.

\bibitem{motivation}
Expert~Market Research.
\newblock Global smart lighting market report and forecast 2024-2032, 2023.

\bibitem{ref93}
Prajnyajit Mohanty, Umesh~C. Pati, Kamalakanta Mahapatra, and Saraju~P.
  Mohanty.
\newblock bslight: Battery-less energy autonomous street light management
  system for smart city.
\newblock {\em IEEE Transactions on Sustainable Computing}, 9(1):100--114,
  2024.

\bibitem{ref94}
Prajnyajit Mohanty, Umesh~C. Pati, Kamalakanta Mahapatra, and Saraju~P.
  Mohanty.
\newblock bslight 2.0: Battery-free sustainable smart street light management
  system.
\newblock {\em IEEE Transactions on Sustainable Computing}, pages 1--15, 2024.

\bibitem{ref118}
Siwar Khemakhem and Lotfi Krichen.
\newblock A comprehensive survey on an iot-based smart public street lighting
  system application for smart cities.
\newblock {\em Franklin Open}, page 100142, 2024.

\bibitem{ref114}
Fouad Agramelal, Mohamed Sadik, Youssef Moubarak, and Saad Abouzahir.
\newblock Smart street light control: A review on methods, innovations, and
  extended applications.
\newblock {\em Energies}, 16(21), 2023.

\bibitem{ref116}
Moabi~K. Manyake and Tebello~N.D. Mathaba.
\newblock An internet of things framework for control and monitoring of smart
  public lighting systems: A review.
\newblock In {\em 2022 International Conference on Artificial Intelligence, Big
  Data, Computing and Data Communication Systems (icABCD)}, pages 1--9, 2022.

\bibitem{ref119}
Amjad Omar, Sara AlMaeeni, Hussain Attia, Maen Takruri, Ahmed Altunaiji, Mihai
  Sanduleanu, Raed Shubair, Moh’d~Sami Ashhab, Maryam Al~Ali, and Ghaya
  Al~Hebsi.
\newblock Smart city: Recent advances in intelligent street lighting systems
  based on iot.
\newblock {\em Journal of Sensors}, 2022(1):5249187, 2022.

\bibitem{ref117}
Eric H.~Grosse Marc~Füchtenhans and Christoph~H. Glock.
\newblock Smart lighting systems: state-of-the-art and potential applications
  in warehouse order picking.
\newblock {\em International Journal of Production Research},
  59(12):3817--3839, 2021.

\bibitem{ref115}
Mohsen Mahoor, Zohreh~S. Hosseini, Amin Khodaei, Aleksi Paaso, and Daniel
  Kushner.
\newblock State-of-the-art in smart streetlight systems: a review.
\newblock {\em IET Smart Cities}, 2(1):24--33, 2020.

\bibitem{ref121}
Héctor~F. Chinchero, J.~Marcos Alonso, and Hugo Ortiz~T.
\newblock Led lighting systems for smart buildings: a review.
\newblock {\em IET Smart Cities}, 2(3):126--134, 2020.

\bibitem{ref122}
Marufa~Yeasmin Mukta, Md~Arafatur Rahman, A.~Taufiq Asyhari, and Md~Zakirul
  {Alam Bhuiyan}.
\newblock Iot for energy efficient green highway lighting systems: Challenges
  and issues.
\newblock {\em Journal of Network and Computer Applications}, 158:102575, 2020.

\bibitem{ref120}
Ivan Chew, Dilukshan Karunatilaka, Chee~Pin Tan, and Vineetha Kalavally.
\newblock Smart lighting: The way forward? reviewing the past to shape the
  future.
\newblock {\em Energy and Buildings}, 149:180--191, 2017.

\bibitem{ref1}
Gul Shahzad, Heekwon Yang, Arbab~Waheed Ahmad, and Chankil Lee.
\newblock Energy-efficient intelligent street lighting system using
  traffic-adaptive control.
\newblock {\em IEEE Sensors Journal}, 16(13):5397--5405, 2016.

\bibitem{ref2}
Philip~Tobianto Daely, Haftu~Tasew Reda, Gandeva~Bayu Satrya, Jin~Woo Kim, and
  Soo~Young Shin.
\newblock Design of smart led streetlight system for smart city with web-based
  management system.
\newblock {\em IEEE Sensors Journal}, 17(18):6100--6110, 2017.

\bibitem{ref3}
Dipanjan Saha, Sk~Mahammad Sorif, and Pallav Dutta.
\newblock Weather adaptive intelligent street lighting system with automatic
  fault management using boltuino platform.
\newblock In {\em 2021 International Conference on ICT for Smart Society
  (ICISS)}, pages 1--6, 2021.

\bibitem{ref4}
Aziera Abdullah, Siti~Hajar Yusoff, Syasya~Azra Zaini, Nur~Shahida Midi, and
  Sarah~Yasmin Mohamad.
\newblock Smart street light using intensity controller.
\newblock In {\em 2018 7th International Conference on Computer and
  Communication Engineering (ICCCE)}, pages 1--5, 2018.

\bibitem{ref5}
Sk~Mahammad Sorif, Dipanjan Saha, and Pallav Dutta.
\newblock Smart street light management system with automatic brightness
  adjustment using bolt iot platform.
\newblock In {\em 2021 IEEE International IOT, Electronics and Mechatronics
  Conference (IEMTRONICS)}, pages 1--6, 2021.

\bibitem{ref6}
F.~Sanchez-Sutil and A.~Cano-Ortega.
\newblock Smart regulation and efficiency energy system for street lighting
  with lora lpwan.
\newblock {\em Sustainable Cities and Society}, 70:102912, 2021.

\bibitem{ref7}
Yuxi Jiang, Yinghong Shuai, Xiaoliang He, Xing Wen, and Liangliang Lou.
\newblock An energy-efficient street lighting approach based on traffic
  parameters measured by wireless sensing technology.
\newblock {\em IEEE Sensors Journal}, 21(17):19134--19143, 2021.

\bibitem{ref8}
Yao-Chung Chang and Ying-Hsun Lai.
\newblock Campus edge computing network based on iot street lighting nodes.
\newblock {\em IEEE Systems Journal}, 14(1):164--171, 2020.

\bibitem{ref9}
Tanumay Halder and Biswanath Roy.
\newblock Design of a weather adoptive and traffic adoptive wireless led street
  light control scheme.
\newblock In {\em 2024 IEEE 3rd International Conference on Control,
  Instrumentation, Energy and Communication (CIEC)}, pages 186--191, 2024.

\bibitem{ref15}
Natthaporn Saokaew, Nuttapol Kitsatit, Thanadol Yongkunawut,
  Piyasawat~Navaratana Na~Ayudhya, Ekkachai Mujjalinvimut, Tirasak Sapaklom,
  Preecha Aregarot, and Jakkrit Kunthong.
\newblock Smart street lamp system using lorawan and artificial intelligence
  part i.
\newblock In {\em 2021 9th International Electrical Engineering Congress
  (iEECON)}, pages 189--192, 2021.

\bibitem{ref10}
Shichao Chen, Gang Xiong, Jia Xu, Shuangshuang Han, Fei-Yue Wang, and Kun Wang.
\newblock The smart street lighting system based on nb-iot.
\newblock In {\em 2018 Chinese Automation Congress (CAC)}, pages 1196--1200,
  2018.

\bibitem{ref11}
Francisco~José Bellido-Outeiriño, Francisco~Javier Quiles-Latorre,
  Carlos~Diego Moreno-Moreno, José~María Flores-Arias, Isabel Moreno-García,
  and Manuel Ortiz-López.
\newblock Streetlight control system based on wireless communication over dali
  protocol.
\newblock {\em Sensors}, 16(5), 2016.

\bibitem{ref12}
Fabio Leccese.
\newblock Remote-control system of high efficiency and intelligent street
  lighting using a zigbee network of devices and sensors.
\newblock {\em IEEE Transactions on Power Delivery}, 28(1):21--28, 2013.

\bibitem{ref13}
Fabio Leccese, Marco Cagnetti, and Daniele Trinca.
\newblock A smart city application: A fully controlled street lighting isle
  based on raspberry-pi card, a zigbee sensor network and wimax.
\newblock {\em Sensors}, 14(12):24408--24424, 2014.

\bibitem{ref14}
Lian Yongsheng, Lin Peijie, and Cheng Shuying.
\newblock Remote monitoring and control system of solar street lamps based on
  zigbee wireless sensor network and gprs.
\newblock In Wensong Hu, editor, {\em Electronics and Signal Processing}, pages
  959--967, 2011.

\bibitem{ref16}
Ravikishore Kodali and Subbachary Yerroju.
\newblock Energy efficient smart street light.
\newblock In {\em 2017 3rd International Conference on Applied and Theoretical
  Computing and Communication Technology (iCATccT)}, pages 190--193, 2017.

\bibitem{ref17}
Anita Gehlot, Sultan~S. Alshamrani, Rajesh Singh, Mamoon Rashid, Shaik~Vaseem
  Akram, Ahmed~Saeed AlGhamdi, and Fahad~R. Albogamy.
\newblock Internet of things and long-range-based smart lampposts for
  illuminating smart cities.
\newblock {\em Sustainability}, 13(11), 2021.

\bibitem{ref18}
Jinsung Byun and Taehwan Shin.
\newblock Design and implementation of an energy-saving lighting control system
  considering user satisfaction.
\newblock {\em IEEE Transactions on Consumer Electronics}, 64(1):61--68, 2018.

\bibitem{ref19}
Samuel Jia~Wei Tang, Vineetha Kalavally, Kok~Yew Ng, Chee~Pin Tan, and Jussi
  Parkkinen.
\newblock Real-time closed-loop color control of a multi-channel luminaire
  using sensors onboard a mobile device.
\newblock {\em IEEE Access}, 6:54751--54759, 2018.

\bibitem{ref20}
Yoonsik Uhm, Insung Hong, Gwanyeon Kim, Byoungjoo Lee, and Sehyun Park.
\newblock Design and implementation of power-aware led light enabler with
  location-aware adaptive middleware and context-aware user pattern.
\newblock {\em IEEE Transactions on Consumer Electronics}, 56(1):231--239,
  2010.

\bibitem{ref21}
Jinsung Byun, Insung Hong, Byoungjoo Lee, and Sehyun Park.
\newblock Intelligent household led lighting system considering energy
  efficiency and user satisfaction.
\newblock {\em IEEE Transactions on Consumer Electronics}, 59(1):70--76, 2013.

\bibitem{ref22}
Francisco~Jose Bellido-Outeirino, Jose~Maria Flores-Arias, Francisco
  Domingo-Perez, Aurora Gil-de Castro, and Antonio Moreno-Munoz.
\newblock Building lighting automation through the integration of dali with
  wireless sensor networks.
\newblock {\em IEEE Transactions on Consumer Electronics}, 58(1):47--52, 2012.

\bibitem{ref23}
Chun-Te Lee, Liang-Bi Chen, Huan-Mei Chu, and Che-Jen Hsieh.
\newblock Design and implementation of a leader-follower smart office lighting
  control system based on iot technology.
\newblock {\em IEEE Access}, 10:28066--28079, 2022.

\bibitem{ref24}
Michele Magno, Tommaso Polonelli, Luca Benini, and Emanuel Popovici.
\newblock A low cost, highly scalable wireless sensor network solution to
  achieve smart led light control for green buildings.
\newblock {\em IEEE Sensors Journal}, 15(5):2963--2973, 2015.

\bibitem{ref148}
Yen~Kheng Tan, Truc~Phuong Huynh, and Zizhen Wang.
\newblock Smart personal sensor network control for energy saving in dc grid
  powered led lighting system.
\newblock {\em IEEE Transactions on Smart Grid}, 4(2):669--676, 2013.

\bibitem{ref145}
Zoltán Nagy, Fah~Yik Yong, Mario Frei, and Arno Schlueter.
\newblock Occupant centered lighting control for comfort and energy efficient
  building operation.
\newblock {\em Energy and Buildings}, 94:100--108, 2015.

\bibitem{ref146}
Jorge Higuera, Wim Hertog, Mariano Perálvarez, Jose Polo, and Josep Carreras.
\newblock Smart lighting system iso/iec/ieee 21451 compatible.
\newblock {\em IEEE Sensors Journal}, 15(5):2595--2602, 2015.

\bibitem{ref138}
Sei~Ping Lau, Geoff~V. Merrett, Alex~S. Weddell, and Neil~M. White.
\newblock A traffic-aware street lighting scheme for smart cities using
  autonomous networked sensors.
\newblock {\em Computers and Electrical Engineering}, 45:192--207, 2015.

\bibitem{ref147}
Ivan Chew, Vineetha Kalavally, Naing~Win Oo, and Jussi Parkkinen.
\newblock Design of an energy-saving controller for an intelligent led lighting
  system.
\newblock {\em Energy and Buildings}, 120:1--9, 2016.

\bibitem{ref141}
Gul Shahzad, Heekwon Yang, Arbab~Waheed Ahmad, and Chankil Lee.
\newblock Energy-efficient intelligent street lighting system using
  traffic-adaptive control.
\newblock {\em IEEE Sensors Journal}, 16(13):5397--5405, 2016.

\bibitem{ref142}
Selcuk Atis and Nazmi Ekren.
\newblock Development of an outdoor lighting control system using expert
  system.
\newblock {\em Energy and Buildings}, 130:773--786, 2016.

\bibitem{ref140}
Eveliina Juntunen, Esa-Matti Sarjanoja, Juho Eskeli, Henrika Pihlajaniemi, and
  Toni Österlund.
\newblock Smart and dynamic route lighting control based on movement tracking.
\newblock {\em Building and Environment}, 142:472--483, 2018.

\bibitem{ref144}
Enrico Petritoli, Fabio Leccese, Stefano Pizzuti, and Francesco Pieroni.
\newblock Smart lighting as basic building block of smart city: An energy
  performance comparative case study.
\newblock {\em Measurement}, 136:466--477, 2019.

\bibitem{ref143}
Marina Bonomolo, Simone Ferrari, and Gaetano Zizzo.
\newblock Assessing the electricity consumption of outdoor lighting systems in
  the presence of automatic control: The ol-bac factors method.
\newblock {\em Sustainable Cities and Society}, 54:102009, 2020.

\bibitem{ref139}
Nikolaos Sifakis, Konstantinos Kalaitzakis, and Theocharis Tsoutsos.
\newblock Integrating a novel smart control system for outdoor lighting
  infrastructures in ports.
\newblock {\em Energy Conversion and Management}, 246:114684, 2021.

\bibitem{stat-3}
Knud~Lasse Lueth.
\newblock Winning in iot: How the enterprise iot market is evolving, aug 2023.

\bibitem{eu-res}
EnABLES.
\newblock Up to 78 million batteries will be discarded daily by 2025,
  researchers warn, jul 2021.

\bibitem{stat-1}
Mark Kenber.
\newblock The big switch: Why it’s time to scale up led street lighting, Sep
  2015.

\bibitem{stat-2}
William Ankréus.
\newblock The global smart street lighting market, nov 2023.

\bibitem{ref111}
Zeeshan Kaleem, Tae~Min Yoon, and Chankil Lee.
\newblock Energy efficient outdoor light monitoring and control architecture
  using embedded system.
\newblock {\em IEEE Embedded Systems Letters}, 8(1):18--21, 2016.

\bibitem{ref112}
Kamiyabhusain Patel, Jayagn Modh, Meet Pandya, Mohit Bhagchandani, and
  Faizmahammad Masi.
\newblock Smart solar street light.
\newblock In {\em 2018 International Conference on Current Trends towards
  Converging Technologies (ICCTCT)}, pages 1--4, 2018.

\bibitem{ref113}
M.~Muhamad and M~I~Md Ali.
\newblock Iot based solar smart led street lighting system.
\newblock In {\em TENCON 2018 - 2018 IEEE Region 10 Conference}, pages
  1801--1806, 2018.

\bibitem{mohanty2}
Prajnyajit Mohanty, Umesh~Chandra Pati, and Kamalakanta Mahapatra.
\newblock Enslight: Energy autonomous lorawan-based, iot-enabled, real-time
  street light management system for smart cities and smart villages.
\newblock In {\em Artificial Intelligence Techniques for Sustainable
  Development}, pages 114--139. CRC Press, 2024.

\bibitem{ref75}
Prajnyajit Mohanty, Umesh~C. Pati, and Kamalakanta Mahapatra.
\newblock Self-powered intelligent street light management system for smart
  city.
\newblock In {\em 2021 IEEE 18th India Council International Conference
  (INDICON)}, pages 1--6, 2021.

\bibitem{scavange}
Manos~M. Tentzeris, Apostolos Georgiadis, and Luca Roselli.
\newblock Energy harvesting and scavenging [scanning the issue].
\newblock {\em Proceedings of the IEEE}, 102(11):1644--1648, 2014.

\bibitem{scavange-2}
Uwakwe~C. Chukwu and Satish~M. Mahajan.
\newblock Harvesting vibration energy from vehicle suspension system for
  mileage improvement.
\newblock In {\em 2018 IEEE Power and Energy Society General Meeting (PESGM)},
  pages 1--8, 2018.

\bibitem{stat-4}
Markets and Markets.
\newblock Energy harvesting system market size and growth, Jul. 2023.

\bibitem{ref25}
Amzar Omairi, Zool~H. Ismail, Kumeresan~A. Danapalasingam, and Mohd Ibrahim.
\newblock Power harvesting in wireless sensor networks and its adaptation with
  maximum power point tracking: Current technology and future directions.
\newblock {\em IEEE Internet of Things Journal}, 4(6):2104--2115, 2017.

\bibitem{ref32}
Santiago Orrego, Kourosh Shoele, Andre Ruas, Kyle Doran, Brett Caggiano, Rajat
  Mittal, and Sung~Hoon Kang.
\newblock Harvesting ambient wind energy with an inverted piezoelectric flag.
\newblock {\em Applied Energy}, 194:212--222, 2017.

\bibitem{ref33}
Zhong~Lin Wang.
\newblock On maxwell's displacement current for energy and sensors: the origin
  of nanogenerators.
\newblock {\em Materials Today}, 20(2):74--82, 2017.

\bibitem{ref34}
Emilio Sardini and Mauro Serpelloni.
\newblock Self-powered wireless sensor for air temperature and velocity
  measurements with energy harvesting capability.
\newblock {\em IEEE Transactions on Instrumentation and Measurement},
  60(5):1838--1844, 2011.

\bibitem{ref31}
Francesca {De Rossi}, Tadeo Pontecorvo, and Thomas~M. Brown.
\newblock Characterization of photovoltaic devices for indoor light harvesting
  and customization of flexible dye solar cells to deliver superior efficiency
  under artificial lighting.
\newblock {\em Applied Energy}, 156:413--422, 2015.

\bibitem{ref30}
Johnny Russo, William Ray, and Marc~S. Litz.
\newblock Low light illumination study on commercially available homojunction
  photovoltaic cells.
\newblock {\em Applied Energy}, 191:10--21, 2017.

\bibitem{ref26}
Yewon Song, Chan~Ho Yang, Seong~Kwang Hong, Sung~Joo Hwang, Jeong~Hun Kim,
  Ji~Young Choi, Seung~Ki Ryu, and Tae~Hyun Sung.
\newblock Road energy harvester designed as a macro-power source using the
  piezoelectric effect.
\newblock {\em International Journal of Hydrogen Energy}, 41(29):12563--12568,
  2016.

\bibitem{ref27}
Giuseppina Monti, Fabrizio Congedo, Paola Arcuti, and Luciano Tarricone.
\newblock Resonant energy scavenger for sensor powering by spurious emissions
  from compact fluorescent lamps.
\newblock {\em IEEE Sensors Journal}, 14(7):2347--2354, 2014.

\bibitem{ref28}
Vishwa, Zhi Ren, Paul Brochu, Miodrag Potkonjak, and Qibing Pei.
\newblock Optimizing the output of a human-powered energy harvesting system
  with miniaturization and integrated control.
\newblock {\em IEEE Sensors Journal}, 14(7):2084--2091, 2014.

\bibitem{ref35}
Hubregt~J. Visser, Adrianus C.~F. Reniers, and Jeroen A.~C. Theeuwes.
\newblock Ambient rf energy scavenging: Gsm and wlan power density
  measurements.
\newblock In {\em 2008 38th European Microwave Conference}, pages 721--724,
  2008.

\bibitem{ref36}
Syed~Tariq Shah, Kae~Won Choi, Syed~Faraz Hasan, and Min~Young Chung.
\newblock Throughput analysis of two-way relay networks with wireless energy
  harvesting capabilities.
\newblock {\em Ad Hoc Networks}, 53:123--131, 2016.

\bibitem{ref37}
Li~Xiao, JunJiang Qian, Zhenggang Lian, Zheng Long, Xiaolong Tang, Tingting
  Chen, Xuesong Du, and Liang Cao.
\newblock Rf energy powered wireless temperature sensor for monitoring
  electrical equipment.
\newblock {\em Sensors and Actuators A: Physical}, 249:276--283, 2016.

\bibitem{ref38}
Miaomiao Yuan, Li~Cheng, Qi~Xu, Weiwei Wu, Suo Bai, Long Gu, Zhe Wang, Jun Lu,
  Huanping Li, Yong Qin, Tao Jing, and Zhong~Lin Wang.
\newblock Biocompatible nanogenerators through high piezoelectric coefficient
  0.5ba(zr0.2ti0.8)o3-0.5(ba0.7ca0.3)tio3 nanowires for in-vivo applications.
\newblock {\em Advanced Materials}, 26(44):7432--7437, 2014.

\bibitem{ref39}
Brijesh Kumar and Sang-Woo Kim.
\newblock Energy harvesting based on semiconducting piezoelectric zno
  nanostructures.
\newblock {\em Nano Energy}, 1(3):342--355, 2012.

\bibitem{ref40}
Ming Yuan, Ziping Cao, Jun Luo, Jinya Zhang, and Cheng Chang.
\newblock An efficient low-frequency acoustic energy harvester.
\newblock {\em Sensors and Actuators A: Physical}, 264:84--89, 2017.

\bibitem{wind-selfpowered}
Di~Liu, Baodong Chen, Jie An, Chengyu Li, Guoxu Liu, Jiajia Shao, Wei Tang, Chi
  Zhang, and Zhong~Lin Wang.
\newblock Wind-driven self-powered wireless environmental sensors for internet
  of things at long distance.
\newblock {\em Nano Energy}, 73:104819, 2020.

\bibitem{vibration-selfpowered}
Hui Huang, Xian Li, Si~Liu, Shiyan Hu, and Ye~Sun.
\newblock Tribomotion: A self-powered triboelectric motion sensor in wearable
  internet of things for human activity recognition and energy harvesting.
\newblock {\em IEEE Internet of Things Journal}, 5(6):4441--4453, 2018.

\bibitem{wu-etal}
Taiyang Wu, Jean-Michel Redouté, and Mehmet~Rasit Yuce.
\newblock A wireless implantable sensor design with subcutaneous energy
  harvesting for long-term iot healthcare applications.
\newblock {\em IEEE Access}, 6:35801--35808, 2018.

\bibitem{ref96}
Fan Wu, Jean-Michel Redouté, and Mehmet~Rasit Yuce.
\newblock We-safe: A self-powered wearable iot sensor network for safety
  applications based on lora.
\newblock {\em IEEE Access}, 6:40846--40853, 2018.

\bibitem{ref78}
S.~R.~Jino Ramson, Walter~D. León-Salas, Zachary Brecheisen, Erika~J. Foster,
  Cliff~T. Johnston, Darrell~G. Schulze, Timothy Filley, Rahim Rahimi, Martín
  Juan Carlos~Villalta Soto, Juan A.~Lopa Bolivar, and Mauricio~Postigo
  Málaga.
\newblock A self-powered, real-time, lorawan iot-based soil health monitoring
  system.
\newblock {\em IEEE Internet of Things Journal}, 8(11):9278--9293, 2021.

\bibitem{ref76}
S.~R.~Jino Ramson, S.~Vishnu, A.~Alfred Kirubaraj, Theodoros Anagnostopoulos,
  and Adnan~M. Abu-Mahfouz.
\newblock A lorawan iot-enabled trash bin level monitoring system.
\newblock {\em IEEE Transactions on Industrial Informatics}, 18(2):786--795,
  2022.

\bibitem{ref79}
Sharafat Ali, Tyrel Glass, Baden Parr, Johan Potgieter, and Fakhrul Alam.
\newblock Low cost sensor with iot lorawan connectivity and machine
  learning-based calibration for air pollution monitoring.
\newblock {\em IEEE Transactions on Instrumentation and Measurement}, 70:1--11,
  2021.

\bibitem{ref77}
Omar~H. Kombo, Santhi Kumaran, and Alastair Bovim.
\newblock Design and application of a low-cost, low- power, lora-gsm, iot
  enabled system for monitoring of groundwater resources with energy harvesting
  integration.
\newblock {\em IEEE Access}, 9:128417--128433, 2021.

\bibitem{NFC-livestock}
Antonio Lazaro, Ramon Villarino, Merce Pacios, Marc Lazaro, Nicolau Canellas,
  David Girbau, and Beatriz Prieto-Simon.
\newblock Battery-less nfc conductivity sensor for bovine mastitis detection in
  farming 4.0.
\newblock {\em IEEE Access}, 12:45824--45838, 2024.

\bibitem{NFC-livestock2}
Marti Boada, Antonio Lazaro, Ramon Villarino, and David Girbau.
\newblock Battery-less nfc sensor for ph monitoring.
\newblock {\em IEEE Access}, 7:33226--33239, 2019.

\bibitem{NFC-livestock3}
Martí Boada, Antonio Lázaro, Ramon Villarino, and David Girbau.
\newblock Battery-less soil moisture measurement system based on a nfc device
  with energy harvesting capability.
\newblock {\em IEEE Sensors Journal}, 18(13):5541--5549, 2018.

\bibitem{NFC-livestock4}
Wan-Young Chung, Giang~Truong Le, Thang~Viet Tran, and Nam~Hoang Nguyen.
\newblock Novel proximal fish freshness monitoring using batteryless smart
  sensor tag.
\newblock {\em Sensors and Actuators B: Chemical}, 248:910--916, 2017.

\bibitem{NFC-livestock5}
Roque Torres-Sánchez, María~Teresa Martínez-Zafra, Noelia Castillejo,
  Antonio Guillamón-Frutos, and Francisco Artés-Hernández.
\newblock Real-time monitoring system for shelf life estimation of fruit and
  vegetables.
\newblock {\em Sensors}, 20(7), 2020.

\bibitem{DC-DC}
Hassan Elahi, Khushboo Munir, Marco Eugeni, Sofiane Atek, and Paolo Gaudenzi.
\newblock Energy harvesting towards self-powered iot devices.
\newblock {\em Energies}, 13(21), 2020.

\bibitem{ref81}
Xihai Zhang, Mingming Zhang, Fanfeng Meng, Yue Qiao, Suijia Xu, and Senghout
  Hour.
\newblock A low-power wide-area network information monitoring system by
  combining nb-iot and lora.
\newblock {\em IEEE Internet of Things Journal}, 6(1):590--598, 2019.

\bibitem{ref82}
Sebastian Sadowski and Petros Spachos.
\newblock Wireless technologies for smart agricultural monitoring using
  internet of things devices with energy harvesting capabilities.
\newblock {\em Computers and Electronics in Agriculture}, 172:105338, 2020.

\bibitem{ref80}
Hari Bhusal, Pankaj Khatiwada, Ajit Jha, J~Soumya, Sagar Koorapati, and
  Linga~Reddy Cenkeramaddi.
\newblock A self-powered long-range wireless iot device based on lorawan.
\newblock In {\em 2020 IEEE International Symposium on Smart Electronic Systems
  (iSES) (Formerly iNiS)}, pages 242--245, 2020.

\bibitem{ref86}
Dhouha El~Houssaini, Carlo Trigona, Abdallah Adawi, Kholoud Hamza, Roberto
  La~Rosa, Salvatore Baglio, and Olfa Kanoun.
\newblock Benchmarking a sensor-less kinetic transducer for inertial
  measurement unit for industrial iot applications.
\newblock In {\em 2022 IEEE 9th International Conference on Computational
  Intelligence and Virtual Environments for Measurement Systems and
  Applications (CIVEMSA)}, pages 1--6, 2022.

\bibitem{ref83}
Roberto~La Rosa, Catherine Dehollain, Andreas Burg, Mario Costanza, and
  Patrizia Livreri.
\newblock An energy-autonomous wireless sensor with simultaneous energy
  harvesting and ambient light sensing.
\newblock {\em IEEE Sensors Journal}, 21(12):13744--13752, 2021.

\bibitem{ref84}
Roberto {La Rosa}, Patrizia Livreri, Catherine Dehollain, Mario Costanza, and
  Carlo Trigona.
\newblock An energy autonomous and battery-free measurement system for ambient
  light power with time domain readout.
\newblock {\em Measurement}, 186:110158, 2021.

\bibitem{ref85}
Roberto~La Rosa, Catherine Dehollain, Mario Costanza, Angelo Speciale, Fabio
  Viola, and Patrizia Livreri.
\newblock A battery-free wireless smart sensor platform with bluetooth low
  energy connectivity for smart agriculture.
\newblock In {\em 2022 IEEE 21st Mediterranean Electrotechnical Conference
  (MELECON)}, pages 554--558, 2022.

\bibitem{ref88}
Stefano Calvo, Mattia Barezzi, Umberto Garlando, Roberto La~Rosa, and Danilo
  Demarchi.
\newblock An energy autonomous and battery-free plant’s electrical impedance
  measurement system.
\newblock In {\em 2024 IEEE International Symposium on Circuits and Systems
  (ISCAS)}, pages 1--4, 2024.

\bibitem{ref87}
Roberto~La Rosa, Lokman Boulebnane, Daniele Croce, Patrizia Livreri, and Ilenia
  Tinnirello.
\newblock An energy-autonomous and maintenance-free wireless sensor platform
  with lora connectivity.
\newblock In {\em 2023 12th International Conference on Renewable Energy
  Research and Applications (ICRERA)}, pages 461--464, 2023.

\bibitem{ref89}
Roberto~La Rosa, Patrizia Livreri, Danilo Demarchi, Catherine Dehollain, and
  Sandro Carrara.
\newblock Self-powered measurement system for remote monitoring of human
  metabolism.
\newblock In {\em 2022 IEEE International Symposium on Medical Measurements and
  Applications (MeMeA)}, pages 1--5, 2022.

\bibitem{ref90}
Sai Nithin~R. Kantareddy, Ian Mathews, Rahul Bhattacharyya, Ian~Marius Peters,
  Tonio Buonassisi, and Sanjay~E. Sarma.
\newblock Long range battery-less pv-powered rfid tag sensors.
\newblock {\em IEEE Internet of Things Journal}, 6(4):6989--6996, 2019.

\bibitem{ref91}
Sai Nithin~R. Kantareddy, Ian Mathews, Shijing Sun, Mariya Layurova, Janak
  Thapa, Juan-Pablo Correa-Baena, Rahul Bhattacharyya, Tonio Buonassisi,
  Sanjay~E. Sarma, and Ian~Marius Peters.
\newblock Perovskite pv-powered rfid: Enabling low-cost self-powered iot
  sensors.
\newblock {\em IEEE Sensors Journal}, 20(1):471--478, 2020.

\bibitem{ref92}
Adrian~I. Petrariu, Alexandru Lavric, Eugen Coca, and Valentin Popa.
\newblock Hybrid power management system for lora communication using renewable
  energy.
\newblock {\em IEEE Internet of Things Journal}, 8(10):8423--8436, 2021.

\bibitem{ref95}
Sergey Mileiko, Oktay Cetinkaya, Darren Mackie, Rishad Shafik, and Domenico
  Balsamo.
\newblock A teg-based non-intrusive ultrasonic system for autonomous water flow
  rate measurement.
\newblock {\em IEEE Transactions on Sustainable Computing}, 8(3):363--374,
  2023.

\bibitem{ref97}
Wai-Kong Lee, Martin J.~W. Schubert, Boon-Yaik Ooi, and Stanley Jian-Qin Ho.
\newblock Multi-source energy harvesting and storage for floating wireless
  sensor network nodes with long range communication capability.
\newblock {\em IEEE Transactions on Industry Applications}, 54(3):2606--2615,
  2018.

\bibitem{ref100}
Bradford Campbell, Branden Ghena, and Prabal Dutta.
\newblock Energy-harvesting thermoelectric sensing for unobtrusive water and
  appliance metering.
\newblock In {\em Proceedings of the 2nd International Workshop on Energy
  Neutral Sensing Systems}, ENSsys '14, page 7–12, New York, NY, USA, 2014.
  Association for Computing Machinery.

\bibitem{ref101}
Eisa Zarepour, Mahbub Hassan, Chun~Tung Chou, and Adesoji~A. Adesina.
\newblock Semon: Sensorless event monitoring in self-powered wireless
  nanosensor networks.
\newblock {\em ACM Trans. Sen. Netw.}, 13(2), may 2017.

\bibitem{ref98}
Antonino Proto, Daniele Bibbo, Martin Cerny, David Vala, Vladimir Kasik, Lukas
  Peter, Silvia Conforto, Maurizio Schmid, and Marek Penhaker.
\newblock Thermal energy harvesting on the bodily surfaces of arms and legs
  through a wearable thermo-electric generator.
\newblock {\em Sensors}, 18(6), 2018.

\bibitem{ref104}
Ambuj Varshney, Andreas Soleiman, Luca Mottola, and Thiemo Voigt.
\newblock Battery-free visible light sensing.
\newblock In {\em Proceedings of the 4th ACM Workshop on Visible Light
  Communication Systems}, VLCS '17, page 3–8, New York, NY, USA, 2017.
  Association for Computing Machinery.

\bibitem{ref103}
Dong Ma, Guohao Lan, Mahbub Hassan, Wen Hu, Mushfika~B. Upama, Ashraf Uddin,
  and Moustafa Youssef.
\newblock Solargest: Ubiquitous and battery-free gesture recognition using
  solar cells.
\newblock In {\em The 25th Annual International Conference on Mobile Computing
  and Networking}, MobiCom '19, New York, NY, USA, 2019. Association for
  Computing Machinery.

\bibitem{ref102}
Muhammad~Moid Sandhu, Sara Khalifa, Kai Geissdoerfer, Raja Jurdak, and Marius
  Portmann.
\newblock Solar: Energy positive human activity recognition using solar cells.
\newblock In {\em 2021 IEEE International Conference on Pervasive Computing and
  Communications (PerCom)}, pages 1--10, 2021.

\bibitem{ref105}
Abdelwahed Khamis, Branislav Kusy, Chun~Tung Chou, Mary-Louise McLaws, and Wen
  Hu.
\newblock Rfwash: a weakly supervised tracking of hand hygiene technique.
\newblock In {\em Proceedings of the 18th Conference on Embedded Networked
  Sensor Systems}, SenSys '20, page 572–584, New York, NY, USA, 2020.
  Association for Computing Machinery.

\bibitem{ref106}
Ju~Wang, Omid Abari, and Srinivasan Keshav.
\newblock Challenge: Rfid hacking for fun and profit.
\newblock In {\em Proceedings of the 24th Annual International Conference on
  Mobile Computing and Networking}, MobiCom '18, page 461–470, New York, NY,
  USA, 2018. Association for Computing Machinery.

\bibitem{ref107}
Swadhin Pradhan, Eugene Chai, Karthikeyan Sundaresan, Lili Qiu, Mohammad~A.
  Khojastepour, and Sampath Rangarajan.
\newblock Rio: A pervasive rfid-based touch gesture interface.
\newblock In {\em Proceedings of the 23rd Annual International Conference on
  Mobile Computing and Networking}, MobiCom '17, page 261–274, New York, NY,
  USA, 2017. Association for Computing Machinery.

\bibitem{ref108}
Guohao Lan, Weitao Xu, Dong Ma, Sara Khalifa, Mahbub Hassan, and Wen Hu.
\newblock Entrans: Leveraging kinetic energy harvesting signal for
  transportation mode detection.
\newblock {\em IEEE Transactions on Intelligent Transportation Systems},
  21(7):2816--2827, 2020.

\bibitem{ref109}
Guohao Lan, Dong Ma, Weitao Xu, Mahbub Hassan, and Wen Hu.
\newblock Capacitor-based activity sensing for kinetic-powered wearable iots.
\newblock {\em ACM Trans. Internet Things}, 1(1), mar 2020.

\bibitem{ref110}
Weitao Xu, Guohao Lan, Qi~Lin, Sara Khalifa, Mahbub Hassan, Neil Bergmann, and
  Wen Hu.
\newblock Keh-gait: Using kinetic energy harvesting for gait-based user
  authentication systems.
\newblock {\em IEEE Transactions on Mobile Computing}, 18(1):139--152, 2019.

\bibitem{ref41}
Charles Leech and Tom~J. Kazmierski.
\newblock Energy efficient multi-core processing.
\newblock {\em Electronics/Elektronika}, 18(1):3, June 2014.

\bibitem{ref42}
Rym Chéour, Sabrine Khriji, Martin Götz, Mohamed Abid, and Olfa Kanoun.
\newblock Accurate dynamic voltage and frequency scaling measurement for
  low-power microcontrollors in wireless sensor networks.
\newblock {\em Microelectronics Journal}, 105:104874, 2020.

\bibitem{ref43}
Xiangyu Li, Nijie Xie, and Xinyue Tian.
\newblock Dynamic voltage-frequency and workload joint scaling power management
  for energy harvesting multi-core wsn node soc.
\newblock {\em Sensors}, 17(2), 2017.

\bibitem{ref44}
Fakhruddin Muhammad~Mahbub {ul Islam}, Man Lin, Laurence~T. Yang, and
  Kim-Kwang~Raymond Choo.
\newblock Task aware hybrid dvfs for multi-core real-time systems using machine
  learning.
\newblock {\em Information Sciences}, 433-434:315--332, 2018.

\bibitem{ref45}
Mohammad Salehi and Alireza Ejlali.
\newblock A hardware platform for evaluating low-energy multiprocessor embedded
  systems based on cots devices.
\newblock {\em IEEE Transactions on Industrial Electronics}, 62(2):1262--1269,
  2015.

\bibitem{ref46}
Josip Zidar, Tomislav Matić, Ivan Aleksi, and Zeljko Hocenski.
\newblock Dynamic voltage and frequency scaling as a method for reducing energy
  consumption in ultra-low-power embedded systems.
\newblock {\em Electronics}, 13(5), 2024.

\bibitem{ref47}
Sabrine Khriji, Rym Chéour, and Olfa Kanoun.
\newblock Dynamic voltage and frequency scaling and duty-cycling for ultra
  low-power wireless sensor nodes.
\newblock {\em Electronics}, 11(24), 2022.

\bibitem{ref48}
Chia-Ming Wu, Ruay-Shiung Chang, and Hsin-Yu Chan.
\newblock A green energy-efficient scheduling algorithm using the dvfs
  technique for cloud datacenters.
\newblock {\em Future Generation Computer Systems}, 37:141--147, 2014.

\bibitem{ref49}
Shaobo Liu, Jun Lu, Qing Wu, and Qinru Qiu.
\newblock Harvesting-aware power management for real-time systems with
  renewable energy.
\newblock {\em IEEE Transactions on Very Large Scale Integration (VLSI)
  Systems}, 20(8):1473--1486, 2012.

\bibitem{ref50}
Mohammadamin HajiKhodaverdian, Hamed Rastaghi, Milad Saadat, and Hamed
  Shah-Mansouri.
\newblock Reinforcement learning-based task scheduling using dvfs techniques in
  mobile devices.
\newblock In {\em 2023 IEEE 34th Annual International Symposium on Personal,
  Indoor and Mobile Radio Communications (PIMRC)}, pages 1--6, 2023.

\bibitem{ref51}
Amir Yeganeh-Khaksar, Mohsen Ansari, Sepideh Safari, Sina Yari-Karin, and
  Alireza Ejlali.
\newblock Ring-dvfs: Reliability-aware reinforcement learning-based dvfs for
  real-time embedded systems.
\newblock {\em IEEE Embedded Systems Letters}, 13(3):146--149, 2021.

\bibitem{ref52}
Liang Zhao and Xin Han.
\newblock Decomposition-based scheduling for parallel real-time tasks on
  multiprocessors.
\newblock {\em Computers and Electrical Engineering}, 97:107644, 2022.

\bibitem{ref53}
Xu~Jiang, Nan Guan, Xiang Long, and Han Wan.
\newblock Decomposition-based real-time scheduling of parallel tasks on
  multicores platforms.
\newblock {\em IEEE Transactions on Computer-Aided Design of Integrated
  Circuits and Systems}, 39(10):2319--2332, 2020.

\bibitem{ref54}
Aijun Liu, Michele Pfund, and John Fowler.
\newblock Scheduling optimization of task allocation in integrated
  manufacturing system based on task decomposition.
\newblock {\em Journal of Systems Engineering and Electronics}, 27(2):422--433,
  2016.

\bibitem{ref55}
Jun Cai, Wei Liu, Zhongwei Huang, and Fei~Richard Yu.
\newblock Task decomposition and hierarchical scheduling for collaborative
  cloud-edge-end computing.
\newblock {\em IEEE Transactions on Services Computing}, pages 1--15, 2024.

\bibitem{ref56}
Andres Gomez, Lukas Sigrist, Michele Magno, Luca Benini, and Lothar Thiele.
\newblock Dynamic energy burst scaling for transiently powered systems.
\newblock In {\em 2016 Design, Automation and Test in Europe Conference and
  Exhibition (DATE)}, pages 349--354, 2016.

\bibitem{ref57}
Xihai Zhang, Mingming Zhang, Fanfeng Meng, Yue Qiao, Suijia Xu, and Senghout
  Hour.
\newblock A low-power wide-area network information monitoring system by
  combining nb-iot and lora.
\newblock {\em IEEE Internet of Things Journal}, 6(1):590--598, 2019.

\bibitem{ref60}
Sharafat Ali, Tyrel Glass, Baden Parr, Johan Potgieter, and Fakhrul Alam.
\newblock Low cost sensor with iot lorawan connectivity and machine
  learning-based calibration for air pollution monitoring.
\newblock {\em IEEE Transactions on Instrumentation and Measurement}, 70:1--11,
  2021.

\bibitem{ref61}
Fan Wu, Jean-Michel Redouté, and Mehmet~Rasit Yuce.
\newblock We-safe: A self-powered wearable iot sensor network for safety
  applications based on lora.
\newblock {\em IEEE Access}, 6:40846--40853, 2018.

\bibitem{ref62}
Mehmet~Erkan Yuksel and Huseyin Fidan.
\newblock Energy-aware system design for batteryless lpwan devices in iot
  applications.
\newblock {\em Ad Hoc Networks}, 122:102625, 2021.

\bibitem{ref63}
Aman Kansal, Dunny Potter, and Mani~B. Srivastava.
\newblock Performance aware tasking for environmentally powered sensor
  networks.
\newblock {\em SIGMETRICS Perform. Eval. Rev.}, 32(1):223–234, jun 2004.

\bibitem{ref64}
Jason Hsu, Sadaf Zahedi, Aman Kansal, Mani Srivastava, and Vijay Raghunathan.
\newblock Adaptive duty cycling for energy harvesting systems.
\newblock In {\em ISLPED'06 Proceedings of the 2006 International Symposium on
  Low Power Electronics and Design}, pages 180--185, 2006.

\bibitem{ref65}
Ons Bouachir, Adel Ben~Mnaouer, Farid Touati, and Damiano Crescini.
\newblock Eamp-aidc - energy-aware mac protocol with adaptive individual duty
  cycle for eh-wsn.
\newblock In {\em 2017 13th International Wireless Communications and Mobile
  Computing Conference (IWCMC)}, pages 2021--2028, 2017.

\bibitem{ref66}
Jing Yang, Xianwen Wu, and Jingxian Wu.
\newblock Adaptive sensing scheduling for energy harvesting sensors with finite
  battery.
\newblock In {\em 2015 IEEE International Conference on Communications (ICC)},
  pages 98--103, 2015.

\bibitem{ref68}
Dong~Kun Noh and Kyungtae Kang.
\newblock Balanced energy allocation scheme for a solar-powered sensor system
  and its effects on network-wide performance.
\newblock {\em Journal of Computer and System Sciences}, 77(5):917--932, 2011.
\newblock PMECT 2009/ICCCN 2009.

\bibitem{ref67}
Jing Yang, Xianwen Wu, and Jingxian Wu.
\newblock Optimal online sensing scheduling for energy harvesting sensors with
  infinite and finite batteries.
\newblock {\em IEEE Journal on Selected Areas in Communications},
  34(5):1578--1589, 2016.

\bibitem{ref69}
Qutaiba~Ibrahim Ali.
\newblock Event driven duty cycling: an efficient power management scheme for a
  solar-energy harvested road side unit.
\newblock {\em IET Electrical Systems in Transportation}, 6(3):222--235, 2016.

\bibitem{ref70}
Sai~Krishna Mothku and Rashmi~Ranjan Rout.
\newblock Fuzzy logic based adaptive duty cycling for sustainability in energy
  harvesting sensor actor networks.
\newblock {\em Journal of King Saud University - Computer and Information
  Sciences}, 34(1):1489--1497, 2022.

\bibitem{syam}
Cherukuri~Syam Sandeep, Prajnyajit Mohanty, and Umesh~C. Pati.
\newblock Machine learning based framework for prediction of photovoltaic
  output power.
\newblock In {\em 2023 IEEE 3rd International Conference on Sustainable Energy
  and Future Electric Transportation (SEFET)}, pages 1--6, 2023.

\bibitem{mohantyifip}
Prajnyajit Mohanty, Umesh~Chandra Pati, and Kamalakanta Mahapatra.
\newblock Deep learning based framework for forecasting solar panel output
  power.
\newblock In Deepak Puthal, Saraju Mohanty, and Baek-Young Choi, editors, {\em
  Internet of Things. Advances in Information and Communication Technology},
  pages 229--239, Cham, 2024. Springer Nature Switzerland.

\bibitem{mohanty1}
Prajnyajit Mohanty, Krityeeprava Subhadarshini, Rashmiranjan Nayak,
  Umesh~Chandra Pati, and Kamalakanta Mahapatra.
\newblock Exploring data-driven multivariate statistical models for the
  prediction of solar energy.
\newblock In {\em Computer Vision and Machine Intelligence for Renewable Energy
  Systems}, Advances in Intelligent Energy Systems, pages 85--101. Elsevier,
  2025.

\bibitem{ref71}
Philipp Sommer, Kai Geissdoerfer, Raja Jurdak, Branislav Kusy, Jiajun Liu, Kun
  Zhao, Adam McKeown, and David Westcott.
\newblock Energy- and mobility-aware scheduling for perpetual trajectory
  tracking.
\newblock {\em IEEE Transactions on Mobile Computing}, 19(3):566--580, 2020.

\bibitem{ref72}
Péter Györke and Béla Pataki.
\newblock Application of energy-harvesting in wireless sensor networks using
  predictive scheduling.
\newblock In {\em 2012 IEEE International Instrumentation and Measurement
  Technology Conference Proceedings}, pages 582--587, 2012.

\bibitem{ref73}
Sohail Sarang, Goran~M. Stojanović, Micheal Drieberg, Stevan Stankovski,
  Kishore Bingi, and Varun Jeoti.
\newblock Machine learning prediction based adaptive duty cycle mac protocol
  for solar energy harvesting wireless sensor networks.
\newblock {\em IEEE Access}, 11:17536--17554, 2023.

\bibitem{ref74}
Nivine Guler and Zied~Ben Hazem.
\newblock Eads-ehwsns: Efficient energy-based adaptive duty cycle scheme for
  energy-harvested wireless sensor networks.
\newblock In {\em 2023 IEEE 8th International Conference on Engineering
  Technologies and Applied Sciences (ICETAS)}, pages 1--6, 2023.

\bibitem{FS-3}
Safa~Ben Atitallah, Maha Driss, Wadii Boulila, and Henda~Ben Ghézala.
\newblock Leveraging deep learning and iot big data analytics to support the
  smart cities development: Review and future directions.
\newblock {\em Computer Science Review}, 38:100303, 2020.

\bibitem{FS-2}
Dong Jin, Christopher Hannon, Zhiyi Li, Pablo Cortes, Srinivasan Ramaraju,
  Patrick Burgess, Nathan Buch, and Mohammad Shahidehpour.
\newblock Smart street lighting system: A platform for innovative smart city
  applications and a new frontier for cyber-security.
\newblock {\em The Electricity Journal}, 29(10):28--35, 2016.

\bibitem{FS-1}
Maha Driss, Daniah Hasan, Wadii Boulila, and Jawad Ahmad.
\newblock Microservices in iot security: Current solutions, research
  challenges, and future directions.
\newblock {\em Procedia Computer Science}, 192:2385--2395, 2021.
\newblock Knowledge-Based and Intelligent Information \& Engineering Systems:
  Proceedings of the 25th International Conference KES2021.

\bibitem{FS-5}
Jacob John, Ganesh Kudva, and N.~S. Jayalakshmi.
\newblock Secondary life of electric vehicle batteries: Degradation, state of
  health estimation using incremental capacity analysis, applications and
  challenges.
\newblock {\em IEEE Access}, 12:63735--63753, 2024.

\end{thebibliography}
\bibliographystyle{unsrt}

\renewcommand{\IEEEiedlistdecl}{\IEEEsetlabelwidth{SONET}}
\printacronyms[sort=true]
\renewcommand{\IEEEiedlistdecl}{\relax}

\section*{Author Biography}
\vskip 0pt plus -1fil
\begin{IEEEbiography}
	[{\includegraphics[height=1.25in,width= 1in,clip,keepaspectratio]{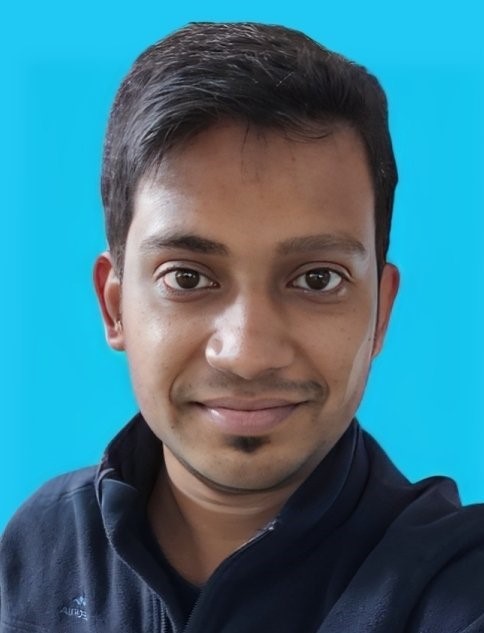}}]{Prajnyajit Mohanty}(Graduate Student Member, IEEE) received B.Tech degree in Electronics and Instrumentation Engineering from National Institute of Science and Technology, Berhampur, in 2017, and M.Tech degree in Control and Instrumentation Engineering from Veer Surendra Sai University of Technology, Burla, in 2019. He is currently pursuing Ph. D. in Electronics and Communication Engineering from National Institute of Technology Rourkela. His research interests include Energy Harvesting Systems, Internet of Things, Low Power Embedded System Design, and Machine Learning.
\end{IEEEbiography} 
\begin{IEEEbiography}
	[{\includegraphics[height=1.25in,width= 1in,clip,keepaspectratio]{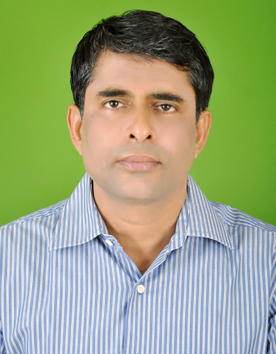}}]{Umesh C. Pati}(Senior Member, IEEE) is a full Professor at the Department of Electronics and Communication Engineering, National Institute of Technology (NIT), Rourkela. He has obtained his B.Tech. degree in Electrical Engineering from NIT, Rourkela, Odisha. He received both M.Tech. and Ph.D. degrees in Electrical Engineering with specialization in Instrumentation and Image Processing respectively from IIT, Kharagpur. His current areas of interest are Image/Video Processing, Computer Vision, Artificial Intelligence, Internet of Things (IoT), Industrial Automation, and Instrumentation Systems. He has authored/edited two books and published over 100 articles in the peer-reviewed international journals as well as conference proceedings. He is a Fellow of The Institution of Engineers (India), and Institution of Electronics and Telecommunication Engineers (IETE). He is also life member of various professional bodies like MIR Labs (USA), Indian Society for Technical Education, Computer Society of India, Instrument Society of India and Odisha Bigyan Academy.  
\end{IEEEbiography}	
\begin{IEEEbiography}
	[{\includegraphics[height=1.25in, width= 1in,clip,keepaspectratio]{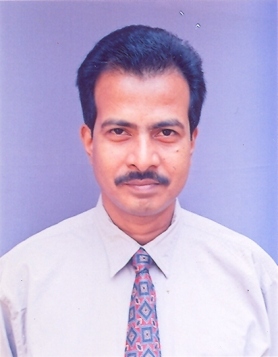}}]{Kamalakanta Mahapatra}(Senior Member, IEEE) is a Professor (HAG) in Electronics and Communication Engineering Department of National institute of Technology, Rourkela. He assumed professor position since February 2004. He obtained his B. Tech degree with Honours from Regional Engineering College, Calicut in 1985, Master’s from Regional Engineering College, Rourkela in 1989 and Ph. D. from IIT Kanpur in 2000. He is a senior member of the IEEE and a fellow of the institution of Engineers (India) in ECE Division. Presently , he is the Chairman of IEEE Rourkela Sub-section and mentor of  IEEE CTSoC chapter of Kolkata section. He has published several research papers in National and International Journals. He received coveted J. C. Bose award for best engineering oriented research in the year 2014. His research interests include Embedded Computing Systems, VLSI Design, Hardware Security and Industrial/Consumer/Power Electronics. He has supervised 24 PhD dissertations and 92 Master’s theses.
\end{IEEEbiography}
\begin{IEEEbiography} 
[{\includegraphics[height=1.25in, keepaspectratio]{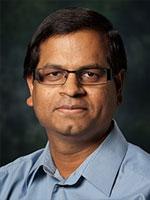}}]
{Saraju P. Mohanty} (Senior Member, IEEE) received the bachelor’s degree (Honors) in electrical engineering from the Orissa University of Agriculture and Technology, Bhubaneswar, in 1995, the master’s degree in Systems Science and Automation from the Indian Institute of Science, Bengaluru, in 1999, and the Ph.D. degree in Computer Science and Engineering from the University of South Florida, Tampa, in 2003. He is a Professor with the University of North Texas. His research is in ``Smart Electronic Systems’’ which has been funded by National Science Foundations (NSF), Semiconductor Research Corporation (SRC), U.S. Air Force, IUSSTF, and Mission Innovation. He has authored 550 research articles, 5 books, and 10 granted and pending patents. His Google Scholar h-index is 58 and i10-index is 269 with 15,000 citations. He is regarded as a visionary researcher on Smart Cities technology in which his research deals with security and energy aware, and AI/ML-integrated smart components. He introduced the Secure Digital Camera (SDC) in 2004 with built-in security features designed using Hardware Assisted Security (HAS) or Security by Design (SbD) principle. He is widely credited as the designer for the first digital watermarking chip in 2004 and first the low-power digital watermarking chip in 2006. He is a recipient of 19 best paper awards, Fulbright Specialist Award in 2021, IEEE Consumer Electronics Society Outstanding Service Award in 2020, the IEEE-CS-TCVLSI Distinguished Leadership Award in 2018, and the PROSE Award for Best Textbook in Physical Sciences and Mathematics category in 2016. He has delivered 30 keynotes and served on 15 panels at various International Conferences. He has been serving on the editorial board of several peer-reviewed international transactions/journals, including IEEE Transactions on Big Data (TBD), IEEE Transactions on Computer-Aided Design of Integrated Circuits and Systems (TCAD), IEEE Transactions on Consumer Electronics (TCE), and ACM Journal on Emerging Technologies in Computing Systems (JETC). He has been the Editor-in-Chief (EiC) of the IEEE Consumer Electronics Magazine (MCE) during 2016-2021. He served as the Chair of Technical Committee on Very Large Scale Integration (TCVLSI), IEEE Computer Society (IEEE-CS) during 2014-2018 and on the Board of Governors of the IEEE Consumer Electronics Society during 2019-2021. He serves on the steering, organizing, and program committees of several international conferences. He is the steering committee chair/vice-chair for the IEEE International Symposium on Smart Electronic Systems (IEEE-iSES), the IEEE-CS Symposium on VLSI (ISVLSI), and the OITS International Conference on Information Technology (OCIT). He has supervised 3 post-doctoral researchers, 17 Ph.D. dissertations, 28 M.S. theses, and 28 undergraduate projects.
\end{IEEEbiography}

\end{document}